\documentclass[a4paper,twocolumn,11pt]{quantumarticle}
\pdfoutput=1
\usepackage[utf8]{inputenc}
\usepackage[english]{babel}
\usepackage[T1]{fontenc}
\usepackage{amsmath}
\usepackage{hyperref}
\usepackage{tikz}

\usepackage[numbers,sort&compress]{natbib}
\usepackage{mathtools}
\usepackage{physics}
\usepackage{amssymb}
\usepackage[capitalise]{cleveref}
\usepackage[normalem]{ulem}

\usepackage{pgfplots}
\DeclareUnicodeCharacter{2212}{−}
\usepgfplotslibrary{groupplots,dateplot}
\usetikzlibrary{patterns,shapes.arrows,patterns.meta}
\pgfplotsset{compat=newest}

\newcommand{\ii}{\mathrm{i}}
\newcommand{\x}{\text{x}}
\newcommand{\y}{\text{y}}
\newcommand{\e}{\text{e}}
\newcommand{\g}{\text{g}}
\newcommand{\cold}{\text{c}}
\newcommand{\hot}{\text{h}}
\newcommand{\com}{\text{com}}
\newcommand{\ex}{\text{exp}}
\newcommand{\cycle}{\text{cycle}}
\newcommand{\osc}{\text{osc}}
\newcommand{\DAMPF}{\text{D}}

\newcommand{\Geff}{\Gamma_\text{eff}}
\newcommand{\thermal}{\text{th}}

\NewDocumentCommand{\Ht}{oo}{
    \IfValueTF{#2}%
        {\hat{H}_\text{#1}(#2)}%
        {\IfValueTF{#1}%
            {\hat{H}_\text{#1}}%
            {\hat{H}}%
        }
}
\NewDocumentCommand{\rhot}{oo}{
    \IfValueTF{#2}%
        {\hat{\rho}_\text{#1}(#2)}%
        {\IfValueTF{#1}%
            {\hat{\rho}_\text{#1}}%
            {\hat{\rho}}%
        }
}
\NewDocumentCommand{\spt}{oo}{
    \IfValueTF{#2}
        {\hat{\sigma}^+_{\text{#1}}(t)}
        {\IfValueTF{#1}
            {\hat{\sigma}^+_{\text{#1}}}
            {\hat{\sigma}^+}
        }
}
\NewDocumentCommand{\smt}{oo}{
    \IfValueTF{#2}
        {\hat{\sigma}^-_{\text{#1}}(t)}
        {\IfValueTF{#1}
            {\hat{\sigma}^-_{\text{#1}}}
            {\hat{\sigma}^-}
        }
}
\NewDocumentCommand{\szt}{o}{
    \IfValueTF{#1}
        {\hat{\sigma}_\text{z}(#1)}
        {\hat{\sigma}_\text{z}}
}
\NewDocumentCommand{\sxt}{o}{
    \IfValueTF{#1}
        {\hat{\sigma}_\text{x}(#1)}
        {\hat{\sigma}_\text{x}}
}

\begin{document}
\title{The Thermodynamic Costs of Pure Dephasing in Quantum Heat Engines: Quasistatic Efficiency at Finite Power}
\author{Raphael Weber}
\email{raphael.weber@uni-ulm.de}
\author{Susana F. Huelga}
\email{susana.huelga@uni-ulm.de}
\orcid{0000-0003-1277-8154}
\author{Martin B. Plenio}
\email{martin.plenio@uni-ulm.de}
\affiliation{Institute of Theoretical Physics \& IQST, Ulm University, 89081 Ulm, Germany}
\orcid{0000-0003-4238-8843}
\maketitle
\begin{abstract}
    Quantum heat engines are commonly believed to achieve their optimal efficiency only when operated quasi-statically. When running at finite power, however, they suffer effective friction due to the generation of coherences and transitions between energy eigenstates. It was noted that it is possible to increase the power of a quantum heat engine using external control schemes or suitable dephasing noise. Here, we investigate the thermodynamic cost associated with dephasing noise schemes using both numerical and analytical methods. Our findings unveil that the observed gain in power is generally not free of thermodynamic costs, as it involves energy costs of the control fields or heat flows between thermal and  dephasing baths. These contributions must be duly accounted for when determining the engine's overall efficiency. Interestingly, we identify a particular working regime where these costs become negligible, demonstrating that quantum heat engines can be operated at any power with an efficiency per cycle that approaches arbitrarily closely that under quasistatic operation.
\end{abstract}

\section{Introduction}
The development of thermodynamics arose from the desire to understand the limitations of heat engines, both in realistic practical settings and under ideal operation. This pursuit led to the formulation of the fundamental laws of macroscopic equilibrium thermodynamics, which have remained remarkably stable over time, particularly in regards to macroscopic systems in equilibrium. These laws are indispensable for our understanding of energy transfer, conversion, and utilisation in science and technology. For instance, the second law of thermodynamics enables us to determine the maximum possible efficiency of heat engines and refrigerators under various possible thermodynamic cycles that power for example Carnot or Otto engines.

While operation close to equilibrium is required to achieve the highest possible efficiency, it also means that the system has to be operated quasi-statically. This leads to vanishing power output. Although this mode of operation is well tailored towards the application of the laws of thermodynamics, in practical applications, machines need to operate at finite power, which leads to departures from the ideal quasi-static regime and results in friction and losses. Thus, we must aim to achieve at the same time high efficiency and high power output, and it is not immediately clear to what extent operating at finite power reduces efficiency. In this general context, Novikov \cite{Novikov1958}, Chambadal \cite{Chambadal1957}, as well as later Curzon and Ahlborn \cite{CurzonA75} analysed the impact of a finite heat conductance from the hot and cold heat baths to the work medium. They discovered a limit, now known as the Curzon-Ahlborn bound, on the attainable efficiency, which, at maximum power, remains independent of the heat conductance value and is considerably more restrictive than the Carnot bound. These bounds, firmly rooted in classical physics and applied to macroscopic heat engines, illustrate the trade-off that can exist between efficiency and power output in thermodynamic systems.

The exploration of microscopic or even atomic-scale heat engines, whose function may depend crucially on the laws of quantum mechanics, was initiated with the work of Scovil and Schulz-DuBois in \cite{Scovil1959}, which developed a thermodynamic model of the maser operation in the form of a quantum mechanical three-level system, and has spawned a vast body of work (see e.g.~\cite{Binder2018} for a recent monograph and \cite{cangemi2023quantum} for a recent review on quantum heat engines). In quantum heat engines, the impact of finite speed of operation on efficiency becomes even more important. A portion of the thermodynamic cycle, which is intended to be executed quasi-statically, corresponds to a mode of operation that meets the conditions of adiabaticity—namely, a slow change of parameters relative to the internal system dynamics, ensuring the system remains in the eigenstate of the instantaneous Hamiltonian throughout the process \cite{wang2016necessary}. A violation of adiabaticity results in the system deviating from its instantaneous eigenstate through the generation of quantum mechanical coherences and transitions to other energy levels. It is well recognised that the creation of these quantum coherences represents a source of inefficiency in the thermodynamic cycle beyond those imposed by the Curzon-Ahlborn bound \cite{kosloff2002discrete,Feldmann2003,rezek2006,wang2012,Cakmak2017}.

To address this issue, researchers have explored external interventions that enable the restoration of adiabaticity, thereby improving efficiency at fixed power or enhancing power at fixed efficiency. For instance, \cite{FeldmannKosloff2006} introduced dephasing noise in the instantaneous eigenbasis of the thermodynamic cycle of an Otto engine to suppress the build-up of coherences and thus deviations from the adiabatic trajectory. Alternatively, shortcuts to adiabaticity \cite{demirplak2003adiabatic,berry2009transitionless,PhysRevE.88.062122,CampoGooldPaternostro2014} which utilise external control fields to maintain the system on the adiabatic trajectory have been examined. Both approaches allow for faster compression and expansion strokes in the quantum heat engine, resulting in larger power at the same level of efficiency.

These types of interventions raise an important question, however: Do they entail hidden thermodynamic costs that offset some, if not all, of the gains in efficiency when analysing a quantum heat engine subject to external controls or dephasing noise? While it is recognised that such costs can exist in the case of counter-adiabatic driving \cite{zheng2016cost,campbell2017trade,abah2018performance} and in specific studies of dephasing noise due to fermionic environments without the presence of a quantum heat engine \cite{PopovicMitchisonGoold2021}, studies of isothermal heat engines suggest that no thermodynamical costs may be incurred at least in some regimes \cite{cangemi2021optimal}. Nevertheless, the question remained whether these costs are necessarily present or whether they can be made negligible for periodically working heat engines at arbitrary power.

In this work, we answer this question by analysing the thermodynamic cost of pure dephasing noise in the instantaneous eigenbasis during the operation of a quantum heat engine. To achieve this, we attach a real bosonic dephasing environment to the heat engine during its thermodynamic cycle. We employ the numerically exact simulation methods of Dissipation-Assisted Matrix Product Factorisation (DAMPF) \cite{SomozaMartyLimHuelgaPlenio2019,MascherpaSS+2020,SomozaLL+2023} and Time Evolving Density matrices using Orthogonal Polynomials Algorithm (TEDOPA) \cite{PriorCH+10,Chin2010,Tamascelli2019} to examine the internal dynamics of the work medium and of the dephasing bath, as well as their mutual interaction. This allows us to track all work and heat flows between the quantum heat engine, its heat baths and the dephasing environments. We utilise these numerical methods to validate an Ansatz that forms the basis of our analytical treatment of the observed phenomenology. On the one hand, our analysis reveals additional heat transport between the thermal baths and the dephasing bath that remains hidden when the dephasing noise is modeled with a Markovian master equation. On the other hand, we demonstrate that for a suitable choice of the parameter regime of the dephasing environment, we can reduce its thermodynamic cost to an arbitrary degree. This allows us to increase the efficiency of the heat engine at any given fixed power to approach that of a quasistatically working heat engine.

The presentation of our work proceeds as follows: \cref{sec:Quantum_thermodynamics} introduces the required quantum thermodynamical concepts and definitions and serves to fix the notation. It also introduces the quantum Otto cycle for a two-level system, which will be examined in the remainder of this work, and defines power and efficiency for this system. We then discuss the QHE with a finite cycle time, which motivates the potential for dephasing noise to reduce frictional losses and to increase the power of the QHE. In \cref{sec:model}, we present a microscopic description of the dephasing bath in our model. The dephasing bath is modeled without uncontrolled approximations, using the numerically exact TEDOPA and DAMPF methods. This approach enables us to track accurately the bath energy of the dephasing bath, which is crucial for analysing the thermodynamic cost of dephasing noise. Moving on to \cref{sec:numerical_results}, we discuss the results of our numerical simulations. These results demonstrate, for a specific set of examples, that the output power can indeed be boosted by dephasing noise. However, this enhancement comes at the expense of increased heat consumption, typically resulting in a decreased efficiency of the QHE due to additional heat exchanges with the dephasing bath. In \cref{sec:analytical_analysis}, we derive an Ansatz for the quantum state of the heat engine. Building upon this approximation, we establish both analytical lower and upper bounds for the additional heat dissipation resulting from the presence of the dephasing bath. These expressions are also confirmed through our numerical investigations. Finally, in \cref{sec:optimising_performance} we employ these analytical results to determine a suitable parameter regime for the quantum heat engine and the dephasing bath that allows for the suppression of dissipated heat such that the efficiency for any fixed power approaches arbitrarily closely the efficiency under quasi-static operation.

\section{Quantum thermodynamics}
\label{sec:Quantum_thermodynamics}

In this section, we establish fundamental concepts and notation that will be used throughout this work by introducing the basic model of a quantum heat engine (QHE), whose work medium consists of a single two-level system. In \cref{sec:Quantum_Otto_cycle}, we define an Otto cycle. After defining power and efficiency in \cref{sec:Power_and_efficiency}, we proceed with a discussion of the effect of finite driving times on the performance of the QHE in \cref{sec:Quantum_friction}.

\subsection{Quantum Otto cycle}
\label{sec:Quantum_Otto_cycle}

\begin{figure}[hbt]
    \centering
    \begin{tikzpicture}
	[scale = 1,
	round/.style={circle, draw=black!100, very thick, shade, left color=green!100, right color=green!75, shading angle=60}, 
	particle/.style={thick, shade, left color=cyan!100, right color=cyan!75, shading angle=60},
	]
	
	\draw[smooth,samples=100,domain=2.5:-2.5,line width = 1] plot(\x,{sin((\x)*1080)/10});
    
    \coordinate (system) at (0,0);
    \node[round] (system) {S};
	
    \draw[draw=white, shade, left color=blue!100, right color=blue!0] (2.75,0.75) .. controls (1.5,0) .. (2.75,-0.75);
    \draw[line width = 1.5] (2.75,0.75) .. controls (1.5,0) .. (2.75,-0.75);
    \draw[draw=white, shade, right color=red!100, left color=red!0] (-2.75,0.75) .. controls (-1.5,0) .. (-2.75,-0.75);
    \draw[line width = 1.5] (-2.75,0.75) .. controls (-1.5,0) .. (-2.75,-0.75);
    
    \node at (-2.25,-1) {\large $T_\hot$};
    \node at (2.25,-1) {\large $T_\cold$};
    \node at (-1,-0.5) {\large $\gamma_\hot(t)$};
    \node at (1,-0.5) {\large $\gamma_\cold(t)$};

	\foreach \angle in {-25,...,-25}
	{
	
	\foreach \x in {0,...,0}
	{
	\foreach \y in {2,...,2}
	{
	\foreach \R in {0.5,...,0.5}
	{
	\foreach \r in {20,...,20}
	{
	\foreach \tooth in {80,...,80}
	{
	\filldraw[rotate=\angle,color=white, shade, top color=black!60, bottom color=black!40](\x,\y) circle (\R);
	\filldraw[rotate=\angle,color=black, fill=white, thick] (\x,\y) circle (\r*\R*0.01);
	\foreach \n in {0,...,15}
	{
	\foreach \deg in {12,...,12}
	{
	\foreach \adddeg in {3,...,3}
	{
	\foreach \inneradddeg in {2,...,2}
	{
	\draw[rotate=\angle,draw=white,fill=white, thick] ({\x+1.1*\R*cos(2*\n*\deg-\adddeg)},{\y+1.1*\R*sin(2*\n*\deg-\adddeg)}) -- ({\x+\tooth*0.01*\R*cos(2*\n*\deg+\inneradddeg)},{\y+\tooth*0.01*\R*sin(2*\n*\deg+\inneradddeg)}) -- ({\x+\tooth*0.01*\R*cos(2*\n*\deg+\deg-\inneradddeg)},{\y+\tooth*0.01*\R*sin(2*\n*\deg+\deg-\inneradddeg)}) -- ({\x+1.1*\R*cos(2*\n*\deg+\deg+\adddeg)},{\y+1.1*\R*sin(2*\n*\deg+\deg+\adddeg)});
	\draw[rotate=\angle, thick] ({\x+\R*cos(2*\n*\deg-\adddeg)},{\y+\R*sin(2*\n*\deg-\adddeg)}) -- ({\x+\tooth*0.01*\R*cos(2*\n*\deg+\inneradddeg)},{\y+\tooth*0.01*\R*sin(2*\n*\deg+\inneradddeg)}) -- ({\x+\tooth*0.01*\R*cos(2*\n*\deg+\deg-\inneradddeg)},{\y+\tooth*0.01*\R*sin(2*\n*\deg+\deg-\inneradddeg)}) -- ({\x+\R*cos(2*\n*\deg+\deg+\adddeg)},{\y+\R*sin(2*\n*\deg+\deg+\adddeg)});
	\draw[rotate=\angle,thick] ({\x+\R*cos(2*\n*\deg+\deg+\adddeg)},{\y+\R*sin(2*\n*\deg+\deg+\adddeg)}) -- ({\x+\R*cos(2*\n*\deg+2*\deg-\adddeg)},{\y+\R*sin(2*\n*\deg+2*\deg-\adddeg)});
	}
	}
	}
	}
	}
	}
	}
	}
	}
	
	\draw[rotate=\angle, line width=1, <->] (-0.125,0.5).. controls(-.25,1) ..(-0.125,1.375);
	\node at (-0.5,1.125) {\large $\Ht[S][t]$};
	
	}
    
\end{tikzpicture}

\vspace{0.5cm}

\begin{tikzpicture}
	[scale = 1.35,
	particle/.style={thick, shade, left color=cyan!100, right color=cyan!75, shading angle=60},
	]
    
	\draw[very thick, <->] (-0.25,3)--(-0.25,0)--(4.75,0);
	\node at (-0.5,3) {\large $\mathcal{S}_{}$};
	\node at (4.75,-0.25) {\large $\varepsilon(t)$};
	
	\draw[very thick] (3.75,0.05)--(3.75,-0.05);
	\node at (3.75,-0.25) {\large $\varepsilon_{\hot}$};
	\draw[very thick] (1.25,0.05)--(1.25,-0.05);
	\node at (1.25,-0.25) {\large $\varepsilon_{\cold}$};
	
	\draw[very thick] (-0.2,2.5)--(-0.3,2.5);
	\node at (-0.5,2.5) {\large $\mathcal{S}_{\hot}$};
	\draw[very thick] (-0.2,1)--(-0.3,1);
	\node at (-0.5,1) {\large $\mathcal{S}_{\cold}$};
	
	\node (P1) at (3.75,1) {};
	\node (P2) at (1.25,1) {};
	\node (P3) at (1.25,2.5) {};
	\node (P4) at (3.75,2.5) {};
	
	\node at (P1)[circle,fill,inner sep=1.5pt]{};
	\node at (P2)[circle,fill,inner sep=1.5pt]{};
	\node at (P3)[circle,fill,inner sep=1.5pt]{};
	\node at (P4)[circle,fill,inner sep=1.5pt]{};
	
	\draw[blue, ultra thick, <-, shorten <=0.15cm, shorten >=0.15cm] (P2) -- (P3);
	\draw[black!60!green, ultra thick, <-, shorten <=0.15cm, shorten >=0.15cm] (P3) -- (P4);
	\draw[red, ultra thick, <-, shorten <=0.15cm, shorten >=0.15cm] (P4) -- (P1);
	\draw[black!60!green, ultra thick, <-, shorten <=0.15cm, shorten >=0.15cm] (P1) -- (P2);
	
	\node at ($ (P1)+(0,-0.375) $) {\large $P_1$};
	\node at ($ (P2)+(0,-0.375) $) {\large $P_{0,4}$};
	\node at ($ (P3)+(0,0.375) $) {\large $P_3$};
	\node at ($ (P4)+(0,0.375) $) {\large $P_2$};

    \node (PS1) at ($ (P1) + (0.75,0) $) {};
    
    \draw[thick] ($ (PS1)+(-0.375,-0.25) $)--($ (PS1)+(0.375,-0.25) $) node[left]{};
    \draw[thick] ($ (PS1)+(-0.375,0.25) $)--($ (PS1)+(0.375,0.25) $) node[left]{};
    
    \draw[particle] ($ (PS1)+(0.25,-0.1875) $) circle [x radius=0.05,y radius=0.05];
    \draw[particle] ($ (PS1)+(0,-0.1875) $) circle [x radius=0.05,y radius=0.05];
    \draw[particle] ($ (PS1)+(-0.125,-0.1875) $) circle [x radius=0.05,y radius=0.05];
    \draw[particle] ($ (PS1)+(0.125,-0.1875) $) circle [x radius=0.05,y radius=0.05];
    \draw[particle] ($ (PS1)+(-0.25,-0.1875) $) circle [x radius=0.05,y radius=0.05];
    
    \draw[particle] ($ (PS1)+(0.0625,-0.079) $) circle [x radius=0.05,y radius=0.05];
    \draw[particle] ($ (PS1)+(-0.0625,-0.079) $) circle [x radius=0.05,y radius=0.05];
    \draw[particle] ($ (PS1)+(0.1875,-0.079) $) circle [x radius=0.05,y radius=0.05];
    \draw[particle] ($ (PS1)+(-0.1875,-0.079) $) circle [x radius=0.05,y radius=0.05];
    
    \draw[particle] ($ (PS1)+(0,0.3125) $) circle [x radius=0.05,y radius=0.05];

    \node (PS1) at ($ (P2) + (-0.75,0) $) {};
    \node (shift) at (0,0.1) {};
    
    \draw[thick] ($ (PS1)+(-0.375,-0.25)+(shift) $)--($ (PS1)+(0.375,-0.25)+(shift) $) node[left]{};
    \draw[thick] ($ (PS1)+(-0.375,0.25)-(shift) $)--($ (PS1)+(0.375,0.25)-(shift) $) node[left]{};
    
    \draw[particle] ($ (PS1)+(0,-0.1875)+(shift) $) circle [x radius=0.05,y radius=0.05];
    \draw[particle] ($ (PS1)+(0.25,-0.1875)+(shift) $) circle [x radius=0.05,y radius=0.05];
    \draw[particle] ($ (PS1)+(0.125,-0.1875)+(shift) $) circle [x radius=0.05,y radius=0.05];
    \draw[particle] ($ (PS1)+(-0.25,-0.1875)+(shift) $) circle [x radius=0.05,y radius=0.05];
    \draw[particle] ($ (PS1)+(-0.125,-0.1875)+(shift) $) circle [x radius=0.05,y radius=0.05];
    
    \draw[particle] ($ (PS1)+(0.0625,-0.079)+(shift) $) circle [x radius=0.05,y radius=0.05];
    \draw[particle] ($ (PS1)+(-0.0625,-0.079)+(shift) $) circle [x radius=0.05,y radius=0.05];
    \draw[particle] ($ (PS1)+(0.1875,-0.079)+(shift) $) circle [x radius=0.05,y radius=0.05];
    \draw[particle] ($ (PS1)+(-0.1875,-0.079)+(shift) $) circle [x radius=0.05,y radius=0.05];
    
    \draw[particle] ($ (PS1)+(0,0.3125)-(shift) $) circle [x radius=0.05,y radius=0.05];

    \node (PS1) at ($ (P3) + (-0.75,0) $) {};
    \node (shift) at (0,0.1) {};
    
    \draw[thick] ($ (PS1)+(-0.375,-0.25)+(shift) $)--($ (PS1)+(0.375,-0.25)+(shift) $) node[left]{};
    \draw[thick] ($ (PS1)+(-0.375,0.25)-(shift) $)--($ (PS1)+(0.375,0.25)-(shift) $) node[left]{};
    
    \draw[particle] ($ (PS1)+(0.1875,-0.1875)+(shift) $) circle [x radius=0.05,y radius=0.05];
    \draw[particle] ($ (PS1)+(0.0625,-0.1875)+(shift) $) circle [x radius=0.05,y radius=0.05];
    \draw[particle] ($ (PS1)+(-0.1875,-0.1875)+(shift) $) circle [x radius=0.05,y radius=0.05];
    \draw[particle] ($ (PS1)+(-0.0625,-0.1875)+(shift) $) circle [x radius=0.05,y radius=0.05];
    
    \draw[particle] ($ (PS1)+(0,-0.079)+(shift) $) circle [x radius=0.05,y radius=0.05];
    \draw[particle] ($ (PS1)+(-0.125,-0.079)+(shift) $) circle [x radius=0.05,y radius=0.05];
    \draw[particle] ($ (PS1)+(0.125,-0.079)+(shift) $) circle [x radius=0.05,y radius=0.05];
    
    \draw[particle] ($ (PS1)+(0,0.3125)-(shift) $) circle [x radius=0.05,y radius=0.05];
    \draw[particle] ($ (PS1)+(0.125,0.3125)-(shift) $) circle [x radius=0.05,y radius=0.05];
    \draw[particle] ($ (PS1)+(-0.125,0.3125)-(shift) $) circle [x radius=0.05,y radius=0.05];

    \node (PS1) at ($ (P4) + (0.75,0) $) {};
    \node (shift) at (0,0) {};
    
    \draw[thick] ($ (PS1)+(-0.375,-0.25)+(shift) $)--($ (PS1)+(0.375,-0.25)+(shift) $) node[left]{};
    \draw[thick] ($ (PS1)+(-0.375,0.25)-(shift) $)--($ (PS1)+(0.375,0.25)-(shift) $) node[left]{};
    
    \draw[particle] ($ (PS1)+(0.1875,-0.1875)+(shift) $) circle [x radius=0.05,y radius=0.05];
    \draw[particle] ($ (PS1)+(0.0625,-0.1875)+(shift) $) circle [x radius=0.05,y radius=0.05];
    \draw[particle] ($ (PS1)+(-0.1875,-0.1875)+(shift) $) circle [x radius=0.05,y radius=0.05];
    \draw[particle] ($ (PS1)+(-0.0625,-0.1875)+(shift) $) circle [x radius=0.05,y radius=0.05];
    
    \draw[particle] ($ (PS1)+(0,-0.079)+(shift) $) circle [x radius=0.05,y radius=0.05];
    \draw[particle] ($ (PS1)+(-0.125,-0.079)+(shift) $) circle [x radius=0.05,y radius=0.05];
    \draw[particle] ($ (PS1)+(0.125,-0.079)+(shift) $) circle [x radius=0.05,y radius=0.05];
    
    \draw[particle] ($ (PS1)+(0,0.3125)-(shift) $) circle [x radius=0.05,y radius=0.05];
    \draw[particle] ($ (PS1)+(0.125,0.3125)-(shift) $) circle [x radius=0.05,y radius=0.05];
    \draw[particle] ($ (PS1)+(-0.125,0.3125)-(shift) $) circle [x radius=0.05,y radius=0.05];
	
\end{tikzpicture}
    \caption{Sketch of a quantum heat engine subject to a time-dependent Hamiltonian $\Ht[S][t]$ and operating between thermal baths at temperature $T_\x$ ($\x \in \{\hot, \cold\}$) with dissipative couplings $\gamma_\x(t)$ (top) and the corresponding ideal Otto cycle in the energy gap-entropy diagram (bottom). The Otto cycle consists of four strokes whose start and end points are labelled $P_i$ and to which we assign the times $\tau_i$ and the states $\rhot_{\text{S}, i}$. From $P_0$ to $P_1$ the work medium is compressed adiabatically, i.e. the energy gap increases from $\varepsilon_\cold$ to $\varepsilon_\hot$ but the populations of the energy eigenstates remain invariant. Work is extracted from the work medium in this stroke. The hot bath heats the work medium whose entropy increases from $\mathcal{S}_\cold$ to $\mathcal{S}_\hot$ between $P_1$ and $P_2$. In the expansion stroke from $P_2$ to $P_3$ work is done on the work medium. Finally, the cold bath couples to the work medium to cool it to the initial state $P_4 = P_0$ and the Otto cycle is completed.} 
    \label{fig:QHE_Otto_cycle}
\end{figure}
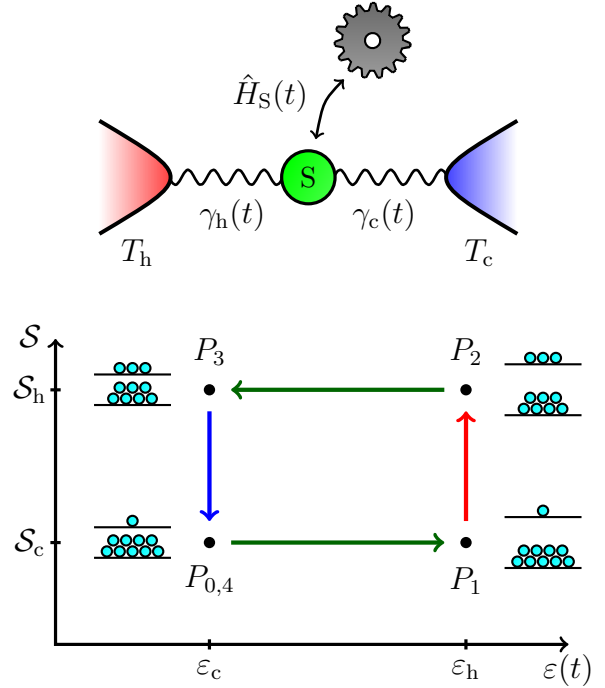

The work medium of the QHE under consideration consists of a two-level system (TLS) driven by an external control field. This driving field induces the time-dependent Hamiltonian ($\hbar=1$)
\begin{align}
\label{eq:H_sys}
    \Ht[S][t] &= \frac{\varepsilon(t)}{2}\left(\ketbra{\e(t)}-\ketbra{\g(t)}\right) \nonumber \\
    &= \frac{\varepsilon(t)}{2} \, \szt[t]
\end{align}
where both the energy gap $\varepsilon(t)$, the instantaneous eigenbasis of the TLS $\{\ket{\g(t)}, \ket{\e(t)}\}$ and the Pauli operator $\szt[t] = \ketbra{\e(t)}-\ketbra{\g(t)}$ in the instantaneous eigenbasis of the work medium are time-dependent. Following \cite{KosloffR17}, we define the Otto cycle in analogy to classical heat engines by relating the reciprocal energy gap of the TLS to the volume of the classical work medium. This identification is reasonable as the wavefunction of quantum particles with a large energy gap is typically strongly confined in space and vice versa (think of the harmonic oscillator as an illustrative example). The Otto cycle for a TLS, which is depicted in \cref{fig:QHE_Otto_cycle}, consists of four strokes \cite{PetersonBT+2019}. 

In the following a positive work (heat) indicates work done on (heat transferred to) the work medium of the heat engine.\\

\noindent{\bf Compression stroke [$P_{0,4}\rightarrow P_1$].} 
In this step $\Delta W_\com^\text{(sys)}<0$ (see \cref{def_W_com} for the mathematical definition), that is, work is extracted from the system\footnote{Note, that we extract work during the compression stroke and do work on the engine during the expansion stroke in this particular QHE. The reason for this behaviour is the symmetrical arrangement of the energy levels, i.e. the ground state energy is lowered when the energy gap is increased. We obtain a QHE where work is extracted during the expansion step by considering the Hamiltonian $\Ht[S][t] = \varepsilon(t) \ketbra{\e(t)}$.} by increasing the energy gap of the work medium from $\varepsilon_\cold$ at time $\tau_0$ to $\varepsilon_\hot > \varepsilon_\cold$ at time $\tau_1$
in a stroke of duration $\tau_\com = \tau_1 - \tau_0$. During this stroke the eigenbasis of the Hamiltonian may also change from $\{\ket{\g_\cold}, \ket{\e_\cold}\}$ at time $\tau_0$ to $\{\ket{\g_\hot}, \ket{\e_\hot}\}$ at time $\tau_1$. In the ideal QHE this process is performed quasi-statically, i.e. adiabatically with $\tau_\com \rightarrow \infty$, such that the populations
\begin{align}
    p_\x(t) = \bra{\x(t)}\rhot[S][t]\ket{\x(t)}
\end{align}
of the energy eigenstates $\x \in \{\g,\e\}$ are conserved throughout the entire stroke.
For finite compression times $\tau_\com$ the extracted work is reduced by quantum friction which we discuss in more detail in \cref{sec:Quantum_friction}.\\

\noindent{\bf Thermalisation with the hot bath [$P_1\rightarrow P_2$].} The work medium is placed in contact with a hot thermal bath at temperature $T_\hot$ for a time interval of duration $\tau_\hot = \tau_2 - \tau_1$ such that $\Delta Q_\hot^\text{(sys)}>0$ (as defined in \cref{def_Q_h}), i.e. heat is transferred from the heat bath to the work medium. The Hamiltonian of the work medium is constant during the thermal strokes and is given by $\Ht[S,\hot] = \varepsilon_\hot (\ketbra{\e_\hot}-\ketbra{\g_\hot})/2$. The interaction of the work medium with the thermal bath is described by the Lindblad equation
\begin{align}
\label{eq:L_hot}
    \mathcal{L}_\hot[\rhot[S]] = \gamma_\hot \, n_\hot \, &\left(\spt[\hot] \, \rhot[S] \, \smt[\hot] - \frac{1}{2}\{\smt[\hot] \spt[\hot], \, \rhot[S]\}\right) \nonumber \\
    + \gamma_\hot \, (n_\hot+1) \, &\left(\smt[\hot] \, \rhot[S] \, \spt[\hot] - \frac{1}{2}\{\spt[\hot] \smt[\hot], \, \rhot[S]\}\right)
\end{align}
with the raising and lowering operators $\spt[\hot] = (\smt[\hot])^\dagger = \ketbra{\e_\hot}{\g_\hot}$, the dissipative coupling constant $\gamma_\hot$ between work medium and heat bath and the mean thermal photon number
\begin{align}
\label{eq:n_hot}
    n_\hot = \frac{1}{\e^{\varepsilon_\hot/T_\hot}-1}
\end{align}
of the bath ($k_\text{B} = 1$) \cite{RivasHuelga2012}.\\

\noindent{\bf Expansion stroke [$P_2\rightarrow P_3$].} In this stroke of duration $\tau_{\exp} = \tau_3 - \tau_2$ the external control field performs work on the work medium, i.e. $\Delta W_\ex^\text{(sys)}>0$ (as defined in \cref{def_W_exp}), by decreasing the energy gap from $\varepsilon_\hot$ to $\varepsilon_\cold$ and rotating the basis of the Hamiltonian from $\{\ket{\g_\hot}, \ket{\e_\hot}\}$ at time $\tau_2$ to $\{\ket{\g_\cold}, \ket{\e_\cold}\}$ at time $\tau_3$. As before, note that friction losses will occur if the stroke is performed in finite time.\\

\noindent{\bf Thermalisation with the cold bath [$P_3\rightarrow P_{0,4}$].} In the final stroke of the Otto cycle the work medium is coupled to a cold thermal bath at temperature $T_\cold < T_\hot$ and heat is transferred from the work medium to the cold thermal bath, i.e. $\Delta Q_\cold^\text{(sys)}<0$ (as defined in \cref{def_Q_c}). The Hamiltonian $\Ht[S,\cold] = \varepsilon_\cold (\ketbra{\e_\cold}-\ketbra{\g_\cold})/2$ is constant during the stroke and, as before, we model the interaction of the work medium with the cold thermal bath by the Lindblad equation
\begin{align}
\label{eq:L_cold}
    \mathcal{L}_\cold[\rhot[S]] = \gamma_\cold \, n_\cold \, &\left(\spt[\cold] \, \rhot[S] \, \smt[\cold] - \frac{1}{2}\{\smt[\cold] \spt[\cold], \, \rhot[S]\}\right) \nonumber \\
    + \gamma_\cold \, (n_\cold+1) \, &\left(\smt[\cold] \, \rhot[S] \, \spt[\cold] - \frac{1}{2}\{\spt[\cold] \smt[\cold], \, \rhot[S]\}\right)
\end{align}
with raising and lowering operators $\spt[\cold] = (\smt[\cold])^\dagger = \ketbra{\e_\cold}{\g_\cold}$, coupling constant $\gamma_\cold$ and mean thermal photon number
\begin{align}
    n_\cold = \frac{1}{\e^{\varepsilon_\cold/T_\cold}-1}
\end{align}
of the bath.

\subsection{Power and efficiency}
\label{sec:Power_and_efficiency}

The calculation of power and efficiency of QHEs requires an explicit definition of work and heat. Following \cite{Alicki1979,Nieuwenhuizen2005}, we define work as
\begin{align}
\label{eq:work_definition}
    W = \int \limits_{t_\ii}^{t_\mathrm{f}} \dd t \, \Tr{\frac{\dd \Ht[S][t]}{\dd t} \rhot[S][t]}
\end{align}
and heat as
\begin{align}
\label{eq:heat_definition}
    Q = \int \limits_{t_\ii}^{t_\mathrm{f}} \dd t \, \Tr{\Ht[S][t] \frac{\dd \rhot[S][t]}{\dd t}}.
\end{align}
As described in \cref{sec:Quantum_Otto_cycle}, the system evolves unitarily during the compression and expansion strokes whence we can integrate \cref{eq:work_definition} by parts and use the equation of motion of the density operator $\rhot[S][t]$ to arrive at an expression that depends only on the states at the beginning and at the end of the stroke. Additionally, there is no heat exchange between the work medium and the heat baths during compression and expansion strokes due to the unitary evolution of the system. Hence, we can express the work done on the system during the compression and expansion strokes as
\begin{align}
\label{def_W_com}
    \Delta W_\com^\mathrm{(sys)} = \Tr{\Ht[S,\hot] \, \rhot[S,1]} - \Tr{\Ht[S,\cold] \, \rhot[S,0]}
\end{align}
and
\begin{align}
\label{def_W_exp}
    \Delta W_\ex^\mathrm{(sys)} = \Tr{\Ht[S,\cold] \, \rhot[S,3]} - \Tr{\Ht[S,\hot] \, \rhot[S,2]}
\end{align}
respectively, where $\rhot_{\text{S}, i}$ is the state of the QHE at the end of stroke $i$, as defined in the caption of \cref{fig:QHE_Otto_cycle}. As described in the previous subsection, we obtain the relations $\Delta W_\com^\text{(sys)} < 0$ (compression) and $\Delta W_\ex^\text{(sys)} > 0$ (expansion) for the ideal quasi-static Otto cycle in \cref{sec:Quantum_Otto_cycle}. During the thermalisation strokes no work is done on the work medium as $\frac{\dd}{\dd t} \Ht[S][t] = 0$, whence the work extracted {\em from} the QHE during a single cycle is given by
\begin{align}
\label{eq:extracted_work_sys}
    \Delta W_\text{ext.}^\text{(sys)} = -(\Delta W_\com^\text{(sys)} + \Delta W_\ex^\text{(sys)}) > 0.
\end{align}
Furthermore, for the thermalisation strokes we can easily integrate \cref{eq:heat_definition} to obtain the heat exchanges
\begin{align}
    \label{def_Q_h}
    \Delta Q_\hot^\text{(sys)} = \Tr{\Ht[S,\hot] \, \rhot[S,2]} - \Tr{\Ht[S,\hot] \, \rhot[S,1]}
\end{align}
and
\begin{align}
\label{def_Q_c}
    \Delta Q_\cold^\text{(sys)} = \Tr{\Ht[S,\cold] \, \rhot[S,0]} - \Tr{\Ht[S,\cold] \, \rhot[S,3]}
\end{align}
during the hot and cold thermalisation stroke respectively, which are given by the difference of the total internal energy in the work medium before and after the stroke. Moreover, the relations $\Delta Q_\hot^\text{(sys)} > 0$ and $\Delta Q_\cold^\text{(sys)} < 0$ hold true for the operation mode where work is extracted from the QHE.
We define the power of the heat engine as
\begin{align}
\label{eq:def_P}
    P^\text{(sys)} = \frac{\Delta W_\text{ext.}^\text{(sys)}}{\tau_\cycle}
\end{align}
with the cycle duration $\tau_\cycle = \tau_\com + \tau_\hot + \tau_\ex + \tau_\cold$. The corresponding efficiency reads
\begin{align}
\label{eq:eta_sys_def}
    \eta^\text{(sys)} = \frac{\Delta W_\text{ext.}^\text{(sys)}}{\Delta Q_\hot^\text{(sys)}}.
\end{align}
The maximal possible efficiency for the particular QHE from \cref{sec:Quantum_Otto_cycle} can be calculated for the ideal quasi-static cycle. 
For this, we assume perfect thermal states
\begin{align}
    \rhot[S,0] = \mathcal{Z}_\cold \e^{-\Ht[S,c]/T_\cold}
\end{align}
and
\begin{align}
    \rhot[S,2] = \mathcal{Z}_\hot \e^{-\Ht[S,h]/T_\hot}
\end{align}
with the partition sums $\mathcal{Z}_\x= \Tr\{\e^{-\Ht[S,h]/T_\x}\}$ after the thermalisation strokes and perfectly adiabatic compression and expansion strokes that fully conserve the populations. A simple evaluation of the expectation values \cref{def_W_com,def_W_exp,def_Q_h,def_Q_c} yields
\begin{align}
    \Delta W_\com^\mathrm{(sys)} &= - \frac{\varepsilon_\hot - \varepsilon_\cold}{2} \tanh{\frac{\varepsilon_\cold}{T_\cold}} < 0,\\
    \Delta W_\ex^\mathrm{(sys)} &= \frac{\varepsilon_\hot - \varepsilon_\cold}{2} \tanh{\frac{\varepsilon_\hot}{T_\hot}} < 0,\\
    \Delta Q_\hot^\text{(sys)} &= \frac{\varepsilon_\hot}{2} \left(\tanh{\frac{\varepsilon_\cold}{T_\cold}} - \tanh{\frac{\varepsilon_\hot}{T_\hot}}\right)
\end{align}
and
\begin{align}
    \Delta Q_\cold^\text{(sys)} = - \frac{\varepsilon_\cold}{2} \left(\tanh{\frac{\varepsilon_\cold}{T_\cold}} - \tanh{\frac{\varepsilon_\hot}{T_\hot}}\right).
\end{align}
Hence, the extracted work per cycle from \cref{eq:extracted_work_sys} reads
\begin{align}
    &\Delta W_\text{ext.}^\text{(sys)} = \frac{\varepsilon_\hot-\varepsilon_\cold}{2} \left(\tanh{\frac{\varepsilon_\cold}{T_\cold}} - \tanh{\frac{\varepsilon_\hot}{T_\hot}}\right)
\end{align}
and $\Delta W_\text{ext.}^\text{(sys)}>0$ if $\frac{\varepsilon_\cold}{\varepsilon_\hot} > \frac{T_\cold}{T_\hot}$.
Under the assumption of a positive work output, the efficiency \cref{eq:eta_sys_def} of this perfect cycle is given by the Otto efficiency
\begin{align}
\label{eq:eta_max_sys}
    \eta_{\max}^\text{(sys)} = 1-\frac{\varepsilon_\cold}{\varepsilon_\hot} \leq  \eta_\text{C}
\end{align}
which is upper bounded by the Carnot efficiency $\eta_\text{C}= 1 - \frac{T_\cold}{T_\hot}$. The constraints of perfect thermalisation and perfect adiabatic processes are fulfilled by assuming infinite stroke durations $\tau_\x \rightarrow \infty$ for all $\x \in \{\com, \hot, \exp, \cold\}$
\cite{Nieuwenhuizen2005}.
However, in this regime the power \cref{eq:def_P} of the heat engine vanishes. In the following section we address the issue of finite cycle times that allow for non-vanishing power output.

\subsection{Effects of finite cycle time}
\label{sec:Quantum_friction}

A QHE that uses strokes of finite duration experiences a reduction in efficiency due to two different effects. First, assuming a finite coupling strength between work medium and thermal bath, the work medium cannot thermalise completely to the temperature of the attached thermal bath. This imperfect equilibration, even when assuming that the work medium is always in a Gibbs state, leads to the well-known Curzon-Ahlborn bound
\begin{align}
    \eta_{\max} \leq \eta_\text{CA} = 1 - \sqrt{\frac{T_\cold}{T_\hot}}
\end{align}
for the efficiency at maximum power, $\eta_{\max}$ \cite{CurzonA75}.
However, as the coupling strength to the thermal baths may, at least in principle, be increased arbitrarily, and therefore the thermalisation time made arbitrarily small, we will not consider the role of imperfect thermalisation in this work.\footnote{While the consequences of strong coupling between work medium and environment in general has been a subject of discussions \cite{pozas2018quantum, wiedmann2020non, newman2020quantum,shirai2021non,kaneyasu2023,latune2023cyclic}, it is possible to decouple the thermal contact of the work medium from the thermal bath if the interaction between the two observes the rotating-wave approximation, i.e. $\hat{\sigma}^{+}\hat{a} + \hat{\sigma}^{-}\hat{a}^{\dagger}$, where $\hat{\sigma}^{+}$ is the raising operator of the spin (i.e. the work medium) while $\hat{a}$ is the annihilation operator of a harmonic oscillator of the environment. In that case, the work medium equilibrates rapidly, and coherences between the work medium and the environment vanish asymptotically. Therefore, the expectation value of the coupling Hamiltonian with the steady state of work medium and environment \cite{latune2023cyclic} vanishes, which, in turn, implies a vanishing thermodynamical cost for decoupling the environment from the work medium. This can be confirmed in both the DAMPF or reaction rate model, as well as in the TEDOPA picture of the interaction between work medium and environment.} Instead we focus on a second cause of loss of efficiency in QHE operated at finite power, quantum friction \cite{FeldmannKosloff2006,Petruccione2021}, which appears in compression and expansion strokes of finite duration.

As described in \cref{sec:Quantum_Otto_cycle}, the eigenbasis of the time-dependent Hamiltonian of the work medium evolves during the compression and expansion strokes. When executed at finite rate, the work medium is not able to follow the energy eigenbasis adiabatically. As a result, coherences between instantaneous energy eigenstates are typically generated which, in turn, lead to changes in the population of the instantaneous energy eigenstates. This process requires additional work therefore reducing the extracted work per cycle.

Quantum friction can be suppressed if coherences are made to vanish and populations remain invariant. In \cite{FeldmannKosloff2006} it was suggested to reduce quantum friction and the associated heat dissipation with the help of pure dephasing in the instantaneous eigenbasis of the work medium. The stronger the dephasing noise, the more effectively it suppresses the creation of coherences between energy eigenstates and forces the state to remain diagonal in the eigenbasis of the Hamiltonian thus following the adiabatic trajectory.

As dephasing can be made arbitrarily strong, this observation seems to suggest that a heat engine can be run at arbitrarily high speed while remaining arbitrarily close to the efficiency of a quasi-static heat engine. In the following chapters, we will examine this statement critically and investigate the effect of dephasing noise on the performance of the QHE defined in \cref{sec:Quantum_Otto_cycle} by taking into account any thermodynamic cost that it may incur.
\section{Model of the dephasing environment}
\label{sec:model}

To determine the efficiency of dephasing-assisted quantum heat engines numerically, we must develop technical tools to compute the thermodynamic cost of pure dephasing in the instantaneous eigenbasis. 
We ensure an unambiguous definition of the energy exchanged between the engine and its dephasing environment by utilising an exact description that considers all degrees of freedom for both the engine and environment.

The numerical treatment of this description is 
achieved in \cref{sec:dephasing_TEDOPA_bath} via the method of the Time Evolving Density matrices
using Orthogonal Polynomials Algorithm (TEDOPA) \cite{Chin2010,PriorCH+10}.
As this approach is computationally expensive we reduce the complexity 
of our numerical studies 
by considering the special case of an effective environment based
on a single damped harmonic oscillator which is based on the
Dissipation-Assisted Matrix Product Factorisation (DAMPF) 
approach \cite{SomozaMartyLimHuelgaPlenio2019,MascherpaSS+2020} similar to applications of the reaction coordinate approach in thermodynamical machines \cite{ivander2022, anto2021strong, nazir2018reaction,AbiusoG2019}.
This method, which we introduce in \cref{sec:dephasing_DAMPF_bath}, 
allows for a compact and numerically efficient description of the 
dynamics of the heat engine and, as we verify via numerically exact TEDOPA simulations in 
\cref{sec:Bath_energy_in_DAMPF_simulations}, its energy exchange with the dephasing environment.

\subsection{Exact dephasing bath using TEDOPA}
\label{sec:dephasing_TEDOPA_bath}

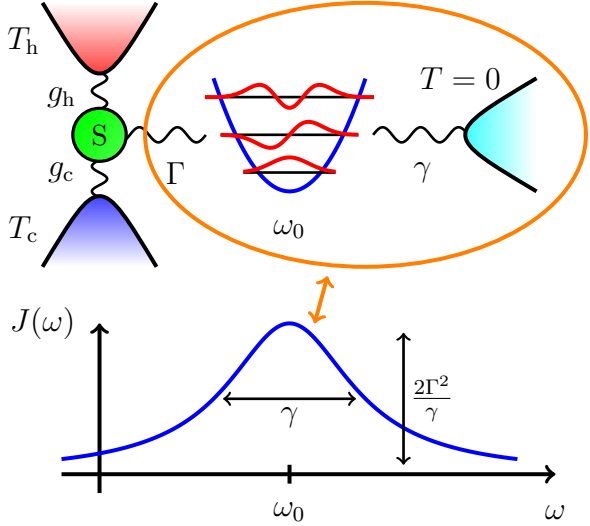
\begin{figure}[t]
    \centering
    \begin{tikzpicture}
	[scale = 1,
	round/.style={circle, draw=black!100, very thick, shade, left color=green!100, right color=green!75, shading angle=60}, 
	]
	
	\draw[smooth,samples=100,domain=0:1.4,line width = 1] plot(\x,{sin((\x-1.4)*720)/10});
	\draw[smooth,samples=100,domain=3.6:5,line width = 1] plot(\x,{sin((\x-3.6)*720)/10});
	\draw[smooth,samples=100,domain=1:-1,line width = 1] plot({sin((\x)*1080)/10},\x);
    
    \coordinate (system) at (0,0);
    \coordinate (mode) at (5,0);
    \node[round] (system) {S};
    
    \draw[color=blue,smooth,samples=100,domain=1.5:3.5,line width = 1.5] plot(\x,{1.5*(\x-2.5)^2-0.75});
    \draw[smooth,samples=100,domain=1.892:3.108,line width = 1] plot(\x,{-0.5});
    \draw[smooth,samples=100,domain=1.593:3.407,line width = 1] plot(\x,{0});
    \draw[smooth,samples=100,domain=1.387:3.613,line width = 1] plot(\x,{0.5});
    \draw[color=red,smooth,samples=100,domain=1.892:3.108,line width = 1.5] plot(\x,{0.2*exp(-(\x-2.5)^2*8)-0.5});
    \draw[color=red,smooth,samples=100,domain=1.593:3.407,line width = 1.5] plot(\x,{0.2*exp(-(\x-2.5)^2*8)/1.41*4*2*(\x-2.5)});
    \draw[color=red,smooth,samples=100,domain=1.387:3.613,line width = 1.5] plot(\x,{0.2*exp(-(\x-2.5)^2*8)/2/1.41*(4*16*(\x-2.5)^2-2)+0.5});
    
    \draw[draw=white, shade, left color=cyan!100, right color=cyan!0] (5.75,0.75) .. controls (4.5,0) .. (5.75,-0.75);
    \draw[line width = 1.5] (5.75,0.75) .. controls (4.5,0) .. (5.75,-0.75);
	
    \draw[draw=white, shade, bottom color=red!100, top color=red!0] (0.75,1.75) .. controls (0,0.5) .. (-0.75,1.75);
    \draw[line width = 1.5] (0.75,1.75) .. controls (0,0.5) .. (-0.75,1.75);
    \draw[draw=white, shade, top color=blue!100, bottom color=blue!0] (0.75,-1.75) .. controls (0,-0.5) .. (-0.75,-1.75);
    \draw[line width = 1.5] (0.75,-1.75) .. controls (0,-0.5) .. (-0.75,-1.75);
    
    \node at (1,-0.5) {\large $\Gamma$};
    \node at (4.25,-0.5) {\large $\gamma$};
    \node at (2.5,-1.25) {\large $\omega_0$};
    \node at (4.75,0.75) {\large $T=0$};
    \node at (-1,1.25) {\large $T_\hot$};
    \node at (-1,-1.25) {\large $T_\cold$};
    \node at (-0.5,0.5) {\large $g_\hot$};
    \node at (-0.5,-0.5) {\large $g_\cold$};
    
    \draw[line width = 1.5, color=orange] (3.5,0) ellipse (2.9 and 1.75);
    
    \draw[line width = 1.5, <->, color=orange] (3.0,-1.875)--(2.83,-2.5);
    
    \draw[->, line width = 1.5] (0,-4.75)--(0,-2.5);
    \draw[->, line width = 1.5] (-0.5,-4.5)--(6,-4.5);
    \draw[line width = 1.5] (2.5,-4.625)--(2.5,-4.375);
    
    \node at (6,-5.0) {\large $\omega$};
    \node at (-0.75,-2.5) {\large $J(\omega)$};
    \node at (2.5,-5.0) {\large $\omega_0$};
    
    \draw[color=blue,smooth,samples=100,domain=-0.5:5.5,line width = 1.5] plot(\x,{2/(1+(\x-2.5)^2)-4.5});
    
    \draw[<->, line width = 1] (1.625,-3.5)--(3.375,-3.5);
    \node at (2.5,-3.75) {\large $\gamma$};
    
    \draw[<->, line width = 1] (4,-4.375)--(4,-2.625);
    \node at (4.375,-3.5) {\large $\frac{2\Gamma^2}{\gamma}$};
    
\end{tikzpicture}
    \caption{A dephasing bath is added to the QHE described in \cref{fig:QHE_Otto_cycle} to reduce quantum friction. This bath is modelled by a single harmonic mode with frequency $\omega_0$ that couples to the work medium with strength $\Gamma$. Additionally, the harmonic mode is damped by a thermal Lindblad dissipator at zero temperature with damping rate $\gamma$. As described in the main text, this particular configuration acts in the same way on the reduced density operator of the work medium as a microscopically described bath with a Lorentzian spectral density $J(\omega)$, centered at $\omega_0$ with width $\gamma$ and height $2\Gamma^2/\gamma$. }
    \label{fig:Dampf_model}
\end{figure}

We begin with the microscopic description of a dephasing bath which is coupled to the work medium. The total microscopic Hamiltonian
\begin{align}
\label{eq:H_micro}
    \Ht(t) = \Ht[S][t] + \Ht[int](t) + \Ht[B]
\end{align}
consists of three terms. The Hamiltonian $\Ht[S][t]$, \cref{eq:H_sys}, describes the internal dynamics of the heat engine. The time-independent bath Hamiltonian
\begin{align}
\label{eq:H_B_micro}
    \Ht[B] = \int \limits_{-\infty}^{\infty} \dd \omega \, \omega \, \hat{a}^\dagger_\omega \hat{a}_\omega
\end{align}
describes the free evolution of the bath that consists of continuously distributed harmonic modes with creation (annihilation) operators $\hat{a}^\dagger_\omega$ ($\hat{a}_\omega$). The interaction of this bosonic environment with the work medium is described by
\begin{align}
    \label{eq:H_int_micro}
    \Ht[int](t) = \szt[t] \otimes \!\!\!\int \limits_{-\infty}^{\infty} \!\dd \omega \, \sqrt{\frac{\mathcal{J}(\omega)}{\pi}} \left(\hat{a}^\dagger_\omega + \hat{a}_\omega\right).
\end{align}
The interaction Hamiltonian $\Ht[int](t)$ which commutes with the Hamiltonian of the work medium at all times thus describes pure dephasing in the instantaneous eigenbasis $\{\ket{\g(t)}, \ket{\e(t)}\}$ of the work medium. The dynamics of the work medium under the action of this environment which is initialised in a thermal state is fully characterised by the spectral density $\mathcal{J}(\omega)$ \cite{RivasHuelga2012,BreuerPetruccione2002}. The continuum of harmonic modes and their coupling structure is not best suited for numerical simulations. The TEDOPA approach overcomes this difficulty by introducing a unitary mode transformation $U_n(\omega)$ to an infinite but discrete set of eigenmodes with annihilation operators
\begin{equation}
    \hat{b}_n = \int \limits_{-\infty}^{\infty} \dd\omega\, U_n(\omega) \, \hat{a}_{\omega},
\end{equation}
for which the free bath Hamiltonian $\Ht[B]$ and the interaction Hamiltonian $\Ht[int](t)$ can be re-expressed as
\begin{align}
\label{eq:H_TEDOPA_chain}
    &\Ht[B] = &\sum_{n=0}^\infty \omega_n \hat{b}^\dagger_n \hat{b}_n + t_n (\hat{b}^\dagger_n \hat{b}_{n+1} + \hat{b}_n \hat{b}^\dagger_{n+1})
\end{align}
and
\begin{align}
    \Ht[int](t) = c_0 \szt[t] \otimes \left(\hat{b}^\dagger_0 + \hat{b}_0\right).
\end{align}
Thus, the new modes describe a one-dimensional semi-infinite harmonic chain with nearest-neighbor interaction, which couples to the work medium exclusively via the first site of the chain - a geometry that is particularly well suited for efficient simulation with tensor networks \cite{Chin2010,PriorCH+10}. Moreover, in TEDOPA we can keep track of the exact state of the environment for the entire duration of the simulation, allowing for the calculation of arbitrary expectation values involving environment operators. These features enable a detailed analysis of correlations and energy flows between the work medium and dephasing environment, which is necessary for the analysis of the thermodynamical cost of dephasing noise. In the numerical simulations, we truncate the dimension of the local Hilbert spaces of each harmonic mode and the length of the semi-infinite chain to a finite size (the incurred error can be bounded \cite{WoodsCP15,MascherpaSH+17} rigorously). The finite length of the chain limits the total simulation time, as excitations from the work medium travel through the chain with finite speed and are reflected at the end of the chain. The time-evolution of the work medium accumulates significant error only after the reflected excitations reach the work medium again. This can be prevented by choosing a sufficiently long chain or the addition of a Markovian closure at the end of a shorter chain \cite{NusselerTS+22}. Despite such improvements, TEDOPA simulations require significant computational resources, especially for strong dephasing and fast operation of the QHE. In these regimes, we observe strong correlations between work medium and bath, as well as highly populated harmonic modes in the initial parts of the TEDOPA chain, requiring accurate description with large local dimensions in the tensor network simulations.

In the next section we address this challenge by specialising the environmental spectral density in a manner that allows us to obtain the numerically exact dynamics of the work medium and, at the same time, allows us to keep track of the energy which is transferred from the work medium to the dephasing bath. This is achieved within the framework of the Dissipation-Assisted Matrix Product Factorisation (DAMPF) approach \cite{SomozaMartyLimHuelgaPlenio2019,MascherpaSS+2020} 
(see also the reaction coordinate approach in thermodynamical machines
\cite{ivander2022, anto2021strong, nazir2018reaction,AbiusoG2019}) which we introduce in the next \cref{sec:dephasing_DAMPF_bath,sec:Bath_energy_in_DAMPF_simulations}.

\subsection{Dephasing DAMPF bath}
\label{sec:dephasing_DAMPF_bath}

The DAMPF approach \cite{SomozaMartyLimHuelgaPlenio2019,MascherpaSS+2020} represents the environmental spectral density as a combination of Lorentzians which correspond to harmonic oscillators that are damped with local dissipators to a Markovian environment. For the following we simplify this description further by restricting attention to a single damped harmonic oscillator with creation (annihilation) operator $\hat{b}^\dagger \, (\hat{b})$ that is coupled to the work medium and, to facilitate analytical expressions, to a heat bath at zero temperature as shown in \cref{fig:Dampf_model}. The finite temperature case will be treated later. This model yields the same bath correlation function as the microscopic description \cref{eq:H_micro} with a Lorentzian spectral density
\begin{align}
\label{eq:spectral_density}
    J(\omega) = \frac{2\Gamma^2 \gamma}{\gamma^2+4(\omega-\omega_0)^2}
\end{align}
and therefore leads to the same dynamics in the work medium \cite{TamascelliSH+2018}. The Hamiltonian of this DAMPF system reads
\begin{align}
\label{eq:H_DAMPF}
    \Ht[\DAMPF][t] = \Ht[S][t] + \Ht[int][t] + \Ht[osc]
\end{align}
with the interaction Hamiltonian
\begin{align}
    \Ht[int][t] = \Gamma \, \szt[t] \otimes \left(\hat{b}^\dagger + \hat{b}\right)
\end{align}
and the free Hamiltonian
\begin{align}
\label{eq:H_osc}
    \Ht[osc] = \omega_0 \, \hat{b}^\dagger \hat{b}
\end{align}
of the harmonic mode. The coupling of this harmonic mode to a zero temperature Markovian bath is described by the Lindbladian
\begin{align}
\label{eq:L_osc}
    \mathcal{L}_\osc[\rhot] = \gamma \, \left(\hat{b} \, \rhot \, \hat{b}^\dagger - \frac{1}{2} \{\hat{b}^\dagger\hat{b}, \rhot\}\right).
\end{align}
The full master equation of the work medium and the harmonic mode reads
\begin{align}
\label{eq:DAMPF_ME}
    \frac{\dd}{\dd t} \rhot[][t] = &-\ii \, \commutator{\Ht[\DAMPF][t]}{\rhot[][t]} + \mathcal{L}_\osc[\rhot[][t]] \\
    &+ \mathcal{L}_\cold(t)[\rhot[][t]] + \mathcal{L}_\hot(t)[\rhot[][t]] \nonumber
\end{align}
where we also include the thermal baths from \cref{sec:Quantum_Otto_cycle}. 

\subsection{Bath energy in DAMPF simulations}
\label{sec:Bath_energy_in_DAMPF_simulations}

\begin{figure*}[t]
    \centering
    \includegraphics[width=\textwidth]{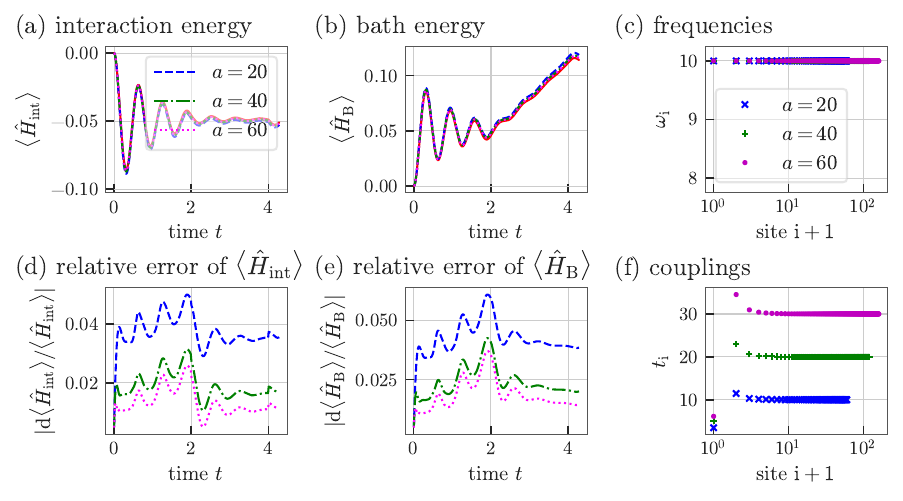}
    \caption{Comparison of the interaction energy $\langle\Ht[int][t]\rangle$ in (a) and the bath energy $\langle\Ht[B][t]\rangle$ in (b) obtained with the numerically exact TEDOPA simulations for different supports $[\omega_0-a,\omega_0+a]$ of the spectral density $J(\omega)$ and with DAMPF simulations (red, solid) via eq. \cref{eq:E_deph}. We chose the parameters $\Gamma = 0.5$, $\omega_0 = 10$, $\gamma = 2$ and $\tau_\x = 2$ for all $\x$. The remaining parameters are chosen as in \cref{sec:numerical_results}. We only show the initial parts of the simulation with factorised initial state $\rhot[0] = \mathcal{Z}^{-1} \e^{-\Ht[S,c]/T_\cold} \otimes \ketbra{0}$ due to the complexity of the TEDOPA simulations. Panels (d) and (e) show that the relative error $\dd \langle{\Ht[x]}\rangle = |\langle{\Ht[x]^\text{(T)}}\rangle - \langle{\Ht[x]^\text{(D)}}\rangle|$ between the interaction and bath energies obtained with TEDOPA $\langle\Ht[x]^\text{(T)}\rangle$
    and DAMPF $\langle\Ht[x]^\text{(D)}\rangle$ rapidly converges to zero in the limit $a \rightarrow \infty$. As discussed in the main text, we are not able to simulate the model for infinite support since the couplings between the TEDOPA modes diverge if the support is increased as depicted in (f) whereas the frequencies in (c) do not change.}
    \label{fig:DAMPF_TEDOPA_Eint_Ebath_ws_ts}
\end{figure*}

In the DAMPF picture we have access only to the effective mode $\hat{b}$ and we are not able to keep track of the state of the continuum of modes that realise the Lorentzian spectral density. Despite of this limitation, we can determine the bath energy $\langle\Ht[B](t)\rangle$ that is required to quantify the thermodynamical cost of dephasing noise by separating the unitary and non-unitary part in the time-evolution of \cref{eq:DAMPF_ME} via the Suzuki-Trotter decomposition \cite{Trotter1959, Suzuki1976}. In leading order, we find that the map that evolves the state unitarily from time $t$ to time $t+\dd t$ can be written as
\begin{align}
    \mathcal{U}_{\dd t} [\rhot[][t]] = \e^{-\ii \Ht[\DAMPF][t] \dd t} \, \rhot[][t] \, \e^{\ii \Ht[\DAMPF][t] \dd t}\, .
\end{align}
The non-unitary part of the time-evolution 
\begin{align}
    \mathcal{G}_{\dd t}^\text{(tot)}[\rhot[][t]] =& \e^{(\mathcal{L}_\osc + \mathcal{L}_\cold(t) + \mathcal{L}_\hot(t)) \dd t}[\rhot[][t]]
\end{align}
can be split into three consecutive maps
\begin{align}
    \mathcal{G}_{\dd t}^\text{(tot)} = \mathcal{G}_{\dd t}^\text{(\osc)} \circ \mathcal{G}_{\dd t}^\text{(\cold)} \circ \mathcal{G}_{\dd t}^\text{(\hot)}
\end{align}
without any Trotter error since $\commutator{\mathcal{L}_\x(t)}{\mathcal{L}_\osc(t)} = 0$ at all times $t$ and for $\x \in \{\hot, \cold\}$ and $\commutator{\mathcal{L}_\hot(t)}{\mathcal{L}_\cold(t)} = 0$ as those two environments do not couple to the work medium at the same time.
We define the formal time-evolution of each Lindbladian $\mathcal{L}_\x(t)$ via the map
\begin{align}
    \mathcal{G}_{\dd t}^\text{(\x)}[\rhot[][t]] = \e^{\mathcal{L}_\x(t) \dd t}[\rhot[][t]].
\end{align}
There is no need for time-ordering as all Lindbladians are piecewise constant. Combining the unitary and the Lindbladian evolutions, the second order Suzuki-Trotter decomposition is given by 
\begin{align}
\label{eq:Suzuki_Trotter_decomposition}
    \rhot[][t+\dd t] = &\left(\mathcal{U}_{\dd t/2} \circ \mathcal{G}_{\dd t}^\text{(th)} \circ \mathcal{G}^{(\osc)}_{\dd t} \circ \mathcal{U}_{\dd t/2}\right) [\rhot[][t]] \nonumber \\
    &+ \order{\dd t^3} \, ,
\end{align}
where the thermal propagator $\mathcal{G}_{\dd t}^\text{(th)} = \mathcal{G}_{\dd t}^\text{(\hot)} \circ \mathcal{G}_{\dd t}^\text{(\cold)}$ accounts for the time-evolution of both thermal baths.
Now, we measure the energy of the system before and after the non-unitary evolution to obtain the differential energy
\begin{align}
\label{eq:dE_x}
    \dd E_\x(t) = \Tr{\Ht[\DAMPF][t] \left(\mathcal{G}^{(\x)}_{\dd t} [\rhot[][t]] - \rhot[][t]\right)}
\end{align}
that is exchanged between the system and the Lindbladian bath $\x \in \{\osc, \hot, \cold\}$ within the time $\dd t$. The total exchanged energy is calculated with a numerical integration of the differential energy. The dephasing bath consists of the harmonic mode and the Lindblad dissipator. Hence, the energy stored in the dephasing bath
\begin{align}
\label{eq:E_deph}
    E_\text{deph}(t) = \Tr{\hat{H}_{\osc}\hat{\rho}(t)} - \int \limits_0^t \dd t^\prime \, \frac{\dd E_\osc(t^\prime)}{\dd t^\prime}
\end{align}
consists of those two energy contributions. The minus sign accounts for the fact that dissipated energy $\dd E_\osc$ from \cref{eq:dE_x} is negative if energy leaves the system to the bath. We have verified the accuracy of \cref{eq:E_deph} in simulations with finite $\dd t$ using the numerical exact TEDOPA simulations of the microscopic description of the dephasing bath from \cref{eq:H_micro} using the python library \textsc{Mpnum} \cite{SuessHolzapfel2017}. The numerical results are presented in \cref{fig:DAMPF_TEDOPA_Eint_Ebath_ws_ts}, where the calculations of the time-dependent interaction energy $\langle\Ht[int][t]\rangle$ are compared in (a) and (d) and the results for the time-dependent bath energy $E_\text{deph}(t) = \langle\Ht[B][t]\rangle$ are compared in (b) and (e) for both numerical methods. We use a finite support $[\omega_0-a,\omega_0+a]$ of the spectral density for the TEDOPA simulations as the couplings $t_i$ of the TEDOPA chain Hamiltonian from \cref{eq:H_TEDOPA_chain} diverge in the limit $a \rightarrow \infty$ as shown in panel (f). Large coupling strengths require significant computational resources as the characteristic time-scale of the interaction between neighboring sites is proportional to the reciprocal coupling strength. Nevertheless, the numerical results of the TEDOPA method converge towards those of DAMPF in the limit $a \rightarrow \infty$ according to panels (d) and (e).

\subsection{Power maximisation}
\label{sec:theo_power_maximization}

In \cref{sec:num_power_maximization}, we investigate the performance of the QHE for different dephasing strengths, specifically focusing on the efficiency at maximum power output. To achieve this, we fix the parameters of the work medium and the dephasing bath, and then optimise the power of the QHE with respect to the stroke durations $\tau_\x$ for $\x \in \{\com, \hot, \ex, \cold\}$. The optimisation is performed by a gradient search in the four search parameters\footnote{This optimisation algorithm does not guarantee to find the global power maximum. However, the results provide a lower bound of the power, which supports the main statement of this work.}. 

There are two different possible optimisation targets for the QHE under the action of a dephasing environment. The first restricts the definition of the power from \cref{eq:def_P} to the energetic changes of the two-level system, which is considered to be the only work medium. In other words, we take the definitions of work and heat exchange from \cref{def_W_com,def_W_exp,def_Q_h,def_Q_c} with the reduced system states $\rhot_{\text{S}, i} = \Tr_\osc\left\{\rhot_i\right\}$ where the indices $i$ label the end points of the strokes in the Otto cycle (see \cref{fig:QHE_Otto_cycle}). 

The alternative is providing a more complete definition of work and heat that also accounts for the energy of the dephasing environment which can be interpreted as an extension of the work medium that has to be taken into account in thermodynamic considerations. Hence, we redefine the work that is extracted during the compression stroke from \cref{def_W_com} by also considering the interaction energy $\langle \Ht[int][t] \rangle$ and the energy $E_\text{deph}(t)$ contained in the dephasing bath from \cref{eq:E_deph}. The final expression of the extracted work with respect to the total energy in work medium and dephasing bath reads 
\begin{align}
\label{eq:Delta_W_com_tot}
    \Delta W_\com^\text{(tot)} =& \Tr{\Ht[\DAMPF,\hot] \, \rhot[1]} - \Tr{\Ht[\DAMPF,\cold] \, \rhot[0]} \nonumber \\
    &- \int \limits_{\tau_0}^{\tau_1} \dd t \, \frac{\dd E_\osc(t)}{\dd t} \, ,
\end{align}
where the times $\tau_0$ and $\tau_1$ are the start and end times of the compression stroke. The work done on the work medium during the expansion stroke
\begin{align}
    \Delta W_\ex^\text{(tot)} =& \Tr{\Ht[\DAMPF,\cold] \, \rhot[3]} - \Tr{\Ht[\DAMPF,\hot] \, \rhot[2]} \nonumber \\
    &- \int \limits_{\tau_2}^{\tau_3} \dd t \, \frac{\dd E_\osc(t)}{\dd t}
\end{align}
is redefined analogously. The heat exchanges during the thermalisation can be expressed as
\begin{align}
\label{eq:Q_tot_h}
    \Delta Q_\hot^\text{(tot)} =& \Tr{\Ht[\DAMPF,\hot] \, \rhot[2]} - \Tr{\Ht[\DAMPF,\hot] \, \rhot[1]} \nonumber \\
    &- \int \limits_{\tau_1}^{\tau_2} \dd t \, \frac{\dd E_\osc(t)}{\dd t}
\end{align}
for the hot bath and as
\begin{align}
    \Delta Q_\cold^\text{(tot)} =& \Tr{\Ht[\DAMPF,\cold] \, \rhot[4]} - \Tr{\Ht[\DAMPF,\cold] \, \rhot[3]} \nonumber \\
    &- \int \limits_{\tau_3}^{\tau_4} \dd t \, \frac{\dd E_\osc(t)}{\dd t}
\end{align}
for the cold bath.
Moreover, we define the total extracted work per cycle
\begin{align}
\label{eq:extracted_work_tot}
    \Delta W_\text{ext.}^\text{(tot)} = -\left(\Delta W_\com^\text{(tot)} + \Delta W_\ex^\text{(tot)}\right) 
\end{align}
analogously to the extracted work $\Delta W_\text{ext.}^\text{(sys)}$ from \cref{eq:extracted_work_sys} that only considers the energy of the work medium. The resulting definition for the power reads
\begin{align} \label{eq:powertot}
    P^\text{(tot)} = \frac{\Delta W_\text{ext}^\text{(tot)}}{\tau_\cycle}
\end{align}
that includes the total energy of work medium and dephasing bath. The corresponding efficiency is given by
\begin{align}
\label{eq:eta_tot}
    \eta^\text{(tot)} = \frac{\Delta W_\text{ext}^\text{(tot)}}{\Delta Q_\hot^\text{(tot)}}.
\end{align}
The optimisation is performed with respect to the power in steady state operation. For that, we simulate the dynamics of the heat engine for some cycles until we see convergence of the results of two consecutive cycles. Our simulation results show that this stationary regime is reached very fast, typically after three cycles.

\subsection{Effective dephasing rate}
\label{sec:effective_dephasing_rate}

The dephasing bath is fully described by the spectral density $J(\omega)$ from \cref{eq:spectral_density} with the parameters $\Gamma$, $\gamma$ and $\omega_0$. However, it will be convenient to define a single parameter that characterises the effect of the bath in order to facilitate the comparison of different dephasing baths. Hence, we calculate an effective dephasing rate $\Geff(\Gamma,\gamma,\omega_0)$ that is then related to the ability of the bath to boost the performance of the QHE. To this end, we analyse the time-evolution of an initial preparation of the form
\begin{align}
\label{eq:dec_psi_0}
    \ket{\psi_0} = \frac{1}{\sqrt{2}} (\ket{\e(t_0)}_\text{S} + \ket{\g(t_0)}_\text{S}) \otimes \ket{0}_\text{B},
\end{align}
that contains coherences between the energy eigenstates $\ket{\e(t_0)}$ and $\ket{\g(t_0)}$, under the time-independent microscopic Hamiltonian $\Ht[][t_0]$ from \cref{eq:H_micro} with fixed time parameter $t_0$. We can solve the time-evolution
\begin{align}
    \ket{\psi(t)} = \mathcal{U}(t) \ket{\psi_0}
\end{align}
analytically as the Hamiltonian $\Ht[][t_0]$ describes the time-independent spin-boson model \cite{BreuerPetruccione2002}. Afterwards, we use this result to calculate the absolute value of the coherences
\begin{align}
    &|c(t)| = |\braket{\psi(t)}{\g}_\text{S}\braket{\e}{\psi(t)}| = \frac{1}{2} \e^{\Gamma(t)}
\end{align}
and the decoherence function $\Gamma(t)$. The factor $\frac{1}{2}$ ensures the initial condition $\Gamma(0) = 0$ for the initial state that was chosen in \cref{eq:dec_psi_0}. The analytical solution of the spin-boson model \cite{BreuerPetruccione2002} yields the decoherence function
\begin{align}
\label{eq:dec_func_integral}
    \Gamma(t) =& -\int \limits_{-\infty}^\infty \dd \omega \, \frac{1}{2}|\alpha_\omega(t)|^2
\end{align}
with the amplitudes
\begin{align}
\label{eq:dec_func_amplitude}
    \alpha_\omega(t) = 2\sqrt{\frac{J(\omega)}{\pi}} \frac{1-\e^{\ii \omega t}}{\omega}.
\end{align}
The decoherence function shows oscillations for small times $t$ as the environment is not Markovian. However, we aim for the calculation of an effective dephasing rate $\Geff$ which corresponds to a linear approximation
\begin{align}
    \Gamma(t) \approx - \Geff \, t
\end{align}
of the decoherence function which is, as we will show now, valid in the limit $t \rightarrow \infty$. For this, we take the spectral density from \cref{eq:spectral_density} and plug it into \cref{eq:dec_func_integral}, which, after substituting $\omega t = x$, yields the integral
\begin{align}
    \Gamma(t) =& -\frac{8\Gamma^2\gamma \, t}{\pi} \int \limits_{-\infty}^\infty \dd x 
    \, \frac{1-\cos(x)}{x^2 \, \left(\gamma^2 +4\left(\frac{x}{t}-\omega_0\right)^2\right)}
\end{align}
that we would like to solve in the limit $t \rightarrow \infty$.
The integrand
\begin{align}
    g(x,t) = \frac{1-\cos(x)}{x^2 \, \left(\gamma^2 +4\left(\frac{x}{t}-\omega_0\right)^2\right)}
\end{align}
converges uniformly
\begin{align}
    g(x,t) \xrightarrow[]{\text{unif.}} g(x) = \frac{1-\cos(x)}{x^2 \, \left(\gamma^2 + 4\omega_0^2\right)}
\end{align}
which implies that we can calculate the integral over $g(x,t)$ in the limit of large times as
\begin{align}
    \lim_{t \rightarrow \infty} \int \limits_{-\infty}^\infty \!\!\dd x \, g(x,t) = \!\int \limits_{-\infty}^\infty \!\!\dd x \, g(x) = \frac{\pi}{\gamma^2 + 4\omega_0^2}.
\end{align}
We use this result to determine the expression for the decoherence function as
\begin{align}
    \Gamma(t) \approx -\frac{8\Gamma^2\gamma}{\gamma^2 + 4\omega_0^2} \, t \, ,
\end{align}
which is valid in the limit of large times $t \rightarrow \infty$. This linear approximation allows us to define the effective decoherence rate
\begin{align}
\label{eq:Gamma_eff}
    \Gamma_\text{eff} = \frac{8\Gamma^2\gamma}{\gamma^2 + 4\omega_0^2}
\end{align}
that characterises the time-scale on which coherences are damped in the QHE. Hence, this quantity determines the ability of the dephasing bath to boost the power of the QHE. For this reason, the effective dephasing rate \cref{eq:Gamma_eff} will be used as the main parameter describing the dephasing bath in the next chapters.
\section{Numerical results}
\label{sec:numerical_results}

In this section we present the numerical results for the QHE discussed in \cref{sec:model} for different effective dephasing rates. First, the explicit time-dependence of the system Hamiltonian $\Ht[S][t]$ and the remaining parameters are defined in \cref{sec:system_parameters}. Then we proceed to maximise the power of the QHE for different effective dephasing rates numerically and present and discuss those results and the corresponding efficiencies in \cref{sec:num_power_maximization}. In the limit of strong dephasing rates, the maximum power surpasses that of the noiseless QHE at the cost of an increased heat consumption from the hot bath that flows into the dephasing bath and which is further analysed in \cref{sec:additional_heat_cost}. 

\subsection{System parameters}
\label{sec:system_parameters}

In our system the time-dependent Hamiltonian of the work medium takes the explicit form
\begin{align}
    \Ht[S][t] = \frac{\omega}{2} \szt^{(0)} + \frac{\Omega(t)}{2} \sxt^{(0)} = \frac{\varepsilon(t)}{2} \, \szt[t]
\end{align}
which describes a TLS with frequency $\omega$ that is subject to an external control field with time-dependent Rabi frequency $\Omega(t)$.
This Rabi frequency $\Omega(t)$ takes the constant values $\Omega(t) = 0$ in the cold thermalisation stroke and $\Omega(t) = \Omega_0$ in the hot thermalisation stroke. It increases (decreases) linearly in the compression (expansion) stroke for a time $\tau_\com$ ($\tau_\ex$) 
starting with $0$ ($\Omega_0$) and ending with $\Omega_0$ (0). 

For definiteness, we fix the parameters of the work medium throughout the rest of the text. We choose $\omega = 1$ and $\Omega_0 = 0.5$ which results in the energy gaps $\varepsilon_\cold = 1$ and $\varepsilon_\hot = \sqrt{\omega^2+\Omega_0^2} \approx 1.118$ during the cold and hot thermalisation strokes. The mean photon numbers of the hot and cold thermal baths are $n_\hot = 0.4857$ and $n_\cold = 0.0524$ which translate into the temperatures $T_\hot = 3$ and $T_\cold = 1$ respectively. Both thermal baths couple with the strength $\gamma_\hot = \gamma_\cold = 0.5$ to the work medium. 
With those parameters we find that the specific heat engine discussed in \cref{sec:Power_and_efficiency} has an efficiency under quasi-static operation of $\eta^\text{(sys)}_{\max} = 1 - \varepsilon_\cold/\varepsilon_\hot = 0.106$. As the Carnot efficiency $\eta_\text{C} = 1 - T_\cold/T_\hot = 0.67$ and the Curzon-Ahlborn efficiency $\eta_\text{CA} = 1 - \sqrt{T_\cold/T_\hot} = 0.42$ are larger we do not analyse the performance of the QHE in the regime of the Carnot and the Curzon-Ahlborn efficiency as this would require a more detailed examination and optimisation over the parameter space of the energy gaps $\varepsilon_\x$ and the temperatures $T_\x$. We will rather focus on the effect of different dephasing rates on the performance of a QHE with fixed energy gaps and temperatures, as detailed in the following sections. 

\subsection{Power maximisation}
\label{sec:num_power_maximization}

\begin{figure*}[t]
    \centering
    \includegraphics[width=1\textwidth]{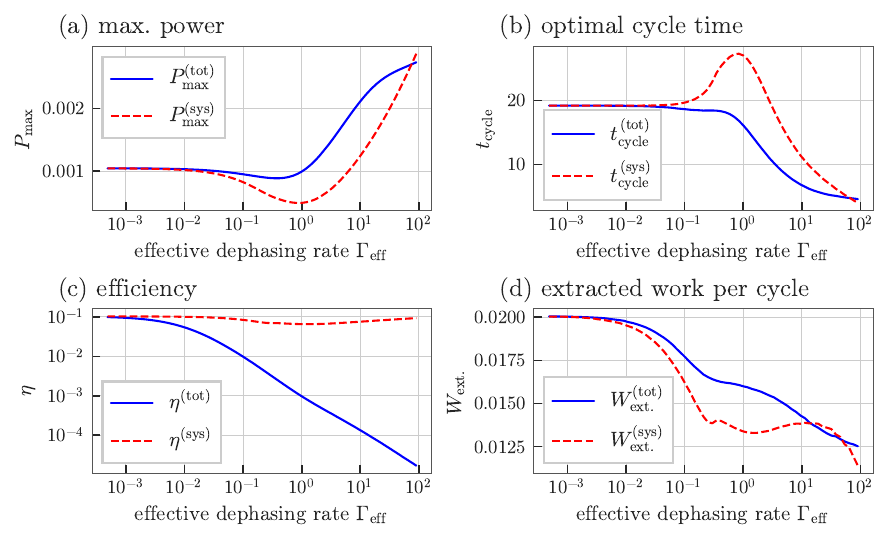}
    \caption{Power maximisation of the QHE for both target functions $P^\text{(tot)}$ (see \cref{eq:powertot}) and $P^\text{(sys)}$ (see \cref{eq:def_P}) for different effective dephasing rates $\Geff$. All quantities and parameters are defined in the main text. The maximum power in (a) is enhanced in the regime of strong dephasing rates whereas it attains a minimum in the regime of moderate dephasing rates for both target functions. The corresponding efficiency $\eta^\text{(tot)}$ in (c) drops if the energy of the dephasing bath is accounted for in the thermodynamical considerations. Otherwise, the efficiency $\eta^\text{(sys)}$ does not change significantly over the whole range of $\Geff$. From panels (b) and (d) we infer that the improvement of the maximum power $P_{\max} = W_\text{ext.}/\tau_\text{cycle}$ is achieved by the ability to run the machine faster within the time $\tau_\text{cycle}$. The gain of speed surpasses the friction-induced losses of the extracted work $W_\text{ext.}$.}
    \label{fig:P_eta_tot}
\end{figure*}

We calculate the maximum power of the QHE for both definitions of the power $P^\text{(tot)}$ (see \cref{eq:powertot}) and $P^\text{(sys)}$ (see (\cref{eq:def_P}). For concreteness, we fix the frequency of the harmonic oscillator $\omega_0 = 1024$ and the coupling constant to the zero temperature dephasing bath to $\gamma = 128$ and only vary the coupling strength $\Gamma$ between the work medium and the harmonic mode. The maximum power for both definitions $P^\text{(\x)}_{\max}$, $\x\in\{\text{tot},\text{sys}\}$, is shown in \cref{fig:P_eta_tot} (a) over the effective dephasing rate $\Geff$ from \cref{eq:Gamma_eff}. These results show that the power of the QHE can be boosted with the help of strong dephasing noise for both definitions of power. 

In our example, we observe an intermediate regime in which the maximum possible power of the QHE is reduced for moderate dephasing rates compared to the dephasing-free QHE. This reduction can be attributed to coherent oscillations of the energy eigenstate populations $p_\e(t) = \bra{\e(t)}\rhot[S][t]\ket{\e(t)}$ and $p_\g(t) = \bra{\g(t)}\rhot[S][t]\ket{\g(t)}$, occurring at finite driving speeds. These oscillations enable a fine-tuning of the compression and expansion stroke durations, $\tau_\com$ and $\tau_\ex$, respectively, so that the final populations closely match the initial populations. This approach approximates a quasi-adiabatic process despite the intermediate non-adiabatic evolution - an effect utilised in the so-called shortcuts to adiabaticity \cite{Petruccione2021,CampoGooldPaternostro2014}. However, the stroke duration is constrained by the characteristic time-scale of these oscillations, as the initial and final populations cannot be made to match if the strokes are shorter than one period of these oscillations. Indeed, the optimal cycle times $\tau_\text{cycle}$ remain essentially constant over a wide range of dephasing rates, as shown in \cref{fig:P_eta_tot} (b). 

In the presence of dephasing noise, those coherent oscillations are damped, leading to losses. As a result, we can no longer obtain a revival of the initial populations at the end of the compression/expansion stroke. The extracted work per cycle is therefore reduced, as shown in \cref{fig:P_eta_tot} (d), and the power drops, as observed in \cref{fig:P_eta_tot} (a). However, the QHE can operate faster in the regime of strong dephasing rates, as the coherent oscillations of the populations are overdamped due to the strong dephasing noise. This overdamping allows for fast, nearly adiabatic compression and expansion strokes. The optimal cycle time drops rapidly in the regime of strong dephasing noise, as shown in \cref{fig:P_eta_tot} (b), and overcompensates the frictional losses. Hence, the maximum power increases, even though less work is extracted from the QHE in each cycle, as observed in \cref{fig:P_eta_tot} (d). 

Finally, we shift our focus to the efficiencies of the QHE, corresponding to the two different power definitions in \cref{fig:P_eta_tot} (c). We observe that if the energy stored in the dephasing bath from \cref{eq:E_deph} is included in the thermodynamical considerations, the corresponding efficiency $\eta^\text{(tot)}$ drops significantly for large dephasing rates. In contrast, the efficiency $\eta^\text{(sys)}$, which only considers the energy changes within the work medium, remains near the maximum efficiency $\eta^\text{(sys)}_{\max} = 0.106$ from \cref{eq:eta_max_sys} over the entire range of effective damping rates.

The major difference in the behaviour of the efficiencies $\eta^\text{(sys)}_{\max}$ and $\eta^\text{(tot)}_{\max}$ indicates that the consideration of the energy of the dephasing bath is crucial for a correct thermodynamical description of dephasing assisted QHEs. If it is ignored, the thermodynamical cost of dephasing noise is hidden and it appears to be possible to increase the power $P_{\max}^\text{(sys)}$ without loss of efficiency $\eta^\text{(sys)}$. In the following \cref{sec:additional_heat_cost} we investigate the origin of the drop of the efficiency $\eta^\text{(tot)}$ in more detail.

\subsection{Additional thermodynamic cost}
\label{sec:additional_heat_cost}

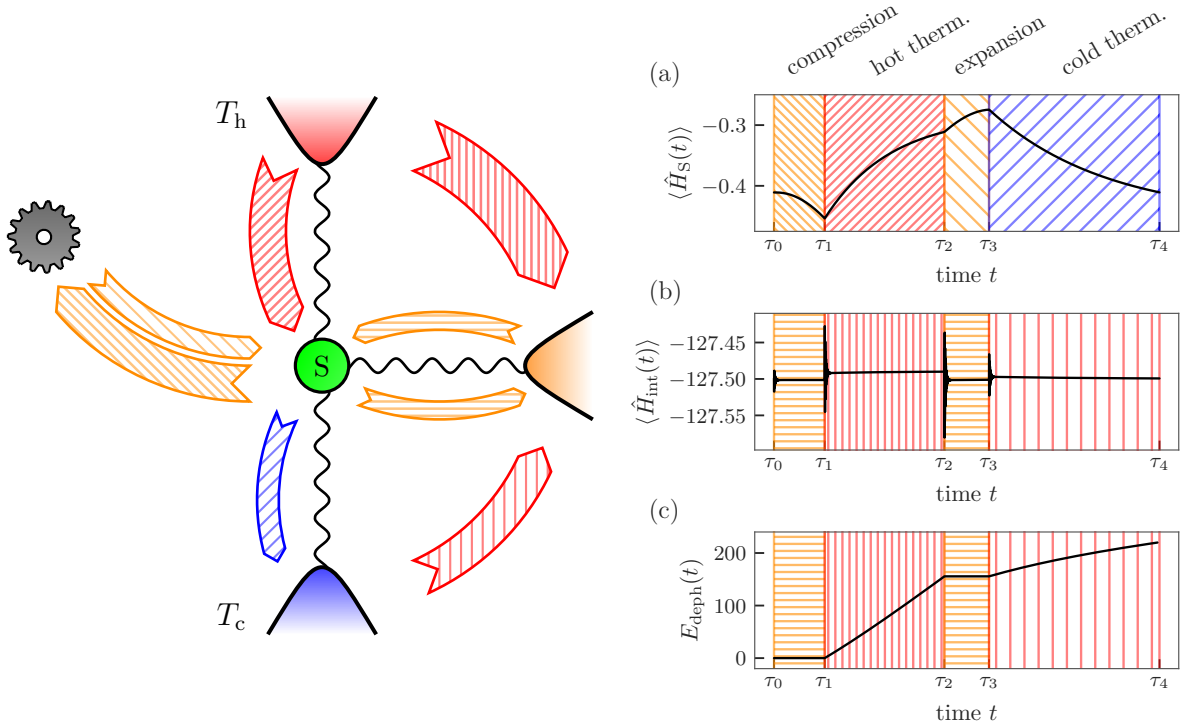
\begin{figure*}[t]
    \centering
    \begin{tikzpicture}
	[scale = 0.95,
	round/.style={circle, draw=black!100, very thick, shade, left color=green!100, right color=green!75, shading angle=60}, 
	]
	
	\draw[color=white] (0,5.25)--(0,-5.25);
	
	\draw[smooth,samples=100,domain=0:3,line width = 1] plot(\x,{sin((\x-1.4)*720)/10});
	\draw[smooth,samples=100,domain=3:-3,line width = 1] plot({sin((\x)*720)/10},\x);
    
    \coordinate (system) at (0,0);
    \node[round] (system) {S};
    
    \draw[draw=white, shade, left color=orange!100, right color=orange!0] (3.75,0.75) .. controls (2.5,0) .. (3.75,-0.75);
    \draw[line width = 1.5] (3.75,0.75) .. controls (2.5,0) .. (3.75,-0.75);
	
    \draw[draw=white, shade, bottom color=red!100, top color=red!0] (0.75,3.75) .. controls (0,2.5) .. (-0.75,3.75);
    \draw[line width = 1.5] (0.75,3.75) .. controls (0,2.5) .. (-0.75,3.75);
    \draw[draw=white, shade, top color=blue!100, bottom color=blue!0] (0.75,-3.75) .. controls (0,-2.5) .. (-0.75,-3.75);
    \draw[line width = 1.5] (0.75,-3.75) .. controls (0,-2.5) .. (-0.75,-3.75);
    
    \node at (-1.25,3.5) {\large $T_\hot$};
    \node at (-1.25,-3.5) {\large $T_\cold$};
    
    
    \pgfmathsetmacro{\alpha}{25}
    \pgfmathsetmacro{\beta}{65}
    \pgfmathsetmacro{\r}{3}
    \pgfmathsetmacro{\R}{.75}
    \pgfmathsetmacro{\Rhalf}{0.5304}
    
    \draw[draw=red,opacity=0.5,fill=red]
    ++(\alpha:\r) arc (\alpha:\beta:\r)-- ++(\beta-45:\Rhalf)-- ++(\beta+45:\Rhalf) arc (\beta:\alpha:\r+\R)
    -- ++(\alpha-135:\Rhalf) -- cycle;
    \foreach \x in {1,...,24}
    {
        \draw[color=white, line width=1.75] (1.25+0.1*\x, 1) -- (1.25+0.1*\x, 3.5);
    }
    \draw[draw=red,line width = 1]
    ++(\alpha:\r) arc (\alpha:\beta:\r)-- ++(\beta-45:\Rhalf)-- ++(\beta+45:\Rhalf) arc (\beta:\alpha:\r+\R)
    -- ++(\alpha-135:\Rhalf) -- cycle;
    
    \pgfmathsetmacro{\alpha}{-25}
    \pgfmathsetmacro{\beta}{-65}
    \pgfmathsetmacro{\r}{3}
    \pgfmathsetmacro{\R}{.5}
    \pgfmathsetmacro{\Rhalf}{0.3536}

    \filldraw[draw=red,opacity=0.5,fill=red]
    ++(\alpha:\r) arc (\alpha:\beta:\r)-- ++(\beta+45:\Rhalf)-- ++(\beta-45:\Rhalf) arc (\beta:\alpha:\r+\R)
    -- ++(\alpha+135:\Rhalf) -- cycle;
    \foreach \x in {1,...,12}
    {
        \draw[color=white, line width=3.25] (1.25+0.15*\x, -0.5) -- (1.25+0.15*\x, -3.5);
    }
    \draw[draw=red,line width = 1]
    ++(\alpha:\r) arc (\alpha:\beta:\r)-- ++(\beta+45:\Rhalf)-- ++(\beta-45:\Rhalf) arc (\beta:\alpha:\r+\R)
    -- ++(\alpha+135:\Rhalf) -- cycle;
    
    \pgfmathsetmacro{\ex}{2}
    \pgfmathsetmacro{\ey}{1.875}
    \pgfmathsetmacro{\alpha}{205}
    \pgfmathsetmacro{\beta}{160}
    \pgfmathsetmacro{\r}{2.5}
    \pgfmathsetmacro{\R}{.5}
    \pgfmathsetmacro{\Rhalf}{0.3536}
    
    \filldraw[draw=red,opacity=0.5,fill=red]
    (\ex,\ey) ++(\alpha:\r) arc (\alpha:\beta:\r)-- ++(\beta+45:\Rhalf)-- ++(\beta-45:\Rhalf) arc (\beta:\alpha:\r+\R)
    -- ++(\alpha+135:\Rhalf) -- cycle;
    \foreach \x in {0,...,15}
    {
        \draw[color=white, line width=1.75] (-0.25, 0.75+0.141*\x) -- (-1.25, -0.25+0.141*\x);
    }
    \foreach \x in {16,...,20}
    {
        \draw[color=white, line width=1.75] (-0.25-0.0707*\x+14*0.0707, 0.75+0.141*\x-0.0707*\x+14*0.0707) -- (-1.25, -0.25+0.141*\x);
    }
    \draw[draw=red,line width = 1]
    (\ex,\ey) ++(\alpha:\r) arc (\alpha:\beta:\r)-- ++(\beta+45:\Rhalf)-- ++(\beta-45:\Rhalf) arc (\beta:\alpha:\r+\R)
    -- ++(\alpha+135:\Rhalf) -- cycle;
    
    \pgfmathsetmacro{\ex}{2}
    \pgfmathsetmacro{\ey}{-1.875}
    \pgfmathsetmacro{\alpha}{195}
    \pgfmathsetmacro{\beta}{155}
    \pgfmathsetmacro{\r}{2.6}
    \pgfmathsetmacro{\R}{.3}
    \pgfmathsetmacro{\Rhalf}{0.2122}
    
    \filldraw[draw=blue,opacity=0.5,fill=blue]
    (\ex,\ey) ++(\alpha:\r) arc (\alpha:\beta:\r)-- ++(\beta+45:\Rhalf)-- ++(\beta-45:\Rhalf) arc (\beta:\alpha:\r+\R)
    -- ++(\alpha+135:\Rhalf) -- cycle;
    \foreach \x in {1,...,10}
    {
        \draw[color=white, line width=3.25] (-0.25, -0.25-0.212*\x) -- (-1.1, -1.1-0.212*\x);
    }
    \draw[draw=blue,line width = 1]
    (\ex,\ey) ++(\alpha:\r) arc (\alpha:\beta:\r)-- ++(\beta+45:\Rhalf)-- ++(\beta-45:\Rhalf) arc (\beta:\alpha:\r+\R)
    -- ++(\alpha+135:\Rhalf) -- cycle;
    
    \definecolor{darkorange}{RGB}{255,140,0}
    
    \pgfmathsetmacro{\ex}{-1}
    \pgfmathsetmacro{\ey}{3}
    \pgfmathsetmacro{\alpha}{270}
    \pgfmathsetmacro{\beta}{220}
    \pgfmathsetmacro{\r}{2.6}
    \pgfmathsetmacro{\R}{.3}
    \pgfmathsetmacro{\Rhalf}{0.2122}
    
    \filldraw[draw=darkorange,opacity=0.5,fill=darkorange]
    (\ex,\ey) ++(\alpha:\r) arc (\alpha:\beta:\r)-- ++(\beta+45:\Rhalf)-- ++(\beta-45:\Rhalf) arc (\beta:\alpha:\r+\R)
    -- ++(\alpha+135:\Rhalf) -- cycle;
    \foreach \x in {1,...,8}
    {
        \draw[color=white, line width=3.25] (-0.25-0.212*\x, 0.0) -- (-1.55-0.212*\x, 1.3);
    }
    \draw[draw=darkorange,line width = 1]
    (\ex,\ey) ++(\alpha:\r) arc (\alpha:\beta:\r)-- ++(\beta+45:\Rhalf)-- ++(\beta-45:\Rhalf) arc (\beta:\alpha:\r+\R)
    -- ++(\alpha+135:\Rhalf) -- cycle;
    
    \pgfmathsetmacro{\ex}{-1}
    \pgfmathsetmacro{\ey}{3}
    \pgfmathsetmacro{\alpha}{270}
    \pgfmathsetmacro{\beta}{220}
    \pgfmathsetmacro{\r}{3}
    \pgfmathsetmacro{\R}{.5}
    \pgfmathsetmacro{\Rhalf}{0.3536}
    
    \filldraw[draw=darkorange,opacity=0.5,fill=darkorange]
    (\ex,\ey) ++(\alpha:\r) arc (\alpha:\beta:\r)-- ++(\beta-45:\Rhalf)-- ++(\beta+45:\Rhalf) arc (\beta:\alpha:\r+\R)
    -- ++(\alpha-135:\Rhalf) -- cycle;
    \foreach \x in {1,...,8}
    {
        \draw[color=white, line width=1.75] (-0.5-0.141*\x, -0.5) -- (-0.5-0.141*\x-0.5-\x*\x*0.0075, \x*\x*0.0075);
    }
    \foreach \x in {9,...,13}
    {
        \draw[color=white, line width=1.75] (-0.5-0.141*\x, -0.5) -- (-0.5-0.141*\x-0.5-1.25, 1.25);
    }
    \draw[draw=darkorange,line width = 1]
    (\ex,\ey) ++(\alpha:\r) arc (\alpha:\beta:\r)-- ++(\beta-45:\Rhalf)-- ++(\beta+45:\Rhalf) arc (\beta:\alpha:\r+\R)
    -- ++(\alpha-135:\Rhalf) -- cycle;
    
    \pgfmathsetmacro{\ex}{1.625}
    \pgfmathsetmacro{\ey}{-2.5}
    \pgfmathsetmacro{\alpha}{70}
    \pgfmathsetmacro{\beta}{110}
    \pgfmathsetmacro{\r}{3}
    \pgfmathsetmacro{\R}{.25}
    \pgfmathsetmacro{\Rhalf}{0.1768}
    
    \filldraw[draw=darkorange,opacity=0.5,fill=darkorange]
    (\ex,\ey) ++(\alpha:\r) arc (\alpha:\beta:\r)-- ++(\beta+45:\Rhalf)-- ++(\beta-45:\Rhalf) arc (\beta:\alpha:\r+\R)
    -- ++(\alpha+135:\Rhalf) -- cycle;
    \foreach \x in {1,...,5}
    {
        \draw[color=white, line width=1.75] (0.25, 0.25+0.1*\x) -- (2.75, 0.25+0.1*\x);
    }
    \draw[draw=darkorange,line width = 1]
    (\ex,\ey) ++(\alpha:\r) arc (\alpha:\beta:\r)-- ++(\beta+45:\Rhalf)-- ++(\beta-45:\Rhalf) arc (\beta:\alpha:\r+\R)
    -- ++(\alpha+135:\Rhalf) -- cycle;
    
    \pgfmathsetmacro{\ex}{1.625}
    \pgfmathsetmacro{\ey}{2.5}
    \pgfmathsetmacro{\alpha}{250}
    \pgfmathsetmacro{\beta}{290}
    \pgfmathsetmacro{\r}{3}
    \pgfmathsetmacro{\R}{.25}
    \pgfmathsetmacro{\Rhalf}{.1768}
    
    \filldraw[draw=darkorange,opacity=0.5,fill=darkorange]
    (\ex,\ey) ++(\alpha:\r) arc (\alpha:\beta:\r)-- ++(\beta+45:\Rhalf)-- ++(\beta-45:\Rhalf) arc (\beta:\alpha:\r+\R)
    -- ++(\alpha+135:\Rhalf) -- cycle;
    \foreach \x in {1,...,5}
    {
        \draw[color=white, line width=1.75] (0.25, -0.25-0.1*\x) -- (2.75, -0.25-0.1*\x);
    }
    \draw[draw=darkorange,line width = 1]
    (\ex,\ey) ++(\alpha:\r) arc (\alpha:\beta:\r)-- ++(\beta+45:\Rhalf)-- ++(\beta-45:\Rhalf) arc (\beta:\alpha:\r+\R)
    -- ++(\alpha+135:\Rhalf) -- cycle;

	\foreach \angle in {65,...,65}
	{
	
	\foreach \x in {0,...,0}
	{
	\foreach \y in {4.25,...,4.25}
	{
	\foreach \R in {0.5,...,0.5}
	{
	\foreach \r in {20,...,20}
	{
	\foreach \tooth in {80,...,80}
	{
	\filldraw[rotate=\angle,color=white, shade, top color=black!60, bottom color=black!40](\x,\y) circle (\R);
	\filldraw[rotate=\angle,color=black, fill=white, thick] (\x,\y) circle (\r*\R*0.01);
	\foreach \n in {0,...,15}
	{
	\foreach \deg in {12,...,12}
	{
	\foreach \adddeg in {3,...,3}
	{
	\foreach \inneradddeg in {2,...,2}
	{
	\draw[rotate=\angle,draw=white,fill=white, thick] ({\x+1.1*\R*cos(2*\n*\deg-\adddeg)},{\y+1.1*\R*sin(2*\n*\deg-\adddeg)}) -- ({\x+\tooth*0.01*\R*cos(2*\n*\deg+\inneradddeg)},{\y+\tooth*0.01*\R*sin(2*\n*\deg+\inneradddeg)}) -- ({\x+\tooth*0.01*\R*cos(2*\n*\deg+\deg-\inneradddeg)},{\y+\tooth*0.01*\R*sin(2*\n*\deg+\deg-\inneradddeg)}) -- ({\x+1.1*\R*cos(2*\n*\deg+\deg+\adddeg)},{\y+1.1*\R*sin(2*\n*\deg+\deg+\adddeg)});
	\draw[rotate=\angle, thick] ({\x+\R*cos(2*\n*\deg-\adddeg)},{\y+\R*sin(2*\n*\deg-\adddeg)}) -- ({\x+\tooth*0.01*\R*cos(2*\n*\deg+\inneradddeg)},{\y+\tooth*0.01*\R*sin(2*\n*\deg+\inneradddeg)}) -- ({\x+\tooth*0.01*\R*cos(2*\n*\deg+\deg-\inneradddeg)},{\y+\tooth*0.01*\R*sin(2*\n*\deg+\deg-\inneradddeg)}) -- ({\x+\R*cos(2*\n*\deg+\deg+\adddeg)},{\y+\R*sin(2*\n*\deg+\deg+\adddeg)});
	\draw[rotate=\angle,thick] ({\x+\R*cos(2*\n*\deg+\deg+\adddeg)},{\y+\R*sin(2*\n*\deg+\deg+\adddeg)}) -- ({\x+\R*cos(2*\n*\deg+2*\deg-\adddeg)},{\y+\R*sin(2*\n*\deg+2*\deg-\adddeg)});
	}
	}
	}
	}
	}
	}
	}
	}
	}
	
	}

\end{tikzpicture}
    \includegraphics[width=0.45\textwidth]{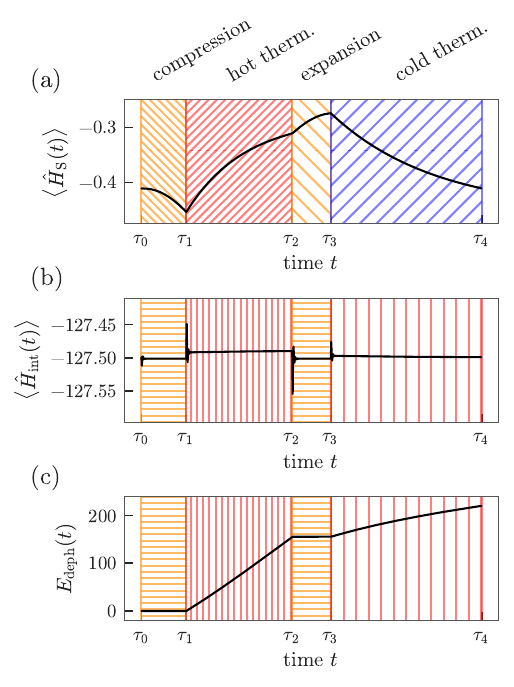}
    \caption{Heat flows within the dephasing assisted QHE with the parameters from \cref{sec:additional_heat_cost} for one cycle. We consider the system, interaction and bath energy separately in order to identify the relevant work and heat flows between the components of the QHE. The work and heat flows are shade coded in order to link the left and right side. The times $\tau_i$ correspond to the times within the Otto cycle as defined in \cref{fig:QHE_Otto_cycle}. The energy changes within the work medium in (a) can be identified as work exchange with the control field and heat exchange with the thermal baths in an Otto cycle (cf. \cref{sec:Quantum_Otto_cycle}). The interaction energy $\langle\Ht[int][t]\rangle$ in (b) shows fast oscillations at the beginning of each stroke that are caused by the instantaneous coupling and decoupling of the thermal baths. However, noticing the scale, we conclude that the energy fluctuations are small compared to the energy fluctuations in the work medium and the dephasing baths. The energy of the dephasing bath in (c) increases during the thermalisation strokes of the Otto cycle. This energy has to originate from the thermal baths as there is no other source of energy during the thermalisation strokes. The transferred heat is much larger than the work and heat exchanges within the work medium and is responsible for the reduction in the efficiency of the QHE.}
    \label{fig:Heat_flows}
\end{figure*}

In \cref{fig:P_eta_tot} (c), we observed that the efficiency of the QHE defined as $\eta^\text{(tot)} = \Delta W_\text{ext.}^\text{(tot)}/\Delta Q_\hot^\text{(tot)}$ drops rapidly with increasing effective dephasing rates $\Geff$. The relatively moderate drop in the extracted work per cycle $\Delta W_\text{ext.}^\text{(tot)}$ that we observed 
in \cref{fig:P_eta_tot} (d) is not sufficient to explain the loss of efficiency and 
it is rather found to be rooted in an enhanced heat consumption $\Delta Q_\hot^\text{(tot)}$
of the QHE. This motivates a closer examination of the internal dynamics of this QHE in 
order to identify the origin of this enhanced heat consumption.

To be specific, we consider a QHE with parameters from \cref{sec:num_power_maximization} and fix the value of $\Gamma = 256$ which corresponds to an effective dephasing rate $\Geff = 15.94$. We chose this noise strength to illustrate a situation where the maximum power output in \cref{fig:P_eta_tot} is improved compared to the noiseless case. In \cref{fig:Heat_flows} we plot the time-dependence of the system energy $\langle\Ht[S][t]\rangle$, the interaction energy $\langle\Ht[int][t]\rangle$ and the dephasing bath energy $E_\text{deph}(t)$ defined in \cref{eq:E_deph}. We identify the energy changes in those quantities and we indicate the work and heat flows between the different components of the QHE graphically on the left hand side. 

First, $\langle\Ht[S][t]\rangle$ in \cref{fig:Heat_flows} (a) describes the energy changes within the system only. One can clearly identify the different strokes of the QHE and the associated work and heat flows which behave as expected for a QHE. \cref{fig:Heat_flows} (b) on the other hand shows that there are only small energy fluctuations in the interaction energy $\langle\Ht[int][t]\rangle$ and \cref{fig:Heat_flows} (c) shows that changes in the bath energy $E_\text{deph}(t)$ remain negligible during the compression and expansion strokes.
The main result of this analysis is the observation of an energy change in the dephasing bath during the thermalisation strokes with both the hot and the cold thermal bath. In both cases we observe that a large amount of heat is transferred from the thermal bath to the dephasing bath during these strokes. The source of this heat must be the thermal baths as the Hamiltonian is time-independent during the thermalisation strokes and there is no other source of energy and no work is being done during either. As this amount of heat is large compared to the heat that is transferred from the thermal bath to the system, the efficiency in \cref{fig:P_eta_tot} (c) drops rapidly if the energy of the dephasing bath is taken into account. 

In summary, the dephasing bath is able to increase the power output by allowing for faster compression and expansion strokes and this process consumes a negligible amount of energy of the dephasing bath. However, there is a large irreversible heat flow to the dephasing bath during the thermalisation strokes which, when accounted for, is responsible for the reduction of the thermodynamical efficiency of the QHE. In the following \cref{sec:analytical_dissipated_heat} we provide an analytical underpinning for this reasoning by computing analytical upper and lower bounds on the  additional heat flow.

\section{Analytical treatment}
\label{sec:analytical_analysis}

In this section we derive analytic expressions for the numerical results of \cref{sec:numerical_results}. In \cref{sec:steady_state} we use numerical evidence to justify an Ansatz for the joint state of heat engine and dephasing environment and then derive an analytic expression of the joint state of work medium and dephasing bath. \cref{sec:analytical_dissipated_heat} uses this expression to calculate bounds on the dissipated heat. The validity of these analytical results is then checked by comparison with the numerical results of \cref{sec:numerical_results}.

\subsection{State of work medium and dephasing bath}
\label{sec:steady_state}

\begin{figure}[t]
    \centering
    \includegraphics[width=0.45\textwidth]{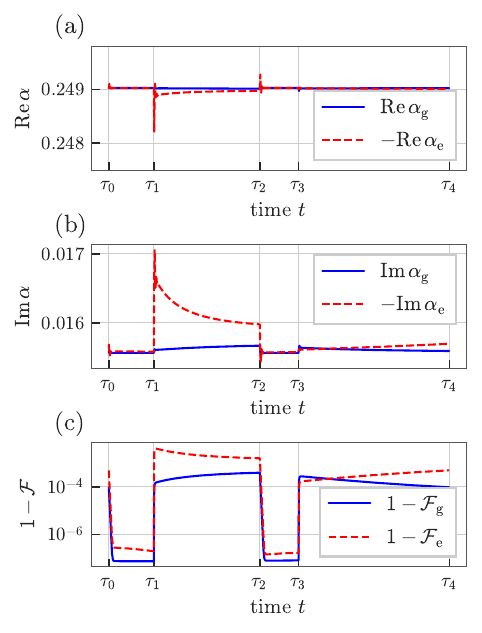}
    \caption{Properties of the state of the dephasing bath for the simulations from \cref{sec:additional_heat_cost} which are used to justify the approximate shape of the state from \cref{eq:state_structure}. In panels (a) and (b), we show the displacements $\alpha_\e$ and $\alpha_\g$ of the oscillator states $\rhot[\osc,\e][t]$ and $\rhot[\osc,\g][t]$ over a whole cycle of the QHE where the times $\tau_i$ have been defined in \cref{fig:QHE_Otto_cycle}. The displacements of both states are nearly constant over the whole cycle and satisfy $\alpha_\g \approx -\alpha_\e \approx \text{const}$. The distance of the states $\rhot[\osc,\e][t]$ and $\rhot[\osc,\g][t]$ to coherent states is shown in panel (c) where we used the fidelity $\mathcal{F}$ from \cref{eq:fidelity}.}
    \label{fig:displacement_fidelity}
\end{figure}

Numerical evaluation of the joint quantum state of system and dephasing environment verifies
that coherences in the system are, as expected, vanishingly small. This justifies and motivates an Ansatz 
\begin{align}
\label{eq:state_structure}
    \rhot[][t] =& \, p_\e(t) \ketbra{\e(t)} \otimes \rhot[\osc,\e][t] \nonumber \\
    +& p_\g(t) \ketbra{\g(t)} \otimes \rhot[\osc,\g][t].
\end{align}
Further numerical calculations show that the states $\rhot[\osc,\x]$ can be approximated well as coherent states with 
amplitude $\alpha_\x(t)$ and we use those to calculate the overlap 
\begin{align}
\label{eq:fidelity}
    \mathcal{F}_\x(t) = \bra{\alpha_\x(t)}\rhot[\osc,\x][t]\ket{\alpha_\x(t)}
\end{align}
of the coherent state $\ket{\alpha_\x(t)}$ and the state $\rhot[\osc,\x][t]$ to quantify the fidelity of the approximation. The results of these calculations are shown in \cref{fig:displacement_fidelity} with the parameters from \cref{sec:additional_heat_cost}. The displacements of two states $\rhot[\osc,\e][t]$ and $\rhot[\osc,\g][t]$ are nearly constant in time over the whole cycle of the QHE. Moreover, both displacements take nearly the same value but with opposite sign, i.e. $\alpha_\e(t) \approx -\alpha_\g(t) \approx - \alpha = \text{const}$. The fidelity thus obtained is very close to unity. As a result we use the excellent approximation 
\begin{align}
\label{eq:ansatz_steady_state}
    \rhot[][t] \approx& \, p_\e(t) \ketbra{\e(t)} \otimes \ketbra{-\alpha} \nonumber \\
    +& p_\g(t) \ketbra{\g(t)} \otimes \ketbra{\alpha}
\end{align}
as an Ansatz for the state of the QHE in the following analytical treatment.

\begin{figure}[t]
    \centering
    \includegraphics[width=0.45\textwidth]{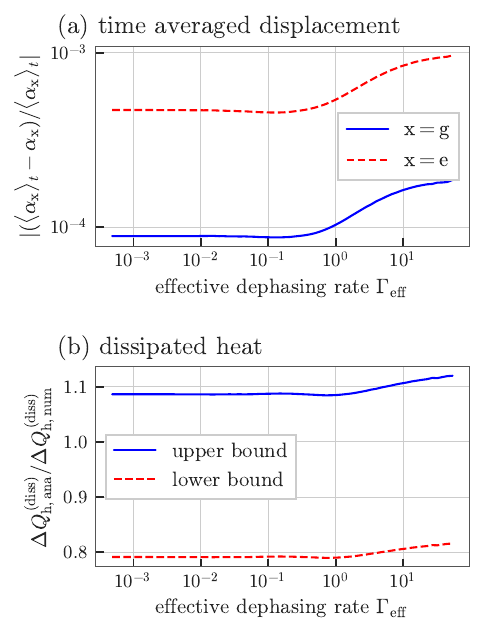}
    \caption{The accuracy of our analytical results for the displacements $\alpha_\g$ and $\alpha_\e$ of the harmonic mode in the steady state and the estimation of the dissipated heat $\Delta Q_\hot^\text{(diss)}$ is shown. We use the simulations from \cref{sec:num_power_maximization} with variable $\Geff$ in both cases. (a) shows that the time-averaged displacement $\langle \alpha_\x \rangle_t$ matches the analytical value of $\alpha_\x$ from \cref{eq:alpha_ss} well for both excited and ground state. (b) shows the ratio between the analytical bounds for the dissipated heat from \cref{eq:diss_heat_estimation} and the actual numerical result, which shows that the upper and lower bounds deviate from the actual result by 10\% and 20\% respectively, which is a quite good result given the wide range of $\Delta Q_\hot^\text{(diss)}$ which is proportional to $\Geff$.}
    \label{fig:alpha_Q_diss}
\end{figure}

The next step is a derivation of the value of $\alpha$ under the assumption that the work medium in contact with the strong dephasing bath is close to being equilibrated, i.e. in steady state. We neglect the effect of the thermal baths and include those in the next section perturbatively. Inserting the Ansatz \cref{eq:ansatz_steady_state} in the corresponding equation for the steady state 
\begin{align}
\label{eq:steady_state_equation}
    0 = - \ii \, \commutator{\Ht[\DAMPF][t_0]}{\rhot[s]} + \mathcal{L}_\osc [\rhot[s]]
\end{align}
for a specific time $t_0$ we find
\begin{align}
    0 = p_\e&(t_0) \ketbra{\e(t_0)} \nonumber \\
    \otimes \Big\{&(\ii \, \Gamma \alpha - \ii \, \Gamma \alpha^\ast + \gamma \abs{\alpha}^2) \,\, \ketbra{-\alpha} \nonumber \\
    +&(-\ii \, \Gamma +\ii \, \omega_0 \alpha + \frac{\gamma}{2} \alpha) \,\, \hat{b}^\dagger \ketbra{-\alpha} \nonumber \\
    +&(\ii \, \Gamma -\ii \, \omega_0 \alpha^\ast + \frac{\gamma}{2} \alpha^\ast) \,\, \ketbra{-\alpha} \hat{b} \Big\} \nonumber \\
    +p_\g&(t_0) \ketbra{\g(t_0)} \nonumber \\
    \otimes \Big\{&(\ii \, \Gamma \alpha - \ii \, \Gamma \alpha^\ast + \gamma \abs{\alpha}^2) \,\, \ketbra{\alpha} \nonumber \\
    +&(\ii \, \Gamma -\ii \, \omega_0 \alpha - \frac{\gamma}{2} \alpha) \,\, \hat{b}^\dagger \ketbra{\alpha} \nonumber \\
    +&(-\ii \, \Gamma +\ii \, \omega_0 \alpha^\ast - \frac{\gamma}{2} \alpha^\ast) \,\, \ketbra{\alpha} \hat{b} \Big\}
\end{align}
which yields
\begin{align}
\label{eq:alpha_ss}
    \alpha = 2 \Gamma \, \frac{2 \omega_0 + \ii \gamma}{4 \omega_0^2 + \gamma^2}.
\end{align}
The populations $p_\e$ and $p_\g$ of the excited and ground state remain free parameters, that are fixed by the time-dependent dynamics of the Otto cycle. In \cref{fig:alpha_Q_diss} (a) we show the time-averaged displacement $\langle\alpha_\x\rangle_t$ from numerical simulations for different values of $\Geff$ and compare those results to the analytical results which show good agreement. Moreover, we calculate the trace distance of the numerically and analytically calculated joint states of work medium and dephasing bath in \cref{fig:increase_power} (d).

\subsection{Dissipated heat}
\label{sec:analytical_dissipated_heat}
\begin{figure*}
    \centering
    \begin{tikzpicture}[scale=0.95]
\draw[->,line width=1] (0,-2) -- (0,2) node[left] {\small Im$\{\alpha\}$} coordinate (y axis);
\draw[->,line width=1] (-2,0) -- (2,0) node[below] {\small Re$\{\alpha\}$} coordinate (x axis);
\draw[->,line width=1] (6,-2) -- (6,2) node[left] {\small Im$\{\alpha\}$} coordinate (y axis);
\draw[->,line width=1] (4,0) -- (8,0) node[below] {\small Re$\{\alpha\}$} coordinate (x axis);
\draw[->,line width=1] (12,-2) -- (12,2) node[left] {\small Im$\{\alpha\}$} coordinate (y axis);
\draw[->,line width=1] (10,0) -- (14,0) node[below] {\small Re$\{\alpha\}$} coordinate (x axis);
\draw[line width=1] (1,1.5) circle[radius=0.5];
\draw[line width=1] (-1,-1.5) circle[radius=0.5];
\node at (1,2.75) {\small $\hat{\rho}_\mathrm{g}^{(0)} = |\alpha\rangle\langle\alpha|$};
\node at (-1,-2.375) {\small $\hat{\rho}_\mathrm{e}^{(0)} = |-\alpha\rangle\langle-\alpha|$};
\draw[line width=1] (13,1.5) circle[radius=0.5];
\draw[line width=1] (11,-1.5) circle[radius=0.5];
\node at (13,2.75) {\small $\hat{\rho}_\mathrm{g}^{(2)} = |\alpha\rangle\langle\alpha|$};
\node at (11,-2.375) {\small $\hat{\rho}_\mathrm{e}^{(2)} = |-\alpha\rangle\langle-\alpha|$};
\draw[dashed,line width=1] (12.8,1.2) circle[radius=0.5];
\draw[dashed,line width=1] (11.2,-1.2) circle[radius=0.5];
\draw[line width=1] (6.8,1.2) circle[radius=0.5];
\draw[line width=1] (5.2,-1.2) circle[radius=0.5];
\draw[dashed,line width=1] (7,1.5) circle[radius=0.5];
\draw[dashed,line width=1] (5,-1.5) circle[radius=0.5];
\node at (7,2.75) {\small $\hat{\rho}_\mathrm{g}^{(1)} = (1-\varepsilon)|\alpha\rangle\langle\alpha|+\varepsilon|-\alpha\rangle\langle-\alpha|$};
\node at (5,-2.5) {\small $\hat{\rho}_\mathrm{e}^{(1)} = (1-\varepsilon)|-\alpha\rangle\langle-\alpha|+\varepsilon|\alpha\rangle\langle\alpha|$};
\draw[->,line width=1] (1.8,0.25) .. controls (3,0.75) .. (4.2,0.25);
\node at (3,1) {\small $\mathcal{G}_{\mathrm{d}t}^\mathrm{(th)}$};
\draw[->,line width=1] (7.8,0.25) .. controls (9,0.75) .. (10.2,0.25);
\node at (9,1) {\small $\mathcal{U}_{\mathrm{d}t} \circ \mathcal{G}_{\mathrm{d}t}^\mathrm{(osc)}$};
\draw[->,line width=1] (-2,-7.5) -- (2,-7.5) node[right] {\small $\hat{\rho}$} coordinate (x axis);
\draw[->,line width=1] (-2,-7.5) -- (-2,-5) node[left] {\small $\langle \hat{H}_\mathrm{S} \rangle$} coordinate (y axis);
\draw[line width=1] (-1.925,-7) -- (-2.075,-7);
\draw[line width=1] (-1.925,-5.5) -- (-2.075,-5.5);
\draw[<->,line width=1] (-2.75,-7) -- (-2.75,-5.5);
\node[fill=white] at (-2.75,-6.25) {\small $\mathrm{d}\langle \hat{H}_\mathrm{S} \rangle$};
\draw[line width=1] (-1.75,-7.425) -- (-1.75,-7.575) node[below] {\small $\hat{\rho}_0$};
\draw[line width=1] (0,-7.425) -- (0,-7.575) node[below] {\small $\hat{\rho}_1$};
\draw[line width=1] (1.75,-7.425) -- (1.75,-7.575) node[below] {\small $\hat{\rho}_2$};
\draw[->,line width=1] (4,-7.5) -- (8,-7.5) node[right] {\small $\hat{\rho}$} coordinate (x axis);
\draw[->,line width=1] (4,-7.5) -- (4,-5) node[left] {\small $\langle \hat{H}_\mathrm{int} \rangle$} coordinate (y axis);
\draw[line width=1] (4.075,-7) -- (3.925,-7);
\draw[line width=1] (4.075,-5.5) -- (3.925,-5.5);
\draw[<->,line width=1] (3.25,-7) -- (3.25,-5.5);
\node[fill=white] at (3.25,-6.25) {\small $\mathrm{d}\langle \hat{H}_\mathrm{int} \rangle$};
\draw[line width=1] (4.25,-7.425) -- (4.25,-7.575) node[below] {\small $\hat{\rho}_0$};
\draw[line width=1] (6,-7.425) -- (6,-7.575) node[below] {\small $\hat{\rho}_1$};
\draw[line width=1] (7.75,-7.425) -- (7.75,-7.575) node[below] {\small $\hat{\rho}_2$};
\draw[->,line width=1] (10,-7.5) -- (14,-7.5) node[right] {\small $\hat{\rho}$} coordinate (x axis);
\draw[->,line width=1] (10,-7.5) -- (10,-5) node[left] {\small $E_\mathrm{deph}$} coordinate (y axis);
\draw[line width=1] (10.075,-7) -- (9.925,-7);
\draw[line width=1] (10.075,-5.5) -- (9.925,-5.5);
\draw[<->,line width=1] (9.25,-7) -- (9.25,-5.5);
\node[fill=white] at (9.25,-6.25) {\small $\mathrm{d}\langle \hat{H}_\mathrm{int} \rangle$};
\draw[line width=1] (10.25,-7.425) -- (10.25,-7.575) node[below] {\small $\hat{\rho}_0$};
\draw[line width=1] (12,-7.425) -- (12,-7.575) node[below] {\small $\hat{\rho}_1$};
\draw[line width=1] (13.75,-7.425) -- (13.75,-7.575) node[below] {\small $\hat{\rho}_2$};
\node[right] at (-2.5,-4.25) {\small (d) system energy};
\node[right] at (-2.5,4) {\small (a) $\hat{\rho}_\mathrm{0}$};
\node[right] at (3.5,-4.25) {\small (e) interaction energy};
\node[right] at (3.5,4) {\small (b) $\hat{\rho}_\mathrm{1} = \mathcal{G}_{\mathrm{d}t}^\mathrm{(th)}[\hat{\rho}_0]$};
\node[right] at (9.5,-4.25) {\small (f) bath energy};
\node[right] at (9.5,4) {\small (c) $\hat{\rho}_\mathrm{2} = \mathcal{U}_{\mathrm{d}t}\circ\mathcal{G}_{\mathrm{d}t}^\mathrm{(osc)}[\hat{\rho}_1]$};
\draw[line width=1,color=red] (-1.75,-7) -- (0,-5.5) -- (1.75,-5.5);
\draw[line width=1,color=red] (4.25,-7) -- (6,-5.5) -- (7.75,-7);
\draw[line width=1,color=red] (10.25,-7) -- (12,-7) -- (13.75,-5.5);
\end{tikzpicture}
    \caption{Mechanism of the heat transfer to the dephasing bath during a single time-step $\dd t$ of the thermalisation stroke. 
    Writing the general state as $\hat{\rho}_i = p_\mathrm{e}^{(i)}|\mathrm{e}\rangle\langle\mathrm{e}|\otimes \hat{\rho}_\mathrm{e}^{(i)} + p_\mathrm{g}^{(i)}|\mathrm{g}\rangle\langle\mathrm{g}|\otimes \hat{\rho}_\mathrm{g}^{(i)}$, we decompose the time-evolution in the first order Suzuki-Trotter decomposition $\rhot[][t+\dd t] = \rhot[2] = \mathcal{U}_{\dd t} \circ \mathcal{G}_{\dd t}^{\text{(osc)}} \circ \mathcal{G}_{\dd t}^\text{(th)}[\rhot[0]]$ with $\rhot[0] = \rhot[][t]$. (a)-(c) shows the evolution of the state of the harmonic mode which starts in the steady state \cref{eq:ansatz_steady_state}. (d)-(f) shows the energy of the system $\langle \Ht[S] \rangle$, the interaction energy $\langle \Ht[int] \rangle$ and the bath energy $E_\text{deph}$ during this time-step. Initially in (a), the states $\rhot[e]^{(0)}$ and $\rhot[g]^{(0)}$, that are correlated to the excited state $\ket{\e}$ and ground state $\ket{\g}$ of the system respectively, are coherent states $\ket{\pm \alpha}$. Moving from (a) to (b) the thermal propagator $\mathcal{G}_{\dd t}^\text{(th)}$ mixes the excited and ground state of the system which, in turn, mixes the states $\rhot[\e]$ and $\rhot[\g]$ of the harmonic mode and effectively displaces them towards the origin. This displacement leads the thermal propagator to change the system energy by $\dd \langle \Ht[S] \rangle$ (see \cref{eq:dH_sys}) in (d) and the interaction energy by $\dd \langle \Ht[int] \rangle$ (see \cref{eq:dH_int}) in (e).  The reduced state and thus the energy of the harmonic mode in (f) does not change during this evolution. From (b) to (c) the unitary evolution and the damping propagator push the states $\rhot[\e]$ and $\rhot[\g]$ back to their original position at $\ket{\pm \alpha}$ as this is the steady state of the system with respect to those two propagators (cf. \cref{sec:steady_state}). The population of the work medium does not change in this evolution due to the $\szt$-coupling to the harmonic mode whence the system energy in (d) does not change. However, the pushback of the states $\rhot[\e]$ and $\rhot[\g]$ lowers the interaction energy in (e) by $\dd \langle \Ht[int] \rangle$ to its initial value. This energy difference is consequently absorbed by the damping term of the harmonic mode as it cannot flow to the work medium as the corresponding Hamiltonians $\Ht[S]$ and $\Ht[int]$ commute.}
    \label{fig:dissipated_heat}
\end{figure*}
In \cref{sec:numerical_results}, we observed that the additional heat consumption of the QHE occurs during the thermal strokes that couple the QHE to the thermal reservoirs. So far we have neglected the effect of the thermal baths and, in the following, we account for those perturbatively to leading order in our analytical treatment starting with the Ansatz \cref{eq:ansatz_steady_state} and \cref{eq:alpha_ss} which, as was demonstrated in Section \cref{sec:steady_state}, is a very good lowest order approximation to the actual state during the entire cycle. To handle the time evolution, we apply a Suzuki-Trotter decomposition \cite{Trotter1959, Suzuki1976} on the time-evolution as described in  \cref{eq:Suzuki_Trotter_decomposition}. For a small time step $\dd t$ the action of the thermal propagator $\mathcal{G}^{(\text{th})}_{\dd t}[\rhot[s]]$ on the steady state can be approximated as
\begin{align}
\label{eq:G_th_approx}
    \rhot[s] + \dd \rhot[s] &= \mathcal{G}^{(\text{th})}_{\dd t}[\rhot[s]]\nonumber\\
    &= \rhot[s] + \mathcal{L}_{\text{th}} [\rhot[s]] \dd t + \order{\dd t^2}
\end{align}
where the thermal Lindbladian $\mathcal{L}_\text{th}$ has been defined in \cref{eq:L_hot,eq:L_cold} for the hot ($\text{th} = \hot$) and the cold ($\text{th} = \cold$) bath respectively. Substituting the explicit form of the steady state \cref{eq:ansatz_steady_state} leads to the differential change of the state
\begin{align}
\label{eq:drho_steadystate}
    \dd \rhot[s] = & \mathcal{L}_{\text{th}} [\rhot[s]] \dd t \nonumber\\
                  = & \gamma_\text{th} \szt[t] \otimes (p_\g \, n_\text{th} \ketbra{\alpha} \\
                  & \hspace*{0.5cm} - p_\e \, (n_\text{th}+1) \ketbra{-\alpha}) \dd t. \nonumber
\end{align}
This, in turn, yields the differential change of the total energy in work medium and dephasing bath
\begin{align}
    \dd \expval{\Ht[\DAMPF](t)} = \Tr{\Ht[\DAMPF](t) \, \dd \rhot[s]}
\end{align}
that can be divided into the differential heat flow to the work medium
\begin{align}
\label{eq:dH_sys}
    \dd \expval{\Ht[S][t]} & = \Tr{\Ht[S][t] \, \dd \rhot[s]} \\
    & = \varepsilon(t) \, \gamma_\text{th} \, (p_\g \, n_\text{th} - p_\e \, (n_\text{th}+1)) \, \dd t\nonumber
\end{align}
and a differential change of the interaction energy
\begin{align}
\label{eq:dH_int}
    \dd \expval{\Ht[int][t]} & = \Tr{\Ht[int][t] \, \dd \rhot[s]} \nonumber \\
    & = 4 \Gamma \gamma_\text{th} \Re{\alpha} (n_\text{th} + p_\e) \dd t
\end{align}
which is positive since $\Re{\alpha}>0$ according to numerical and analytical results from \cref{sec:steady_state}. The energy of the harmonic mode does not change, $\dd \langle \Ht[osc][t] \rangle = 0$, because $\szt[t]$ in the differential change of the state \cref{eq:drho_steadystate} is traceless. By integrating the differential heat flow to the work medium from \cref{eq:dH_sys} we obtain the total heat
\begin{align}
\label{eq:Delta_Q_sys_thermal}
    \Delta Q^\text{(sys)}_\thermal &= \int \limits_{\tau_i}^{\tau_{i+1}} \dd t \, \frac{\dd \expval{\Ht[S][t]}}{\dd t}
\end{align}
that is exchanged between the thermal bath and the work medium during a thermalisation stroke from $\tau_i$ to $\tau_{i+1}$.
The result from \cref{eq:dH_int} shows that the thermal bath does not only exchange the heat with the work medium but also with the pure dephasing bath via the interaction energy. This energy is lost to the work medium and cannot be converted to useful work as the interaction Hamiltonian commutes with the Hamiltonian of the work medium. The details of the mechanism which leads to this heat transfer are explained in more detail in \cref{fig:dissipated_heat}.

As a result of this transfer, energy is dissipated to the dephasing bath and we integrate $\dd \expval{\Ht[int][t]}$ from \cref{eq:dH_int} to obtain the total dissipated heat
\begin{align}
\label{eq:DeltaQ_diss}
    \Delta Q^\text{(diss)}_\hot &= \int \limits_{\tau_1}^{\tau_2} \dd t \,\, \frac{\dd \expval{\Ht[int][t]}}{\dd t} \nonumber \\
    &= 4 \Gamma \gamma_\hot \int \limits_{\tau_1}^{\tau_2} \dd t \Re{\alpha(t)} (n_\hot + p_\e(t))
\end{align}
which is transferred from the hot thermal bath to the dephasing bath during the hot thermalisation stroke from $\tau_1$ to $\tau_2$. Analogously, we calculate the heat
\begin{align}
\label{eq:DeltaQ_diss_cold}
    \Delta Q^\text{(diss)}_\cold &= \int \limits_{\tau_3}^{\tau_4} \dd t \,\, \frac{\dd \expval{\Ht[int][t]}}{\dd t} \nonumber \\
    &= 4 \Gamma \gamma_\cold \int \limits_{\tau_3}^{\tau_4} \dd t \Re{\alpha(t)} (n_\cold + p_\e(t))
\end{align}
that is transferred from the cold bath to the dephasing bath during the cold thermalisation stroke. Note, that the amount of heat is positive for both the hot and the cold thermalisation stroke. We can further simplify these equations with the assumption that the displacement $\alpha$ takes the constant value from \cref{eq:alpha_ss} which is supported by \cref{sec:steady_state}. Moreover, the population of the excited state is bounded by the excited state population of the hot and cold thermal equilibrium states $p_{\e,\hot} = \frac{n_\hot}{2 n_\hot +1}$ and $p_{\e,\cold} = \frac{n_\cold}{2 n_\cold +1}$ because the temperature of the work medium lies between the two temperatures of the thermal baths throughout the whole Otto cycle. Hence, making use of the effective dephasing rate $\Geff(\Gamma,\gamma,\omega_0) = 8\Gamma^2\gamma/(\gamma^2 + 4\omega_0^2)$ (see \cref{eq:Gamma_eff}) we can upper and lower bound the dissipated heat
\begin{widetext}
\begin{align}
    \label{eq:diss_heat_estimation}
    \frac{2 \omega_0 \gamma_\hot \tau_\hot}{\gamma} \Geff(\Gamma,\gamma,\omega_0)
    \left(n_\hot + \frac{n_\cold}{2 n_\cold +1}\right)
    \leq \Delta Q^\text{(diss)}_\hot \leq 
    \frac{2 \omega_0 \gamma_\hot \tau_\hot}{\gamma}
    \Geff(\Gamma,\gamma,\omega_0)  \left(n_\hot + \frac{n_\hot}{2 n_\hot +1}\right)
\end{align}
\end{widetext}
which is transferred to the dephasing bath during hot thermalisation.
An analogous expression can be derived for the heat that is transferred during the cold thermalisation. The result from \cref{eq:diss_heat_estimation} is supported by the numerical results in \cref{fig:alpha_Q_diss} (b).

\section{Optimising heat engine performance}
\label{sec:optimising_performance}

The additional thermodynamic cost introduced by the dephasing bath, as examined numerically in \cref{sec:additional_heat_cost} and analytically in \cref{sec:analytical_dissipated_heat}, typically has a negative impact on the heat engine's performance. While it may  increase its power it will, at the same time, lower its efficiency. Here, we turn our attention to whether the engine’s thermodynamic efficiency, at a specific power level and under the influence of dephasing noise, can be improved to approach the efficiency of quasi-static operation, taking into account all associated thermodynamic costs.

We explore two potential strategies. In the first approach, \cref{sec:increasing_the_efficiency}, the dephasing environment remains continuously connected to the system throughout the entire cycle. As discussed earlier, this induces additional heat flow between the thermal baths and the dephasing environment during the thermalisation strokes, which is the source of the extra thermodynamic cost. We constrain these heat flows analytically and then show that, by adjusting the parameters of the spectral density, it is possible to maintain the engine's power while reducing the additional thermodynamic cost arbitrarily.

The second strategy, \cref{sec:decouple_deph_bath}, recognises that the dephasing bath primarily enhances efficiency during the expansion and compression strokes, whereas the harmful heat dissipation, which reduces efficiency, occurs during the thermalisation strokes. To mitigate this, we decouple and recouple the dephasing bath before and after each thermal stroke. We quantify the thermodynamic cost in the decoupling and recoupling steps and show, again, that it can be made to vanish in a suitable parameter limit.

\subsection{Continuously connected dephasing bath}
\label{sec:increasing_the_efficiency}

In this section, we analyse the thermodynamic efficiency of a quantum Otto engine that remains continuously connected to a dephasing heat bath. We present the special case of $T=0$ in \cref{sec:increasing_the_power} and discuss certain pathologies that disappear in the finite-T case that we discuss in this section.
For finite temperatures the damping term of the DAMPF mode from \cref{eq:L_osc} generalises to
\begin{align}
    \mathcal{L}_\osc[\rhot] =& \, \gamma \, (1 + n_\osc) \, \left(\hat{b} \, \rhot \, \hat{b}^\dagger - \frac{1}{2} \{\hat{b}^\dagger\hat{b}, \rhot\}\right) \nonumber \\
    +& \, \gamma \, n_\osc \, \left(\hat{b}^\dagger \, \rhot \, \hat{b} - \frac{1}{2} \{\hat{b}\,\hat{b}^\dagger, \rhot\}\right)
\end{align}
with the mean photon number $n_\osc = (\e^{\omega_0 \beta} - 1)^{-1}$ at inverse temperature
$\beta=T^{-1}$. The remaining parts of the equations of motion from \cref{sec:dephasing_DAMPF_bath} 
are unaffected by the introduction of the finite temperature in the dephasing bath. Given the environmental spectral density $J(\omega)$, \cref{eq:spectral_density}, the corresponding thermalised spectral density of the finite-temperature dephasing environment \cite{SomozaMartyLimHuelgaPlenio2019,Tamascelli2019} is given by
\begin{align}
\label{eq:spectrum_thermalized_DAMPF}
    J_\beta(\omega) =& \, \frac{1}{2} J(\omega) \left(\coth{\frac{\beta \omega_0}{2}} + 1\right)\nonumber \\
    +& \, \frac{1}{2} J(-\omega) \left(\coth{\frac{\beta \omega_0}{2}} - 1\right)\, .
\end{align}
Note, that \cite{Tamascelli2019} defines the thermal spectral density slightly different as
\begin{align}
\label{eq:thermalised_spectrum_alt}
    J_\beta^\prime(\omega) = \text{sgn}(\omega) \, J^\prime(|\omega|) \, \frac{\coth \frac{\omega \beta}{2} + 1}{2}
\end{align}
with the spectral density
\begin{align}
\label{eq:spectrum_alt}
    J^\prime(\omega) = \frac{2\Gamma^2 \gamma \, \omega_0 \, \omega}{\gamma^2 \omega^2 + (\omega^2 - \omega_0^2)^2}.
\end{align}
In the limit $\frac{\gamma}{\omega_0} \ll 1$ that is relevant to our discussion, both $J^\prime(\omega) \rightarrow J(\omega)$ and $J^\prime_\beta(\omega) \rightarrow J_\beta(\omega)$ converge towards the previously defined spectral densities. The condition $\frac{\gamma}{\omega_0} \ll 1$ is typically fulfilled if we restrict ourselves to the physical regime of a weakly damped DAMPF mode.

In the limit $\beta \rightarrow \infty$, i.e. $\coth\frac{\beta \omega_0}{2} \rightarrow 1$, the thermalised spectral 
density of \cref{eq:spectrum_thermalized_DAMPF} converges to the zero-temperature spectral density $J(\omega)$ as expected, whereas in the 
high-temperature limit, where $\coth\frac{\beta \omega_0}{2} \gg 1$, we obtain two Lorentzian peaks 
with very similar shape located at $\pm \omega_0$. As we will see in the following, the symmetry of 
the spectrum turns out to be a very convenient property for the minimisation of the heat dissipation rate.

Because \cref{eq:diss_heat_estimation} is expressed in terms of the effective dephasing rate that we wish to keep constant, we now  calculate the effective dephasing rate for the thermalised spectral density based on the 
decoherence function defined in \cref{eq:dec_func_integral,eq:dec_func_amplitude} and the spectral density \cref{eq:spectral_density} to find
\begin{align}
\label{eq:decoherence_function_thermal}
    \Gamma(t) =& -\frac{4}{\pi} \int \limits_{-\infty}^\infty \dd \omega \, J_\beta(\omega) \frac{1-\cos(\omega t)}{\omega^2} \nonumber \\
    \cong & -\frac{4}{\pi} \coth{\frac{\beta \omega_0}{2}} \int \limits_{-\infty}^\infty \dd \omega \, J(\omega) \frac{1-\cos(\omega t)}{\omega^2}\nonumber\\
    \cong &-\left[\frac{8\Gamma^2\gamma}{\gamma^2 + 4\omega_0^2} \, \coth{\frac{\beta \omega_0}{2}}\right] t
\end{align}
in the limit of large $t$ where we exploited the symmetry 
of the thermalised spectral density $J_\beta(\omega)$ from \cref{eq:spectrum_thermalized_DAMPF} and 
the calculation in \cref{sec:effective_dephasing_rate}. Hence, the effective dephasing rate of a single 
DAMPF oscillator at finite inverse temperature $\beta$ reads
\begin{align}
\label{eq:Geff_beta}
    \Geff^\beta = \frac{8\Gamma^2\gamma}{\gamma^2 + 4\omega_0^2} \coth{\frac{\beta \omega_0}{2}} \, .
\end{align}

With this analytical result, we can now demonstrate that the efficiency of the QHE at constant power can be increased by reducing the dissipated heat to any desired degree through an increase in the temperature of the dephasing bath. The constancy of the power is maintained by keeping both the extracted work per cycle and the stroke durations fixed. To this end, we fix the choice of the internal parameters of the QHE, including the stroke durations $\tau_\x$ and the thermalisation rates $\gamma_\thermal$, the peak location $\omega_0$ and the width $\gamma$ of the spectral density. The only remaining degrees of freedom are the coupling strength $\Gamma$ between the work medium and the harmonic mode and the inverse temperature $\beta$, which can be adjusted to reduce the dissipated heat
\begin{align}
\label{eq:Delta_Q_diss_beta_scaling}
    0\le \Delta Q^{\text{(diss)}}_{\thermal, \beta} \le
    \frac{4\omega_0}{\gamma}\frac{\Geff^{\beta}}{\coth\frac{\beta\omega_0}{2}}
    \gamma_\thermal \tau_\thermal n_\hot,
\end{align}
which is derived in \cref{sec:heat_dissipation_rate_general}, and to keep the effective dephasing rate $\Geff^\beta$ constant at the same time.
The latter ensures a constant amount of extracted work per cycle $\Delta W_\text{ext.}^\text{(tot)}$ 
and thus a constant power output $P^\text{(tot)}(\beta)$.
Given those restrictions, the dissipated heat $\Delta Q_{\hot, \beta}^\text{(diss)}$ vanishes in the limit of high temperatures. Thus, its contribution to the efficiency of the engine, \cref{eq:eta_tot}, vanishes in this limit, such that the efficiency approaches the value
\begin{align}
    \lim_{\beta\rightarrow 0}\eta^\text{(tot)} &= \lim_{\beta\rightarrow 0} \frac{\Delta W_\text{ext.}^\text{(tot)}}{\Delta Q_\hot^\text{(sys)} + \Delta Q_{\hot, \beta}^\text{(diss)}}\nonumber\\
    & = \frac{\Delta W_\text{ext.}^\text{(tot)}}{\Delta Q_\hot^\text{(sys)}},
\end{align}
where we used an analogue reasoning and analogue definitions as in \cref{sec:analytical_dissipated_heat} to split the total heat transfer $\Delta Q_\hot^\text{(tot)} = \Delta Q_\hot^\text{(sys)} + \Delta Q_{\hot, \beta}^\text{(diss)}$ into the heat flow that actually heats the work medium and the dissipated heat.
It is important to note that this maximum value is not the absolute
maximum achievable under quasi-static operation, as the operation speed, effective dephasing 
rate and coupling strengths to the thermal baths remain finite. To reach this limit, the
dephasing rate must be increased to suppress quantum friction, and the coupling to the thermal
baths must be increased to ensure rapid thermalisation. By carefully adjusting these parameters,
we can ensure that the dissipated heat $\Delta Q^{\text{(diss)}}_{\hot}$ vanishes, allowing us 
to approach the quasi-static efficiency arbitrarily closely. Thus, for any finite power, the
heat engine can reach an efficiency that is bound solely by the properties of the hot and cold
bath.
\cref{sec:increasing_the_power} demonstrates numerically the capability to increase the power of the studied QHE at constant efficiency and vice versa.

\subsection{Decouple the dephasing bath during thermalisation strokes}
\label{sec:decouple_deph_bath}
As the dephasing environment does not assist the QHE during the thermal strokes, and 
most of the additional heat dissipation occurs during these strokes, one may ask whether 
the thermodynamic costs associated with heat dissipation could be avoided by decoupling 
the dephasing bath from the work medium during the thermalisation strokes to prevent 
heat leakage from the thermal baths into the dephasing bath. 
In this case, we have to spend work to decouple (recouple) 
the dephasing bath from (to) the system at the beginning (end) of the thermalisation cycle. We bound this work and show that it can be reduced arbitrarily.

\subsubsection{Decoupling the dephasing bath from the work medium}
\label{sec:decouple_a_generalised_bath}
Decoupling the bath from the work medium requires time-dependent terms in the 
Hamiltonian that are associated with additional work on the work medium and the 
dephasing bath. To describe the decoupling of the dephasing bath from the working 
medium at the start time $t_i \, (i \in \{1, 3\})$ of a thermalisation stroke (see \cref{fig:QHE_Otto_cycle}), which happens faster than any timescale of the system, we consider the Hamiltonian from \cref{eq:H_micro} for an arbitrary spectral density $\mathcal{J}$
with
\begin{align}
    \Ht(t_i) = \Ht[S](t_i) + \Ht[int](t_i) + \Ht[B]
\end{align}
and without the coupling term
\begin{align}
    \Ht^\text{(dec)}(t_i) = \Ht[S](t_i) + \Ht[B].
\end{align}
As the decoupling is assumed to be rapid, the state $\rhot[][t_i]$ does not change 
during the decoupling step. Therefore, the work involved in the decoupling of system
and dephasing environment is calculated as the difference 
\begin{align}
\label{eq:E_dec_general}
    \Delta E^\text{(dec)}_{\mathcal{J}} =& \expval{\Ht^\text{(dec)}(t_i)}_{\rhot[s]} - \expval{\Ht(t_i)}_{\rhot[s]} \nonumber \\
    =& - \expval{\Ht[int](t_i)}_{\rhot[s]}\, .
\end{align}
assuming the system and dephasing environment to be in a steady state $\rhot[s]$ with respect to the Hamiltonian $\Ht[][t_i]$ from \cref{eq:H_micro}.
Using \cref{eq:general_expval_xx_H_int_dec_xx} for $\tau = 0$ computed in \cref{sec:exp_val_H_int_arbitrary_J}, the decoupling energy for an arbitrary spectral density $\mathcal{J}$ can be written as 
\begin{align}
\label{eq:dec_work_arbitrary_J}
    \Delta E^\text{(dec)}_\mathcal{J} =& - \expval{\ketbra{\e}\Ht[int][t_i]\ketbra{\e}}_{\text{s}} \nonumber \\
    &- \expval{\ketbra{\g}\Ht[int][t_i]\ketbra{\g}}_{\text{s}} \nonumber \\
    =& \frac{2}{\pi} \int \limits_{-\infty}^\infty \dd \omega \, \frac{\mathcal{J}(\omega)}{\omega}.
\end{align}

\subsubsection{Recoupling the dephasing bath to the work medium}
\label{sec:recouple_a_generalised_bath}

As the presence of the dephasing environment is required during the compression and expansion strokes, we need to recouple the 
dephasing environment to the system after each thermalisation stroke of duration $\tau_\thermal$. The evolution of work medium and dephasing bath can be
factorised during the thermalisation. The equation of motion for the work medium reads
\begin{align}
    \frac{\text{d}\rhot[S]}{\text{d}t} = -\ii \left[\Ht[S,\thermal], \rhot[S]\right] + \mathcal{L}_\thermal \left[\rhot[S]\right]
\end{align}
as defined in \cref{eq:L_hot,eq:L_cold} and its solution can be expressed through a Kraus channel with Kraus operators $\hat{K}_{\text{S}, i}(\tau_\thermal)$.
The free evolution of the bath is defined by
\begin{align}
    \frac{\dd \rhot[B]}{\dd t} = - \ii \left[\Ht[B], \rhot[B]\right]\, .
\end{align}
As in the previous \cref{sec:decouple_a_generalised_bath}, we assume that the state of work medium and dephasing bath is the steady state $\rhot[s]$ at the time $t_i$ where the bath got decoupled. Hence, at time $t_i + \tau_\thermal$, the joint state of work medium and dephasing bath reads
\begin{align}
\label{eq:rho_dec_tau}
    &\rhot[dec][\tau_\thermal] \nonumber \\
    =& \sum_i \e^{-\ii \Ht[B]\tau_\thermal}\, \hat{K}_{\text{S}, i}(\tau_\thermal) \, \rhot[s]\, \hat{K}_{\text{S}, i}^\dagger(\tau_\thermal) \, \e^{\ii \Ht[B]\tau_\thermal},
\end{align}
and we measure the expectation value of the interaction Hamiltonian on this state to obtain the recoupling energy
\begin{align}
\label{eq:formal_recoupling_energy}
    &\Delta E^{\text{(rec)}}_\mathcal{J}(\tau_\thermal) = \expval{\Ht[int](t_i + \tau_\thermal)}_{\rhot[dec][\tau_\thermal]}.
\end{align}
\cref{sec:rec_energy_arbitrary_spectrum} demonstrates that \cref{eq:formal_recoupling_energy} can be upper bounded by
\begin{align}
\label{eq:bound}
    \left|\Delta E^{\text{(rec)}}_\mathcal{J}(\tau_\thermal)\right| \leq& \left|\frac{2}{\pi} \int \limits_{-\infty}^\infty \dd \omega \, \frac{\mathcal{J}(\omega)}{\omega} \cos{\omega \tau_\thermal}\right| .
\end{align}
Note, that we have not put any constraints on the arbitrary spectral density $\mathcal{J}(\omega)$ yet.

\paragraph{De- and recoupling work --}
Now, we use the result of the decoupling work $\Delta E^\text{(dec)}_\mathcal{J}$ for a general spectral density $\mathcal{J}$ from \cref{eq:dec_work_arbitrary_J} to calculate the work that needs to be invested to decouple a dephasing bath with spectral density $J_\beta(\omega)$ given in \cref{eq:spectrum_thermalized_DAMPF} from the work medium. The result reads
\begin{align}
    \Delta E^\text{(dec)}_{J_\beta} = &\, \frac{2}{\pi} \int \limits_{-\infty}^\infty \dd \omega \, \frac{J_\beta(\omega)}{\omega} \nonumber \\
    =& \, \frac{2}{\pi} \int \limits_{-\infty}^\infty \dd \omega \, \frac{J(\omega)}{\omega}
\end{align}
where we used the symmetry of the thermalised spectral density $J_\beta(\omega)$. The remaining integral has been computed in \cref{sec:heat_dissipation_rate_general} and we can use the final result from \cref{eq:odd_integral_result} to identify
\begin{align}
\label{eq:heat_dissipation_rate_beta}
    \Delta E^\text{(dec)}_{J_\beta} = \Delta E^\text{(dec)}_{J} = \frac{8 \Gamma^2 \omega_0}{4\omega_0^2 + \gamma^2}
\end{align}
for the decoupling work.
The recoupling work from \cref{eq:bound} can be bounded from above and below for the special case of the thermalised spectral density $J_\beta(\omega)$. To this end we exploit the identity $\int_{-\infty}^\infty \dd \omega f(\omega) = \int_0^\infty \dd \omega \left\{f(\omega) + f(-\omega)\right\}$ and the symmetry of the spectral density $J_\beta(\omega)$ to arrive at
\begin{align}
    &\left|\Delta E^{\text{(rec)}}_{J_\beta}(\tau_\thermal)\right| \nonumber \\
    \leq& \left|\frac{2}{\pi} \int \limits_{0}^\infty \dd \omega \, \left\{J_\beta(\omega) - J_\beta(-\omega)\right\} \frac{\cos{\omega \tau_\thermal}}{\omega}\right| \nonumber \\
    =& \left|\frac{2}{\pi} \int \limits_{0}^\infty \dd \omega \, \left\{J(\omega) - J(-\omega)\right\} \frac{\cos{\omega \tau_\thermal}}{\omega}\right| .
\end{align}
Next, we apply the triangle inequality and exploit $J(\omega) - J(-\omega) \geq 0$ for the spectral density $J(\omega)$ from \cref{eq:spectral_density}, which allows to bound the recoupling work as
\begin{align}
\label{eq:bounded_heat_dissipation_rate_beta}
    &\left|\Delta E^\text{(rec)}_{J_\beta}(\tau_\thermal)\right| \nonumber \\
    \leq& \frac{2}{\pi} \int \limits_{0}^\infty \dd \omega \, \left\{J(\omega) - J(-\omega)\right\} \left|\frac{\cos{\omega \tau_\thermal}}{\omega}\right| \nonumber \\
    \leq& \frac{2}{\pi} \int \limits_{0}^\infty \dd \omega \, \frac{J(\omega) - J(-\omega)}{\omega} \nonumber \\
    =& \frac{2}{\pi} \int \limits_{-\infty}^\infty \dd \omega \, \frac{J(\omega)}{\omega} = \Delta E^\text{(dec)}_{J_\beta} .
\end{align}
Finally, we express the additional thermodynamic cost, the de- and recoupling work
\begin{align}
    \left|\Delta E^\text{(rec)}_{J_\beta}(\tau_\thermal)\right| \leq \Delta E^\text{(dec)}_{J_\beta} = \frac{\omega_0}{\gamma} \frac{\Geff^\beta}{\coth{\frac{\beta \omega_0}{2}}},
\end{align}
in terms of the effective dephasing rate $\Geff^\beta$ from \cref{eq:Geff_beta}.

Now we are able to study how the additional energy costs of de- and recoupling affect the overall efficiency
\begin{align}
    \eta^\text{(tot)} &= \frac{\Delta W_\text{ext.}^\text{(tot)}}{\Delta Q_\hot^\text{(sys)} + \Delta E_{J_\beta}^\text{(dec)} + \Delta E_{J_\beta}^\text{(rec)}}
\end{align}
of the QHE, where those additional energy costs are treated as additional energy inputs to the heat engine that impact the efficiency negatively, just as the heat flow $\Delta Q_\hot^\text{sys}$ from the hot bath to the work medium. For this, we follow the procedure in \cref{sec:increasing_the_efficiency} and note that keeping the effective dephasing rate $\Geff^\beta$ constant allows for a constant extracted work per cycle $\Delta W_\text{ext.}^\text{(tot)}$ as well as a constant total power $P^\text{(tot)}$. As discussed extensively in \cref{sec:increasing_the_efficiency}, we fix all parameters except for the inverse temperature $\beta$ and the coupling strength $\Gamma$ such that the extracted work is kept constant\footnote{Note, that only one of the two parameters $\beta$ and $\Gamma$ is a free parameter as they are correlated by demanding the effective dephasing rate to be constant.}. In the limit of high temperatures, i.e. $\beta \rightarrow 0$, both the de- and recoupling energies vanish such that
\begin{align}
    \lim_{\beta \rightarrow 0} \eta^\text{(tot)} = \frac{\Delta W_\text{ext.}^\text{(tot)}}{\Delta Q_\hot^\text{(sys)}}.
\end{align}
This result implies, as discussed as well in \cref{sec:increasing_the_efficiency}, that the parameters of the heat engine can be chosen such that both efficiency and power are not affected by quantum friction in the first place as well as the dephasing bath used to suppress quantum friction.

\section{Discussion}
While our study shows that efficiency can be increased to that of a quasistatic heat engine, some questions deserve further attention. One such question concerns the 
relationship between efficiency, power and output fluctuations in terms of thermodynamical uncertainty relations. Relations bounding these quantities have been proven for classical systems \cite{ShiraishiST2016,PietzonkaS2018} and also found attention in quantum systems \cite{GuarnieriLCG2019,VanVuS2022}. Although it has been recognised that quantum systems may violate the classical bounds, these violations are typically related to coherences that are suppressed in our heat engine by means of dephasing noise and are therefore likely to be suppressed. Another source of power fluctuations may arise from the imperfect timing of the strokes of our heat engine, which requires an external control device. Variations in the length of compression and expansion strokes lead to fluctuations in the extracted work. Furthermore, the switch between strokes requires increasingly fast control of unitary operations with the rising speed of the heat engine's operation. In principle, such operations may induce back action on the control device and may incur associated thermodynamic costs. \cite{WoodsH2023} have derived conditions on the quality of such a quantum control device to ensure that the laws of thermodynamics do not change at finite rate of operation. While it seems that these conditions can also be met for dephasing assisted quantum heat engines, a careful examination of the resulting conditions will be of interest. The exploration of thermodynamical uncertainty relations for our dephasing assisted quantum heat engines and the conditions on the quality of quantum control methods to avoid thermodynamic costs will form the subject of a future work.

\section{Conclusions}
The fundamental model of a thermodynamic heat engine considers an engine connected 
to two heat baths: one hot and one cold. Through judicious cyclic operation, the 
engine can extract work by transferring heat from the hot bath to the cold bath. 
The ultimate efficiency of such an engine is constrained by the laws of thermodynamics 
and is achieved through quasi-static operation, that is, operating at zero power. 
These limitations also extend to quantum heat engines.

However, operating at a finite power introduces additional quantum friction due to 
the creation of coherences during the operation of the QHE. While the laws of 
thermodynamics do not explicitly dictate fundamental limits on efficiency at 
finite power, it has been suggested that applying appropriate dephasing mechanisms 
could enhance efficiency and power. Nonetheless, subjecting the machine to a dephasing 
environment introduces in effect a third heat bath, potentially leading to additional 
heat flows. These additional flows must be considered when accurately assessing the 
thermodynamic efficiency.

Our study demonstrates that dephasing might increase power, often but not always, at 
the expense of efficiency. The central outcome of our work is a meticulous evaluation 
of the efficiency of a quantum heat engine under dephasing and a proof that, under 
specific conditions, a quantum heat engine can achieve for any finite power an efficiency 
approaching that under quasi-static operation. 

\acknowledgements We acknowledge discussions with Alexander N{\"u}ßeler and Mark Mitchison at early stages of this work, {Micha\l} Horodecki and the {Gda\'nsk} team for discussions during the Ulm-{Gda\'nsk} workshop in April 2023 and Eric Lutz and Mark Mitchison for comments on this manuscript. This work 
was supported by the ERC Synergy grant HyperQ (grant no. 856432), the DFG via QuantERA project 
ExtraQt (grant no. 500314265), the state of Baden-W{\"u}rttemberg through bwHPC and the German 
Research Foundation (DFG) through grant no INST 40/575-1 FUGG (JUSTUS 2 cluster).

\bibliographystyle{unsrtnat}
\bibliography{bibliography}
\onecolumn\newpage
\appendix
\section{Expectation value of interaction Hamiltonian for arbitrary spectral densities}
\label{sec:exp_val_H_int_arbitrary_J}

Here, we determine the expectation values
\begin{align}
    \expval{\ketbra{\x} \Ht[int] \ketbra{\x}}_{\rhot[dec](\tau)} = \Tr{\ketbra{\x} \Ht[int] \ketbra{\x} \rhot[dec](\tau)}
\end{align}
for $\ket{\x} \in \{\ket{\e}, \ket{\g}\}$ with the interaction Hamiltonian from \cref{eq:H_int_micro}
\begin{align}
    \Ht[int] = \szt \otimes \int \limits_{-\infty}^\infty \dd \omega \sqrt{\frac{\mathcal{J}(\omega)}{\pi}} \left(\hat{a}^\dagger_\omega + \hat{a}_\omega \right)
\end{align}
and the state $\rhot[dec](t)$ which is defined as
\begin{align}
    \rhot[dec](\tau) = \e^{-\ii \left(\Ht[S] + \Ht[B]\right)\tau}\, \rhot[s]\, \e^{\ii \left(\Ht[S] + \Ht[B]\right)\tau}
\end{align}
where $\rhot[s]$ is the steady state with respect to the total time-independent Hamiltonian $\Ht = \Ht[S] + \Ht[int] + \Ht[B]$ from \cref{eq:H_micro}\footnote{Here, we are interested in the dynamics during the thermalisation strokes. During those strokes, the Hamiltonian is time-independent, which implies explicitly $\szt[t] = \szt = \text{const}$.}.
This describes the situation where the dephasing bath is decoupled instantaneously at $t = 0$ and the system continues evolving under the time-independent Hamiltonians $\Ht[S]$ and $\Ht[B]$ for time $\tau$. Making use of $\left[\Ht[S], \Ht[B]\right] = 0$ and $\left[\Ht[S], \Ht[int]\right] = 0$ we rewrite the expectation value as
\begin{align}
    \expval{\ketbra{\x} \Ht[int] \ketbra{\x}}_{\rhot[dec](\tau)} &= \Tr{\e^{\ii \left(\Ht[S] + \Ht[B]\right)\tau}\ketbra{\x} \Ht[int] \ketbra{\x} \e^{-\ii \left(\Ht[S] + \Ht[B]\right)\tau}\, \rhot[s]}\\
    & = \Tr{ \ketbra{\x} \e^{\ii \Ht[S] \tau} \,\e^{\ii \Ht[B] \tau}\, \Ht[int]\, \e^{-\ii \Ht[S] \tau}\, \e^{-\ii \Ht[B] \tau} \ketbra{\x} } \\
    & = \Tr{ \ketbra{\x} \e^{\ii \Ht[B] \tau}\, \Ht[int] \,\e^{-\ii \Ht[B] \tau} \ketbra{\x}}
\end{align}
where we used $\left[\ketbra{\x}, \Ht[S]\right] = \left[\ketbra{\x}, \Ht[B]\right] = 0$ ($\ket{\x} \in \{\ket{\e}, \ket{\g}\}$) as the system Hamiltonian $\Ht[S]$ is diagonal in the basis spanned by $\ket{\e}$ and $\ket{\g}$. The expression $\e^{\ii \Ht[B] \tau}\, \Ht[int] \,\e^{-\ii \Ht[B] \tau}$ can be simplified using the Baker-Campbell-Hausdorff formula as
\begin{align}
\label{eq:free_evolved_H_int}
    \Ht[int, dec][\tau] = \e^{\ii \Ht[B] \tau}\, \Ht[int] \,\e^{-\ii \Ht[B] \tau} = \szt \otimes \int \limits_{-\infty}^\infty \dd \omega \sqrt{\frac{\mathcal{J}(\omega)}{\pi}} \left(\hat{a}^\dagger_\omega \, \e^{\ii \omega \tau} + \hat{a}_\omega \, \e^{-\ii \omega \tau}\right).
\end{align}
Using this expression, the expectation value of the interaction Hamiltonian $\Ht[int]$ with respect to the decoupled state $\rhot[dec][\tau]$ can be expressed as the expectation value
\begin{align}
    \expval{\ketbra{\x} \Ht[int] \ketbra{\x}}_{\rhot[dec](\tau)} = \expval{\ketbra{\x} \Ht[int, dec][\tau] \ketbra{\x}}_{\rhot[s]}
\end{align}
on the steady state $\rhot[s]$ with respect to the Hamiltonian $\Ht$.

We determine the steady state by starting in an arbitrary state $\rhot[0]$ which we evolve with the full Hamiltonian $\Ht$ for a sufficiently long time $t$, i.e.
\begin{align}
    \rhot[s] = \lim_{t \rightarrow \infty}  \e^{-\ii \Ht t} \, \rhot[0] \, \e^{\ii \Ht t} \, .
\end{align}
Here, the excited state population of the work medium $p_\e$ is preserved under the time-independent purely dephasing Hamiltonian $\Ht$ from \cref{eq:H_micro}. Hence, this population can be chosen arbitrarily in the initial state $\rhot[0]$ which will be exploited in the end of this calculation. We use this representation of the steady state to derive the time-dependent expectation value
\begin{align}
\label{eq:general_E_int}
    \expval{\ketbra{\x}\Ht[int, dec][\tau]\ketbra{\x}}_{\rhot[s]} &= \lim_{t \rightarrow \infty} \Tr{\ketbra{\x}\Ht[int, dec][\tau]\ketbra{\x} \, \e^{-\ii \Ht t} \, \rhot[0] \, \e^{\ii \Ht t}}\nonumber\\
    & = \lim_{t \rightarrow \infty} \Tr{\ketbra{\x} \e^{\ii \Ht t} \, \Ht[int, dec][\tau] \, \e^{-\ii \Ht t} \ketbra{\x} \, \rhot[0]}
\end{align}
where we have used the cyclic property of the trace and $[\ketbra{\x}, \Ht] = 0$ for $\x \in \{\e, \g\}$ to arrive at the time-dependent interaction Hamiltonian in the Heisenberg picture
\begin{align}
    \Ht[int, H][\tau, t] &= \e^{\ii \Ht t} \, \Ht[int, dec][\tau]\, \e^{-\ii \Ht t} = \sum \limits_{k=0}^{\infty} \frac{(\ii t)^k}{k!} \left[\Ht, \Ht[int, dec][\tau]\right]_k.
\end{align}
In the last step, we have used the well-known Baker-Campbell-Hausdorff formula and the recursively defined commutators $[X,Y]_{k+1} = [X,[X,Y]_k]$ and $[X,Y]_0 = Y$. With the explicit expressions of the Hamiltonian $\Ht$ from \cref{eq:H_micro,eq:H_B_micro,eq:H_int_micro} and $\Ht[int, dec][\tau]$ from \cref{eq:free_evolved_H_int} one can easily check the general expressions for the $(2k-1)$th order
\begin{align}
\label{eq:H_int_H_odd}
    \left[\Ht, \Ht[int, dec][\tau]\right]_{2k-1} =&\,\szt \otimes \int \limits_{-\infty}^\infty \dd \omega \, \omega^{2k-1} \, \sqrt{\frac{\mathcal{J}(\omega)}{\pi}} \left(\hat{a}^\dagger_\omega \, \e^{\ii \omega \tau} - \hat{a}_\omega \, \e^{-\ii \omega \tau}\right) \nonumber \\
    &+ \int \limits_{-\infty}^{\infty} \dd \omega \, \frac{\mathcal{J}(\omega)}{\pi} \, \omega^{2k-2} \, \left(\e^{\ii \omega \tau} - \e^{-\ii \omega \tau}\right)
\end{align}
and the $2k$th order
\begin{align}
\label{eq:H_int_H_even}
    \left[\Ht, \Ht[int, dec][\tau]\right]_{2k} =&\, \szt \otimes \int \limits_{-\infty}^{\infty} \dd \omega \, \omega^{2k} \, \sqrt{\frac{\mathcal{J}(\omega)}{\pi}} \left(\hat{a}_\omega^\dagger \, \e^{\ii \omega \tau} + \hat{a}_\omega \, \e^{-\ii \omega \tau}\right) \nonumber \\
    &+\, 2 \int \limits_{-\infty}^\infty \dd \omega \, \omega^{2k-1} \, \frac{\mathcal{J}(\omega)}{\pi} \,\left(\e^{\ii \omega \tau} + \e^{-\ii \omega \tau}\right)
\end{align}
commutators for $k > 0$ whereas $\big[\Ht, \Ht[int, dec][\tau]\big]_0 = \Ht[int, dec][\tau]$ as defined before. We combine \cref{eq:H_int_H_odd,eq:H_int_H_even} to obtain 
\begin{align}
\label{eq:H_int_H}
    \Ht[int, H][\tau, t] =& \,\frac{2}{\pi} \int \limits_{-\infty}^\infty \dd \omega \, \frac{\mathcal{J}(\omega)}{\omega} \, \left(\cos{\omega \tau} \, (\cos{\omega t}-1) -\sin{\omega \tau} \, \sin{\omega t}\right) \nonumber \\
    &+\,\szt \otimes \int \limits_{-\infty}^\infty \dd \omega \, \sqrt{\frac{\mathcal{J}(\omega)}{\pi}} \left(\hat{a}_\omega^\dagger \, \e^{\ii \omega t} + \hat{a}_\omega \, \e^{-\ii \omega t}\right)
\end{align}
where we used the Taylor expansion of the cosine and exponential function. Now, we evaluate \cref{eq:general_E_int} to obtain the expectation value
\begin{align}
\label{eq:time_dependent_x_H_int_x}
    \Tr{\ketbra{\x}\Ht[int, H][\tau, t]\ketbra{\x} \, \rhot[0]}
    =& \, \frac{2 p_\x}{\pi} \int \limits_{-\infty}^\infty \dd \omega \, \frac{\mathcal{J}(\omega)}{\omega} \, \left(\cos{\omega \tau} \, (\cos{\omega t}-1) - \sin{\omega \tau} \, \sin{\omega t}\right)\nonumber\\
    & + 2\bra{\x}\szt \ket{\x} \int \limits_{-\infty}^\infty \dd \omega \, \sqrt{\frac{\mathcal{J}(\omega)}{\pi}}\, \Re{\expval{\ketbra{\x} \otimes \hat{a}_\omega^\dagger}_{\rhot[0]}\, \e^{\ii \omega t}}\, .
\end{align}
The initial population $p_\x$ of the work medium is defined as $p_\x = \Tr_\text{B}\{\bra{\x}\rhot[0]\ket{\x}\} = \Tr_\text{B}\{\bra{\x}\rhot[s]\ket{\x}\}$ and coincides with the population of work medium in the steady state as it is a constant of motion under the evolution under $\Ht$. From \cref{eq:general_E_int,eq:time_dependent_x_H_int_x} we obtain the steady state expectation value in the limit $t \rightarrow \infty$ resulting in
\begin{align}
\label{eq:general_expval_xx_H_int_dec_xx}
    \expval{\ketbra{\x}\Ht[int, dec][\tau]\ketbra{\x}}_{\text{s}} = -\frac{2 p_\x}{\pi} \int \limits_{-\infty}^\infty \dd \omega \, \frac{\mathcal{J}(\omega)}{\omega} \cos{\omega \tau}
\end{align}
because the integral over the fast oscillating cosine and exponential functions vanishes if both the spectral density $\mathcal{J}(\omega)$ and the expectation value $\langle \ketbra{\x} \otimes \hat{a}_\omega^\dagger\rangle_{\rhot[0]}$ are analytic functions. Both conditions can be met trivially by choosing an appropriate spectral density and by choosing the state $\rhot[0]$ to be diagonal in the Fock basis, e.g. a thermal state.

\section{Heat dissipation rate for arbitrary spectral densities}
\label{sec:heat_dissipation_rate_general}

In the following section, we derive an analytical expression for the heat dissipation rate, applicable to baths with arbitrary spectral densities $\mathcal{J}(\omega)$ defined on the whole real axis. This is relevant for the thermalised spectral density of our model from \cref{eq:spectrum_thermalized_DAMPF} and also for general baths where the temperature is encoded into a thermalised spectral density \cite{Tamascelli2019}. Similar to \cref{sec:analytical_dissipated_heat}, we assume that the work medium and dephasing bath are in a steady state with respect to their joint dynamics that is governed by the Hamiltonian from \cref{eq:H_micro,eq:H_B_micro,eq:H_int_micro}. Now, we study the effect of the thermal Lindbladian from \cref{eq:L_cold,eq:L_hot} perturbatively, focusing solely on the first order of the time evolution
\begin{align}
    \dd \rhot = \mathcal{L}_\thermal \left[\rhot[s]\right] \, \dd t,
\end{align}
and analyse the associated energy change of work medium and dephasing bath. The differential change of the system energy is given by
\begin{align}
    \frac{\dd Q^\text{(sys)}_\thermal}{\dd t} = \frac{\dd \expval{\Ht[S]}}{\dd t} = \Tr{\Ht[S] \, \mathcal{L}_\thermal \left[\rhot[s]\right]}
\end{align}
which is equivalent to \cref{eq:dH_sys} and contributes to the heat
\begin{align}
    \Delta Q^\text{(sys)}_\thermal = \int \limits_{\tau_0}^{\tau_0 + \tau_\thermal} \dd t \, \frac{\dd Q^\text{(sys)}_\thermal}{\dd t}
\end{align}
that can be converted to useful work in the Otto cycle. The bath energy $\expval{\mathbf{1}_\text{S} \otimes \Ht[B]}$ does not change as $\Tr_\text{S}{\dd \rhot} = 0$. However, the expectation value of the interaction Hamiltonian
\begin{align}
\label{eq:heat_dissipation_rate_general}
    \frac{\dd Q^\text{(diss)}_\thermal}{\dd t} = \Tr{\Ht[int] \, \mathcal{L}_\mathrm{th}\left[\rhot[s]\right]}
\end{align}
is non-zero leading to a finite amount of dissipated heat
\begin{align}
\label{eq:integrated_diss_heat_appendix}
    \Delta Q^\text{(diss)}_\thermal = \int \limits_{\tau_0}^{\tau_0+\tau_\thermal} \dd t \frac{\dd Q^\text{(diss)}_\thermal}{\dd t}
\end{align}
that cannot be converted in work any more as explained in \cref{sec:analytical_dissipated_heat}.
First, we focus on the heat dissipation rate from \cref{eq:heat_dissipation_rate_general} which cannot be calculated straightforwardly as there is no analytic solution of the steady state $\rhot[s]$ for the microscopic spin-boson model from \cref{eq:H_micro,eq:H_B_micro,eq:H_int_micro}. To solve it anyway, we rewrite \cref{eq:heat_dissipation_rate_general} as
\begin{align}
\label{eq:general_dQ_diss}
    &\frac{\dd Q^\text{(diss)}_\thermal}{\dd t} = \Tr{\Ht[int] \, \mathcal{L}_\mathrm{th}\left[\rhot[s]\right]} \nonumber \\
    = &\Tr{\Ht[int] \gamma_\thermal \left[n_\thermal \left(\spt[th] \rhot[s] \smt[th] - \frac{1}{2} \{\smt[th] \spt[th], \rhot[s]\}\right) + (n_\thermal +1) \left(\smt[th] \rhot[s] \spt[th] - \frac{1}{2} \{\spt[th] \smt[th], \rhot[s]\}\right)\right]} \nonumber \\
    = &\Tr{\gamma_\thermal \left[n_\thermal \left(\smt[th] \Ht[int] \spt[th] - \frac{1}{2} \{\smt[th] \spt[th], \Ht[int]\}\right) + (n_\thermal +1) \left(\spt[th] \Ht[int] \smt[th] - \frac{1}{2} \{\spt[th] \smt[th], \Ht[int]\}\right)\right] \rhot[s]} \nonumber \\
    = &\Tr{\mathcal{L}_\thermal^\ast\left[\Ht[int]\right] \, \rhot[s]}
\end{align}
to obtain the dual channel $\mathcal{L}_\thermal^\ast$ that transforms the interaction Hamiltonian from \cref{eq:H_int_micro} as
\begin{align}
    \mathcal{L}_\thermal^\ast\left[\Ht[int]\right] =&- 2\gamma_\thermal \, (n_\thermal+1) \ketbra{\e} \Ht[int] \ketbra{\e} -2\gamma_\thermal \, n_\thermal \ketbra{\g} \Ht[int] \ketbra{\g}
\end{align}
where we used the identities
\begin{align}
    \smt[th] \Ht[int] \spt[th] = -\ketbra{\g} \Ht[int] \ketbra{\g}
\end{align}
and
\begin{align}
    \spt[th] \Ht[int] \smt[th] = -\ketbra{\e} \Ht[int] \ketbra{\e}.
\end{align}
Hence, we can rewrite the heat dissipation rate
\begin{align}
\label{eq:heat_diss_rate_eHe_gHg}
    \frac{\dd Q^\text{(diss)}_\thermal}{\dd t} = &- 2\gamma_\thermal \, (n_\thermal+1) \expval{\ketbra{\e}\Ht[int]\ketbra{\e}}_\text{s} - 2\gamma_\thermal \, n_\thermal \expval{\ketbra{\g} \Ht[int] \ketbra{\g}}_\text{s}
\end{align}
in terms of steady state expectation values of projections of the interaction Hamiltonian $\ketbra{\x} \Ht[int] \ketbra{\x}$ on the eigenstates of the work medium $\x \in \{\e,\g\}$.
Those expectation values have been computed in \cref{sec:exp_val_H_int_arbitrary_J}.
Using the result from \cref{eq:general_expval_xx_H_int_dec_xx} for $\tau = 0$, we can express the dissipated heat rate from \cref{eq:heat_diss_rate_eHe_gHg} as
\begin{align}
\label{eq:diss_heat_rate_general}
    \frac{\dd Q^\text{(diss)}_\thermal}{\dd t} = &\, 4 \gamma_\thermal \, \frac{n_\thermal+p_\e}{\pi} \int \limits_{-\infty}^\infty \dd \omega \, \frac{\mathcal{J}(\omega)}{\omega}
\end{align}
for a general spectral density.
\paragraph{Application to relevant spectra of the main text --}
We proceed to compute the dissipated heat for the thermalised spectral density
\begin{align}
    J_\beta(\omega) = \, \frac{1}{2} J(\omega) \left(\coth{\frac{\beta \omega_0}{2}} + 1\right)
    + \, \frac{1}{2} J(-\omega) \left(\coth{\frac{\beta \omega_0}{2}} - 1\right)
\end{align}
from \cref{eq:spectrum_thermalized_DAMPF}, where
\begin{align}
    J(\omega) = \frac{2\Gamma^2 \gamma}{\gamma^2+4(\omega-\omega_0)^2}
\end{align}
as defined in \cref{eq:spectral_density}. For the integral in \cref{eq:diss_heat_rate_general}, only the odd part of $J_\beta(\omega)$ matters due to symmetry. Hence, the dissipated heat
\begin{align}
    \frac{\dd Q^\text{(diss)}_\thermal}{\dd t} = \, 4 \gamma_\thermal \, \frac{n_\thermal+p_\e}{\pi} \int \limits_{-\infty}^\infty \dd \omega \, \frac{J_\beta(\omega)}{\omega} = \, 4 \gamma_\thermal \, \frac{n_\thermal+p_\e}{\pi} \int \limits_{-\infty}^\infty \dd \omega \, \frac{J(\omega) - J(-\omega)}{2\omega}
\end{align}
does not depend on the temperature any more and can be further simplified as
\begin{align}
    \frac{\dd Q^\text{(diss)}_\thermal}{\dd t} = \, 4 \gamma_\thermal \, \frac{n_\thermal+p_\e}{\pi} \int \limits_{-\infty}^\infty \dd \omega \, \frac{J(\omega)}{\omega}
\end{align}
using further symmetry considerations. This result describes the situation of a single DAMPF mode at zero temperature that we already investigated in \cref{sec:analytical_dissipated_heat}. We solve the integral
\begin{align}
    \int \limits_{-\infty}^{\infty} \dd \omega \frac{J(\omega)}{\omega}=&  \int \limits_{-\infty}^\infty \dd \omega \, \frac{2 \Gamma^2 \gamma}{\gamma^2+4\left(\omega- \omega_0\right)^2} \frac{1}{\omega}
\end{align}
by rewriting as
\begin{align}
    \int \limits_{-\infty}^{\infty} \dd \omega \frac{J(\omega)}{\omega}=& \frac{2 \Gamma^2}{\gamma} \int \limits_{-\infty}^\infty \dd \omega \, \frac{1}{1+4\left(\frac{\omega- \omega_0}{\gamma}\right)^2} \frac{1}{\omega}.
\end{align}
Substituting $\gamma x = \omega - \omega_0$ yields
\begin{align}
    \int \limits_{-\infty}^{\infty} \dd \omega \frac{J(\omega)}{\omega}=& \frac{2 \Gamma^2}{\gamma} \int \limits_{-\infty}^\infty \dd x \, \frac{1}{1+4 x^2} \frac{1}{x + \frac{\omega_0}{\gamma}} = \frac{2 \Gamma^2}{\gamma} \int \limits_{-\infty}^\infty \dd x \, \frac{1}{1+4 \frac{\omega_0^2}{\gamma^2}} \left( \frac{4 \frac{\omega_0}{\gamma} - 4 x}{1+4 x^2} + \frac{1}{x + \frac{\omega_0}{\gamma}} \right).
\end{align}
The integrals over the odd parts vanish, whence the remaining even part yields the result
\begin{align}
\label{eq:odd_integral_result}
    \int \limits_{-\infty}^{\infty} \dd \omega \frac{J(\omega)}{\omega}=& \frac{8 \Gamma^2 \omega_0}{\gamma^2+4 \omega_0^2} \int \limits_{-\infty}^\infty \dd x \, \frac{1}{1+4 x^2} = \frac{4 \Gamma^2 \omega_0 \pi}{\gamma^2+4 \omega_0^2} \, .
\end{align}
Similar boundaries as in \cref{eq:diss_heat_estimation} apply also here after integrating the heat dissipation rate for the duration of the thermalisation stroke, resulting in the estimation
\begin{align}
    0
    \leq \Delta Q^\text{(diss)}_{\thermal, \beta} \leq 
    \frac{16 \Gamma^2 \omega_0 \gamma_\thermal \tau_\thermal}{4 \omega_0^2 + \gamma^2} \left(n_\thermal + 1\right)
\end{align}
which can be expressed equivalently in terms of the effective dephasing rate $\Geff^\beta$ defined in \cref{eq:Geff_beta}
\begin{align}
    \label{eq:diss_heat_estimation_thermal}
    0
    \leq \Delta Q^\text{(diss)}_{\thermal, \beta} \leq 
    \frac{2 \omega_0 \gamma_\thermal \tau_\thermal}{\gamma} \frac{\Geff^\beta}{\coth{\frac{\beta \omega_0}{2}}} \left(n_\thermal + 1\right)\, .
\end{align}

\section{Upper bound on the recoupling energy for arbitrary spectral densities}
\label{sec:rec_energy_arbitrary_spectrum}

This appendix derives \cref{eq:bound}. During thermalisation, the 
work medium evolves from time $t_i$ to $t_{i+1}$ with $i \in \{1, 3\}$ and $t_{i+1} - t_i = \tau_\thermal$ (see \cref{fig:QHE_Otto_cycle}) under the equation of motion
\begin{align}
    \frac{\text{d}\rhot[S]}{\text{d}t} = -\ii \left[\Ht[S,\thermal], \rhot[S]\right] + \mathcal{L}_\thermal \left[\rhot[S]\right]
\end{align}
defined in \cref{eq:L_hot,eq:L_cold}
can be expressed as a Kraus channel with up to four ($i \in \{1, 2, 3, 4\}$) Kraus operators
\begin{align}
    \hat{K}_{\text{S}, i} = \kappa_{\e \e, i} \ketbra{\e} + \kappa_{\e \g, i} \ketbra{\e}{\g}
    + \kappa_{\g \e, i} \ketbra{\g}{\e} + \kappa_{\g \g, i} \ketbra{\g}.
\end{align}
Using $\sum_i \hat{K}_{\text{S}, i}^\dagger \hat{K}_{\text{S}, i} = \mathbf{1}$, we identify the following relations
\begin{align}
\label{eq:kraus_condition_0}
    \sum_i |\kappa_{\e \e, i}|^2 + |\kappa_{\g \e, i}|^2 &= 1, \\
\label{eq:kraus_condition_1}
    \sum_i |\kappa_{\g \g, i}|^2 + |\kappa_{\e \g, i}|^2 &= 1, \\
    \sum_i \kappa_{\e \e, i}^\ast \kappa_{\e \g, i} + \kappa_{\g \e, i}^\ast \kappa_{\g \g, i} &= 0.
\end{align}
Note that the Kraus channel commutes with the free evolution of the bath, i.e. $\forall i:\left[\hat{K}_{\text{S}, i}, \e^{-\ii \Ht[B]\tau_\thermal} \right] = 0$ and, thus, the evolved state can be written as
\begin{align}
    \rhot[dec][\tau_\thermal] = \sum_i \e^{-\ii \Ht[B]\tau_\thermal}\, \hat{K}_{\text{S}, i} \, \rhot[s]\, \hat{K}_{\text{S}, i}^\dagger \, \e^{\ii \Ht[B]\tau_\thermal}
\end{align}
(see \cref{eq:rho_dec_tau}).
This yields
\begin{align}
\label{eq:recoupling_energy_kraus}
    \Delta E^{\text{(rec)}}_\mathcal{J}(\tau) &= \expval{\Ht[int](t_0)}_{\rhot[dec][\tau_\thermal]} \nonumber \\
    &= \sum_i \Tr{\Ht[int] \, \e^{-\ii \Ht[B]\tau_\thermal}\, \hat{K}_{\text{S}, i} \, \rhot[s]\, \hat{K}_{\text{S}, i}^\dagger \, \e^{\ii \Ht[B]\tau_\thermal}} \nonumber \\
    &= \sum_i \Tr{\Ht[int, dec][\tau_\thermal] \, \hat{K}_{\text{S}, i} \, \rhot[s]\, \hat{K}_{\text{S}, i}^\dagger}
\end{align}
where we used the definition of $\Ht[int, dec][\tau_\thermal]$ from \cref{eq:free_evolved_H_int}.
Note that both $\Ht[int, dec][\tau_\thermal]$ and $\rhot[s]$ are diagonal in the computational subspace, i.e.
\begin{align}
    \Tr{\bra{\e}\Ht[int, dec][\tau_\thermal]\ket{\g}} &= 0 \, \; \text{and}\,\; \Tr{\bra{\e}\rhot[s]\ket{\g}} = 0.
\end{align}
The latter is satisfied because $\rhot[s]$ is the steady state of a pure dephasing spin-boson model \cite{BreuerPetruccione2002}.
Hence, we can rewrite \cref{eq:recoupling_energy_kraus} as
\begin{align}
    \Delta E^{\text{(rec)}}_\mathcal{J}(\tau_\thermal) =& \sum_i |\kappa_{\e \e, i}|^2 \expval{\ketbra{\e} \Ht[int, dec](\tau_\thermal) \ketbra{\e}}_\text{s} + \sum_i |\kappa_{\e \g, i}|^2 \expval{\ketbra{\g}{\e} \Ht[int, dec](\tau_\thermal) \ketbra{\e}{\g}}_\text{s} \nonumber \\
    +& \sum_i |\kappa_{\g \e, i}|^2 \expval{\ketbra{\e}{\g} \Ht[int, dec](\tau_\thermal) \ketbra{\g}{\e}}_\text{s} + \sum_i |\kappa_{\g \g, i}|^2 \expval{\ketbra{\g} \Ht[int, dec](\tau_\thermal) \ketbra{\g}}_\text{s}.
\end{align}
Using
\begin{align}
    \ketbra{\g}{\e} \Ht[int, dec](\tau_\thermal) \ketbra{\e}{\g} =& - \ketbra{\g} \Ht[int, dec](\tau_\thermal) \ketbra{\g}
\end{align}
and
\begin{align}
    \ketbra{\e}{\g} \Ht[int, dec](\tau_\thermal) \ketbra{\g}{\e} =& - \ketbra{\e} \Ht[int, dec](\tau_\thermal) \ketbra{\e},
\end{align}
we arrive at
\begin{align}
    \Delta E^{\text{(rec)}}_\mathcal{J}(\tau_\thermal) =& \sum_i \left(|\kappa_{\e \e, i}|^2 - |\kappa_{\g \e, i}|^2\right) \expval{\ketbra{\e} \Ht[int, dec](\tau_\thermal) \ketbra{\e}}_\text{s} \nonumber \\
    +& \sum_i \left(|\kappa_{\g \g, i}|^2 - |\kappa_{\e \g, i}|^2\right) \expval{\ketbra{\g} \Ht[int, dec](\tau_\thermal) \ketbra{\g}}_\text{s}.
\end{align}
Using the triangle inequality, \cref{eq:kraus_condition_0,eq:kraus_condition_1} and \cref{eq:general_expval_xx_H_int_dec_xx} , we find the upper bound
\begin{align}
    \left|\Delta E^{\text{(rec)}}_\mathcal{J}(\tau_\thermal)\right| \leq& \left|\expval{\ketbra{\e} \Ht[int, dec](\tau_\thermal) \ketbra{\e}}_\text{s}\right| + \left|\expval{\ketbra{\g} \Ht[int, dec](\tau_\thermal) \ketbra{\g}}_\text{s}\right| \nonumber \\
    =& \left|\frac{2}{\pi} \int \limits_{-\infty}^\infty \dd \omega \, \frac{\mathcal{J}(\omega)}{\omega} \cos{\omega \tau_\thermal}\right| \, .
\end{align}

\section{Power and efficiency of zero-temperature quantum with a dephasing bath}
\label{sec:increasing_the_power}
The following section will examine two distinct scenarios: the first, the enhancement of a heat engine's efficiency at a constant power output, and the second, the increase of a heat engine's power at a constant efficiency.

\subsection{Increasing efficiency at constant power}
\label{sec:increasing_efficiency_at_constant_power}
In order to achieve the theoretical maximum efficiency of a quasi-static Otto engine given in \cref{eq:eta_max_sys}, it is necessary to adjust the parameters of the zero-temperature dephasing 
bath while maintaining a constant power output. The total efficiency of the QHE 
\begin{align}
\label{eq:eta_tot_with_dissipated_heat}
    \eta^\text{(tot)} = \frac{\Delta W_\text{ext.}^\text{(tot)}}{\Delta Q_\hot^\text{(tot)}} = \frac{\Delta W_\text{ext.}^\text{(tot)}}{\Delta Q_\hot^\text{(sys)} + \Delta Q_\hot^\text{(diss)}}
\end{align}
splits the heat transfer $\Delta Q_\hot^\text{(tot)}$ into the heat that is purely transferred to the work medium $\Delta Q_\hot^\text{(sys)}$ (see \cref{eq:Delta_Q_sys_thermal}) and the heat that is dissipated to the dephasing bath $\Delta Q_\hot^\text{(diss)}$ (see \cref{eq:DeltaQ_diss}). $\Delta Q_\hot^\text{(diss)}$ represents an additional thermodynamic cost that lowers the efficiency of the QHE when compared to the case of a quasi-static engine in the absence of dephasing.
The upper bound on $\Delta Q_\hot^\text{(diss)}$ in \cref{eq:diss_heat_estimation} depends on the three parameters $\Gamma$, $\gamma$ and $\omega_0$ which may be varied to reduce the dissipated heat while keeping all other parameters of the work medium and the thermal baths unchanged. In what follows, we keep the effective dephasing rate of the bath
\begin{align}
    \Geff(\Gamma,\gamma,\omega_0) = \frac{8\Gamma^2\gamma}{\gamma^2 + 4\omega_0^2}
\end{align}
at a constant value which is crucial to maintain the ability of the QHE to suppress frictional losses during the compression and expansion strokes. Indeed, numerical simulations have confirmed that the power of the QHE remains essentially constant when the bath parameters are varied in a way that ensures the condition $\Geff = \text{const}$.
We vary $\gamma$ and $\omega_0$ for fixed $\Gamma$, $\bar{\gamma}$ and $\bar{\omega}_0$ and we fix the corresponding stroke durations $\tau_\x$
which leaves only one free parameter, $\gamma$, as the demand for a constant effective dephasing rate, i.e.
\begin{align}
\label{eq:condition_Geff}
    \Geff(\Gamma,\gamma,\omega_0) = \Geff(\Gamma,\bar{\gamma},\bar{\omega_0}),
\end{align}
fixes
\begin{align}
\label{eq:condition_Geff_omega}
    \omega_0(\gamma) = \frac{1}{2}\sqrt{\frac{4\bar{\omega_0}^2+\bar{\gamma}^2}{\bar{\gamma}}\gamma - \gamma^2}.
\end{align}
This relation restricts the values of $\gamma$ to the domain $[0,\gamma_0]$ with
\begin{align}
\label{eq:condition_Geff_gamma_0}
    \gamma_0 = \frac{4\bar{\omega_0}^2+\bar{\gamma}^2}{\bar{\gamma}}.
\end{align}
Moreover, we observe that $\omega_0(\gamma)$ vanishes as $\gamma$ approaches $\gamma_0$. Now, we use $\Geff = \text{const}$ to rewrite the analytical upper bound for the dissipated heat of \cref{eq:diss_heat_estimation} as
\begin{align}
    \Delta Q^\text{(diss)}_\hot \leq \, &2\Geff(\Gamma,\bar{\gamma},\bar{\omega_0}) \frac{\omega_0(\gamma)}{\gamma} 
    \gamma_\hot \tau_\hot \left(n_\hot + \frac{n_\hot}{2 n_\hot +1}\right).
\end{align}
The explicit expression of $\omega_0(\gamma)$ and in particular the limit $\omega_0(\gamma) \rightarrow 0$ for $\gamma \rightarrow \gamma_0$ implies that the upper bound tends to zero in the limit $\gamma \rightarrow \gamma_0$. This leads us to the remarkable conclusion that the thermodynamic cost of adding the dephasing bath can be made to vanish while allowing the heat engine to run at finite speed.

\begin{figure*}[ht]
    \centering
    \includegraphics[width=1\textwidth]{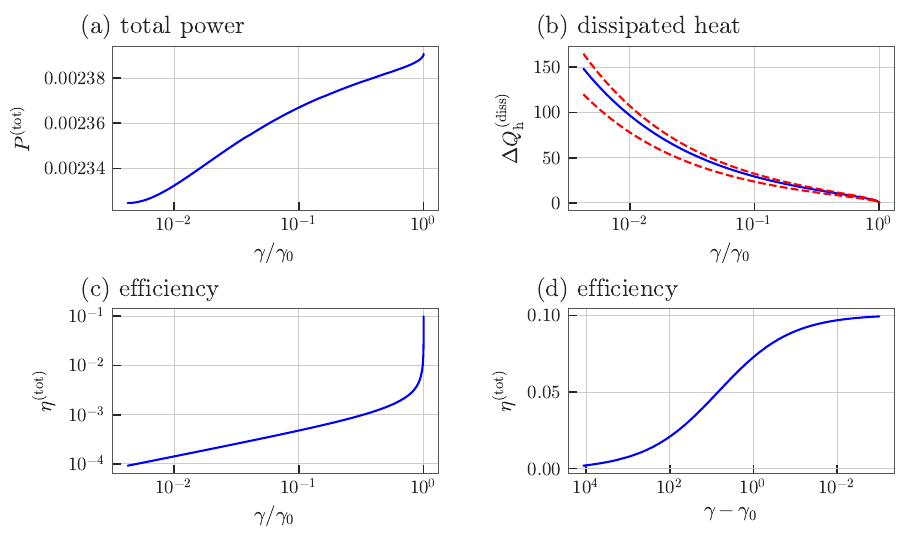}
    \caption{The performance of the QHE under constant effective dephasing rate $\Geff$ and fixed stroke durations $\tau_\x$, considering the parameter $\gamma$. The total power of the QHE in panel (a) exhibits an increasing trend for all considered values of $\gamma$, approaching $\gamma_0$, which is a necessary condition for the further analysis. In panel (b), the dissipated heat decreases significantly and tends towards zero as $\gamma$ approaches $\gamma_0$, as shown for the simulations (blue) and the analytical bounds (red, dashed) from \cref{eq:diss_heat_estimation}. This indicates an improvement in the efficiency of the QHE, which is illustrated in panels (c) and (d). Particularly, in panel (d), we observe that the efficiency of the QHE can be boosted to values close to the analytic limit $\eta^\text{(sys)}_{\max} \approx 0.106$ of the ideal QHE without dephasing noise from \cref{eq:eta_max_sys} as $\gamma$ approaches $\gamma_0$.}
    \label{fig:Ptot_etatot_Geff}
\end{figure*}

We illustrate these analytic considerations by numerical simulations of the QHE that are shown in \cref{fig:Ptot_etatot_Geff} where we chose $\Gamma = 256$ and $\bar{\gamma} = 128$ and $\bar{\omega_0} = 1024$ and fix the corresponding stroke durations $\tau_\x$ 
to the values that optimised the total power of the QHE for $\Gamma = 256$ in \cref{sec:num_power_maximization}; those are in particular $\tau_\com = 0.78125$, $\tau_\hot = 1.84375$, $\tau_\ex = 0.6875$ and $\tau_\cold = 2.625$. 
As expected in panel (a) we observe that for fixed effective dephasing rate $\Geff$ the total power of the QHE is nearly constant within the entire interval $\gamma \in [32,\gamma_0=32896]$. Moreover, panel (b) shows that the dissipated heat vanishes as $\gamma$ approaches $\gamma_0$ which yields the increase of the efficiency in panel (c). The last panel (d) shows that the efficiency converges to the maximum value of $\eta_0 \approx 0.1$ in the limit $\gamma \rightarrow \gamma_0$. This value is comparable to the analytic result, \cref{eq:eta_max_sys}, for the maximum possible efficiency $\eta_{\max}^\text{(sys)} = 1 - \frac{\varepsilon_\cold}{\varepsilon_\hot} \approx 0.106$ under quasi-static operation. This shows, remarkably, that it is indeed possible to increase the power of the QHE by adding dephasing noise and still operate in a regime near to the optimal theoretical efficiency under quasi-static operation. 

As a word of caution we note that in the regime of most efficient operation of the QHE, 
$\gamma \rightarrow \gamma_0$, the position of the maximum of the Lorentzian spectral density \cref{eq:spectral_density}
given by $\omega_0(\gamma)$ approaches zero-frequency while the width $\gamma$ of the spectral density
is much larger than the position of the maximum, i.e. $\gamma\gg \omega_0(\gamma)$, and the harmonic
oscillator experiences ultrastrong coupling to its environment. While it is formally possible to define 
the DAMPF Hamiltonian, its microscopic derivation from a physical bath is not immediate and leaves a question mark regarding the validity of these considerations. Fortunately, as we have seen in the main text, these issues do not arise for the finite temperature case.

\subsection{Increasing power at constant efficiency} We proceed to find a parameter scaling that allows to increase the total power of the QHE
\begin{align}
    P^\text{(tot)} = \frac{\Delta W_\text{ext.}^\text{(tot)}}{\tau_\text{cycle}}
\end{align}
while keeping the total efficiency $\eta^\text{(tot)}$ in \cref{eq:eta_tot_with_dissipated_heat}
at a constant level. To this end, the stroke durations of the machine
\begin{align}
\label{eq:condition_stroke_durations}
    \tau_\x^\prime = \frac{1}{\lambda} \tau_\x
\end{align}
are reduced by the factor $\lambda > 1$ for all $\x \in \{\com, \hot, \exp, \cold\}$ resulting in the shorter cycle time $\tau_\text{cycle}^\prime = \lambda^{-1} \tau_\text{cycle}$. At the same time, we keep the heat exchange between system and thermal bath $\Delta Q^{\text{(sys)}}_\thermal$ constant by ensuring that the total thermalisation of the work medium, characterised by
\begin{align}
\label{eq:condition_gamma_thermal}
    \gamma_\thermal^\prime \, \tau_\thermal^\prime = \text{const},
\end{align}
does not change, i.e. the coupling to the thermal baths increases as $\gamma_\thermal^\prime = \lambda \gamma_\thermal$. Next, we consider the conditions that have to be fulfilled in order to keep $\Delta W^{\text{(tot)}}_\text{ext.}$ constant. In addition to condition \cref{eq:condition_gamma_thermal}, that fixes the initial population before compression and expansion, one has to consider that faster compression and expansion strokes lead to stronger quantum friction in the work medium, that has to be suppressed with stronger dephasing noise. The extracted work per cycle does not change if a bigger effective dephasing rate $\Geff^\prime = \lambda \Geff$ compensates for the faster strokes such that
\begin{align}
\label{eq:condition_Geff_B}
    \Geff^\prime \, \tau_\x^\prime = \text{const},
\end{align}
as the effective dephasing rate determines the ability of the machine to suppress quantum friction. We explicitly focus on the Lorentzian spectrum as this can be simulated easily. The corresponding effective dephasing rate from \cref{eq:Gamma_eff} scales as
\begin{align}
\label{eq:Geff_scaling}
    \Geff^\prime = \frac{8\Gamma^{\prime2}\,\gamma^\prime}{\gamma^{\prime2} + 4\omega_0^{\prime2}} \propto \lambda
\end{align}
according to \cref{eq:condition_stroke_durations,eq:condition_Geff_B}. We encode this scaling in the three free parameters via
\begin{align}
\label{eq:condition_Gamma}
    \Gamma^\prime = \sqrt{\lambda} \Gamma
\end{align}
\begin{figure*}[t]
    \centering
    \includegraphics[width=1\textwidth]{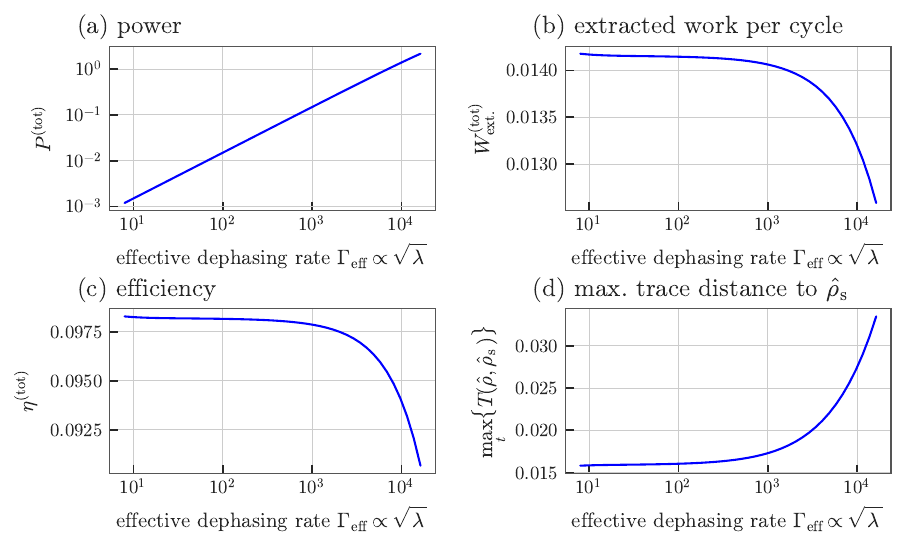}
    \caption{In the main text, we introduce a scaling factor $\lambda$ to increase the power of the QHE while maintaining its efficiency at a constant value near the maximum possible efficiency. Here, we present the simulation results for the parameters given in the main text. (a) By reducing the cycle time of the QHE to $\tau_\text{cycle}^\prime = \tau_\text{cycle}/\lambda$, we effectively increase its power. (b) We ensure a constant amount of extracted work per cycle by a proper rescaling of all the other parameters as explained in the main text. A constant amount of extracted work is necessary to ensure the linear power increase in (a). (c) By suppressing the dissipated heat, we are able to run the thermal heat engine near the optimal efficiency $\eta_\text{max}^\text{(sys)} = 0.106$. (d) However, for large scaling factors $\gamma$, the analytical description of the state from \cref{eq:ansatz_steady_state} begins to deviate from the actual state in the simulations. In this regime, the extracted work per cycle and the efficiency begin to decline.}
    \label{fig:increase_power}
\end{figure*}
which leaves the condition
\begin{align}
    \frac{\gamma^\prime}{\gamma^{\prime2} + 4\omega_0^{\prime2}} = \frac{\gamma}{\gamma^{2} + 4\omega_0^{2}} = \text{const.}
\end{align}
for $\gamma^\prime$ and $\omega^\prime_0$. We eliminate one variable by solving for $\omega_0^\prime$
\begin{align}
\label{eq:omega_of_gamma}
    \omega_0^\prime(\gamma^\prime) = \frac{1}{2}\sqrt{\gamma^\prime \left(\gamma_0 - \gamma^\prime\right)}
\end{align}
with
\begin{align}
    \gamma_0 = \frac{4\omega_0^2+\gamma^2}{\gamma} = 32896
\end{align}
which is equivalent to \cref{eq:condition_Geff_omega,eq:condition_Geff_gamma_0}. Finally, we aim for the scaling of the variable $\gamma^\prime$ by enforcing constant dissipated heat during the hot thermalisation
\begin{align}
\label{eq:condition_dQ_diss}
    \Delta Q_\hot^\text{(diss)} \propto \Geff^\prime \frac{\omega_0^\prime(\gamma^\prime)}{\gamma^\prime} \gamma_\hot^\prime \tau_\hot^\prime = \text{const.}
\end{align}
which is obtained from \cref{eq:diss_heat_estimation} by using the explicit expression of the effective dephasing rate \cref{eq:Gamma_eff}. This expression reduces to
\begin{align}
\label{eq:condition_Q_diss_constant}
    \lambda \sqrt{\frac{\gamma_0 - \gamma^\prime}{\gamma^\prime}} = \text{const}
\end{align}
with the help of \cref{eq:condition_gamma_thermal,eq:Geff_scaling,eq:omega_of_gamma}. We introduce the new variable $\Delta \gamma = \gamma_0 - \gamma$ and we notice that the dissipated heat from \cref{eq:condition_dQ_diss} vanishes in the limit $\Delta \gamma \rightarrow 0$ as $\omega_0^\prime(\gamma^\prime) \rightarrow 0 \, (\gamma^\prime \rightarrow 0)$. We are in particular interested in the regime $\Delta \gamma \ll \gamma_0$ and expand \cref{eq:condition_Q_diss_constant} to the lowest order to obtain
\begin{align}
    \sqrt{\frac{\gamma_0 - \gamma^\prime}{\gamma^\prime}} = \sqrt{\frac{\Delta \gamma^\prime}{\gamma_0}}\left(1 + \mathcal{O}\left(\frac{\Delta \gamma^\prime}{\gamma_0}\right)\right),
\end{align}
which implies the scaling
\begin{align}
    \Delta \gamma^\prime = \frac{1}{\lambda^2} \Delta \gamma.
\end{align}

For the simulations, we use the same parameters as in \cref{sec:increasing_efficiency_at_constant_power} and fix the damping rate $\gamma$ to a value close to the critical $\gamma_0$ by setting $\gamma = \gamma_0 - \Delta \gamma$ where $\Delta \gamma = 2^{-8}$. The choice of $\gamma$ automatically determines the energy gap of the harmonic mode through the function $\omega_0(\gamma)$. Now, we rescale all parameters with the scaling factor $\lambda$ according to the previously derived rules. The results of the simulations are shown in \cref{fig:increase_power}. 

We observe that the extracted work per cycle and the efficiency of the QHE remain nearly constant over a large range of the scaling parameter $\lambda$, covering more than two orders of magnitude. Consequently, the power of the engine scales linearly with $\lambda$, given that the cycle time is inversely proportional to $\lambda$. However, as the speed of the engine and the effective dephasing rate increase, the accuracy of the analytical calculations decreases. We quantify this by computing the maximum trace distance between the analytically calculated state from \cref{eq:ansatz_steady_state} and the numerical results during a full cycle at steady state operation. In this regime, the performance of the QHE also diminishes slightly, as the extracted work per cycle drops slightly. The trend of the trace distance suggests that the QHE's performance will deteriorate further if larger values of $\lambda$ are considered. As a result the scaling 
relations used here, which are based on our analytical Ansatz for the state of the QHE, will need to be adapted to achieve full
convergence to the quasi-static efficiency of the QHE. Nonetheless, by utilising a dephasing bath, we manage to increase the power 
of the QHE by a factor of $~ 10^3$ compared to the maximum power of a QHE without dephasing noise, $P_\text{max}^\text{(tot)} \approx 0.001$, as shown in \cref{fig:P_eta_tot}. Moreover, the right choice of parameters of the dephasing bath allows for operation close to the maximum efficiency $\eta_\text{max}^\text{(sys)} = 0.106$ from \cref{eq:eta_max_sys}.

\end{document}